%% file: Semi-Third-Revision.tex
\documentclass{statsoc}
\usepackage[a4paper]{geometry}
\usepackage{amsmath, amssymb}
\usepackage{amsfonts, multirow, epsfig, subfig, floatrow}
\usepackage{graphicx, pdflscape, verbatim, enumerate, colortbl, setspace}
\usepackage{setspace, color,bm}
\usepackage[normalem]{ulem}
\usepackage{cite}
\usepackage{multirow}
\usepackage{booktabs,array}

 \usepackage[linesnumbered,ruled]{algorithm2e}

\newtheorem{Corollary}{Corollary}

\newtheorem{Lemma}{Lemma}

\newtheorem{Theorem}{Theorem}
 
\newtheorem{Remark}{Remark}

\newcommand{\argmin}{\mathop{\rm arg\min}}

\newcommand{\Md}{M_1}
\newcommand{\Mv}{M_2}

\newcommand{\R}{\mathbb{R}}

\newcommand{\E}{{\mathbb{E}}}
\newcommand{\PP}{{\mathbb{P}}}
\newcommand{\RR}{{\mathbf{R}}}
\newcommand{\LL}{{\mathbf{L}}}

\newcommand{\HH}{{\mathcal{H}}}

\newcommand{\TV}{L_1}
\newcommand{\T}{{\rm Q}}

\newcommand{\suppx}{{\rm supp}(x)}

\newcommand{\xa}{X^{\left(1\right)}}
\newcommand{\xb}{X^{\left(2\right)}}

\newcommand{\ya}{y^{\left(1\right)}}

\newcommand{\W}{X}
\newcommand{\Q}{{\rm Q}}
\newcommand{\Sigmahalf}{\Sigma^{\frac{1}{2}}}

\newcommand{\eb}{\epsilon^{\left(2\right)}}

\newcommand{\na}{n_1}
\newcommand{\nb}{n_2}

\newcommand{\thetab}{{{\theta}}}

\newcommand{\cip}{\overset{p}{\to}}
\newcommand{\cid}{\overset{d}{\to}}
\newcommand{\hprob}{1-\gamma(n)-C(p^{-c}+\exp(-c{N})+e^{-ct^2})}
\newcommand{\hprobsup}{1-\gamma(n)-C(p^{-c}+\exp(-ct^2))}
\newcommand{\lprobsup}{C(p^{-c}+\exp(-c{N})+e^{-ct^2})+\gamma(n)}
\newcommand{\lprob}{C(p^{-c}+e^{-ct^2})+\gamma(n)}

\newcommand{\SigmaLa}{\frac{1}{n}\sum_{i=1}^{n} X_{i\cdot} X_{i\cdot}^{\intercal}}
\newcommand{\SNR}{{\rm SNR}}
\def\limsup{\mathop{\overline{\rm lim}}}
\def\liminf{\mathop{\underline{\rm lim}}}

\title[Semi-supervised Inference for Explained Variance]{{Semi-supervised Inference for Explained Variance in High-dimensional Linear Regression and Its Applications}}

\author{T. Tony Cai}
\address{University of Pennsylvania, Philadelphia, USA}
\author[Cai $\&$ Guo]{Zijian Guo}
\address{Rutgers University, Piscataway, USA}

\begin{document}

\maketitle

\begin{abstract}
This paper considers statistical inference for the explained variance $\beta^{\intercal}\Sigma \beta$ under the  high-dimensional linear model $Y=X\beta+\epsilon$ in the semi-supervised setting, where $\beta$ is the regression vector and $\Sigma$ is the design covariance matrix. A calibrated estimator, which efficiently integrates both labelled and unlabelled data, is proposed. It is shown that the estimator achieves the minimax optimal  rate of convergence in the general semi-supervised framework. The optimality result characterizes how the unlabelled data contributes to the estimation accuracy. Moreover, the limiting distribution for the proposed estimator is established and the unlabelled data has also proven useful in reducing the length of the confidence interval for the explained variance. The proposed method is extended to the semi-supervised inference for the unweighted quadratic functional, $\|\beta\|_2^2$. 
The obtained inference results are then applied to a range of high-dimensional statistical problems, including signal detection and global testing, prediction accuracy evaluation, and confidence ball construction. The numerical improvement of incorporating the unlabelled data is demonstrated through simulation studies and an analysis of estimating heritability for a yeast segregant data set with multiple traits. 
\end{abstract}


\section{Introduction}
\label{introduction.sec}

High-dimensional linear models are ubiquitous in contemporary statistical modeling with a wide range of applications in many scientific fields. The early focus has been mainly on developing methods for the recovery of the whole regression vector via penalized or constrained $\ell_1$ minimization approaches. Examples include the Lasso \citep{tibshirani1996regression}, Dantzig Selector \citep{candes2007dantzig}, MCP \citep{zhang2010nearly}, square-root Lasso \citep{belloni2011square}, and scaled Lasso \citep{sun2012scaled}. There have been significant recent interests in statistical inference for low-dimensional functionals, including confidence intervals and hypothesis testing for individual regression coefficients \citep{zhang2014confidence,van2014asymptotically,javanmard2013hypothesis,javanmard2014confidence}, minimaxity and adaptivity of  confidence intervals  for general linear functionals \citep{cai2017confidence}, estimation of the signal-to-noise-ratio \citep{verzelen2016adaptive,janson2015eigenprism}, inference for the $\ell_{q}$ accuracy of a given estimator \citep{cai2016accuracy},  and estimation of quadratic functionals \citep{janson2015eigenprism,guo2016optimal}. Motivated by a range of applications, the present paper considers the semi-supervised inference problem in high dimensions, where the main statistical goal is to integrate both the labelled and unlabelled data, and propose efficient point and interval estimators. 


\subsection{Problem Formulation and Motivations}
\label{sec: motivation}

We consider the high-dimensional linear model with a random design,
\begin{equation}
y_{i}=X_{i\cdot}^{\intercal} \beta+\epsilon_{i}, \quad \text{for}\; 1\leq i\leq n
\label{eq: high dim linear model}
\end{equation}
where $y_i\in \R$ and $X_{i\cdot}\in \R^{p}$ denote respectively the outcome and the measured covariates of the $i$-th observation, $\epsilon_i$ denotes the error and  $\beta\in \R^{p}$ denotes the high-dimensional regression vector. 
The covariates $X_{i\cdot} $ are i.i.d. $p$-dimensional random vectors with mean $0$ and covariance matrix $\Sigma$ and the errors $\{\epsilon_i\}_{1\leq i\leq n}$ are i.i.d  random variables with mean $0$ and variance $\sigma^2$ and independent of $\{X_{i\cdot}\}_{1\leq i\leq n}$. 
The explained variance under the regression model \eqref{eq: high dim linear model} is represented by $
\Q={\rm Var}(X_{i\cdot}^{\intercal} \beta)=\beta^{\intercal}\Sigma \beta. 
$
We focus on the semi-supervised setting, where the data is a combination of the labelled data $\{y_{i},X_{i\cdot}\}_{1\leq i\leq n}$ in the regression model \eqref{eq: high dim linear model} and the unlabelled data $\{X_{i\cdot}\}_{n+1\leq i\leq n+N}$. Here the measured covariates of both the labelled and unlabelled data are assumed to be independent and  follow the same distribution.  
The more conventional supervised setting is treated as a special case with no additional unlabelled data.  

The setting of semi-supervised learning is commonly seen in applications where the outcomes are more expensive to collect than the covariates. For example, in the analysis of Electronic Health Records (EHR) databases, the covariates are easy to be automatically extracted while labelling of the outcomes is costly and time-consuming \citep{chakrabortty2017efficient,gronsbell2017semi}. In addition, semi-supervised learning naturally arises in the integrative analysis of multiple (genetics) data sets where the covariates are the same across all data sets but the outcomes measured vary from study to study  due to the specific purposes of individual studies \citep{van2017extending}. This can be naturally formulated as semi-supervised learning, where the pre-specified outcome is only measured over one or several (but not all) data sets while the covariates are measured across all data sets. 


The construction of the optimal estimator and confidence intervals for $\Q= \beta^{\intercal}\Sigma\beta$  in the semi-supervised and high-dimensional setting is not only of significant interest on its own right, but is also closely connected to several other important statistical problems. 

\begin{enumerate}
\item {\bf Heritability.} 
Heritability is among the most important genetics concepts. 
Under the model \eqref{eq: high dim linear model} with the outcome normalized to have unit variance, $\beta^{\intercal}\Sigma \beta$ is a measure of heritability, which quantifies the total variance explained by genetic variants \citep{owen2012quasi,guo2016optimal,janson2015eigenprism,verzelen2016adaptive}.

\item {\bf Signal-to-Noise Ratio (SNR) and Proportion-of-Variance Explained (PVE).} 
SNR and PVE  are important statistics concepts and are defined respectively as ${\beta^{\intercal}\Sigma\beta}/({\beta^{\intercal}\Sigma\beta+\sigma^2})$ and ${\beta^{\intercal}\Sigma\beta}/{\sigma^2}$ under model \eqref{eq: high dim linear model}. 
Together with a good estimator of $\sigma^2$ \citep{
sun2012scaled,belloni2011square}, the results for $\beta^{\intercal}\Sigma \beta$ established in this paper are useful for inference of SNR and PVE.

\item {\bf Signal Detection and Global Testing.} Inference for the explained variance can be applied to testing the global hypothesis $H_0: \beta=\beta^{\rm null}$ for $\beta^{\rm null}\in \R^{p}$, which includes signal detection as a special case with $\beta^{\rm null}={0}$.
The connection is revealed in the adjusted linear model, 
$
y_i-X_{i\cdot}^{\intercal} \beta^{\rm null}=X_{i\cdot}^{\intercal} \left(\beta-\beta^{\rm null}\right)+\epsilon_i$ for $1\leq i\leq n$,
where testing for $H_0: \beta=\beta^{\rm null}$ is recast as testing the hypotheses  $H_0: \left(\beta-\beta^{\rm null}\right)^{\intercal}\Sigma\left(\beta-\beta^{\rm null}\right)= 0$ versus $H_1: \left(\beta-\beta^{\rm null}\right)^{\intercal}\Sigma\left(\beta-\beta^{\rm null}\right) > 0$.


\item {\bf Prediction Accuracy Assessment.} Accuracy assessment is of significant importance in applications. Let $\check{\beta}$ denote a given estimator based on the training data. We define the out-of-sample prediction accuracy for a given observation $x_{\rm new}$ as $\E_{x_{\rm new}} \left(x_{\rm new}^{\intercal}(\check{\beta}-\beta)\right)^2=(\check{\beta}-\beta)^{\intercal}\Sigma(\check{\beta}-\beta)$. We introduce the following adjusted linear model for the independent test data $\{X_{i\cdot},y_{i}\}_{1\leq i\leq n},$ 
\begin{equation}
y_i-X_{i\cdot}^{\intercal}\check{\beta}= X^{\intercal}_{i\cdot}\left(\beta-\check{\beta}\right)+\epsilon_i \quad \text{for}\; 1\leq i\leq n.
\label{eq: connection to accuracy assessment}
\end{equation}
Inference results developed for the explained variance can be applied to \eqref{eq: connection to accuracy assessment} to obtain the corresponding results for the prediction accuracy $\E_{x_{\rm new}} \left(x_{\rm new}^{\intercal}(\check{\beta}-\beta)\right)^2$.

\item {\bf Confidence Ball for $\beta$.} Construction of confidence balls for $\beta$ is another important application. Based on \eqref{eq: connection to accuracy assessment}, a confidence interval $\left(L(Z), U(Z)\right)$ for $(\check{\beta}-\beta)^{\intercal}\Sigma(\check{\beta}-\beta)$ leads to a confidence ball for $\beta$ centering at $\check{\beta}$, 
$
\left\{\beta: \|\beta-\check{\beta}\|_2^2\leq U(Z)/\lambda_{\min}(\Sigma)\right\},
$
where $\lambda_{\min}(\Sigma)$ denotes the smallest eigenvalue of $\Sigma$. 
\end{enumerate}
More detailed discussions about these statistical applications are present in Section \ref{sec: stat applications}.
%
\subsection{Results and Contributions}
A central question in semi-supervised learning is {\em how to efficiently use both labelled and unlabelled data} \citep{chakrabortty2017efficient,gronsbell2017semi}. We introduce a novel two-step estimator, Calibrated High-dimensional Inference for Variance Explained (CHIVE), where the first step is to plug in the estimators of $\beta$ and $\Sigma$, denoted by $\widehat{\beta}$ and $\widehat{\Sigma}$, respectively,  and the second step is to calibrate this plug-in estimator $\widehat{\beta}^{\intercal}\widehat{\Sigma}\widehat{\beta}$ through estimating a dominating term in its error decomposition. The second step is  to rebalance the bias and variance and improve the estimation accuracy. Different forms of $\widehat{\beta}$ and $\widehat{\Sigma}$ can be taken as inputs of CHIVE method and this flexibility is useful in integrating the unlabelled data to estimate $\Sigma$ more accurately. This idea is then extended to semi-supervised inference for the unweighted quadratic functional $\|\beta\|_2^2$, where the additional unlabelled data facilitates the estimation of $\Sigma^{-1}$. 

Another important question is whether the unlabelled data has been efficiently utilized in semi-supervised learning. We address this question by establishing the minimax optimal rate of convergence for estimating $\beta^{\intercal}\Sigma\beta$, where the optimal rate is ${M}/{\sqrt{n}}+{M^2}/{\sqrt{N+n}}+{k\log p}/{n},$ with $p$, $n$, $N$, $k$, and $M$  denoting respectively the dimension, the size of the labelled data,   the size of the unlabelled data, the sparsity, and the $\ell_2$ norm of $\beta$. The proposed CHIVE estimator achieves this optimal rate, which justifies the efficient use of the unlabelled data. The optimal rate is not just achieved for the case where there is a large amount of unlabelled data, but is also for any given amount of unlabelled data. The minimax optimal rate characterizes the fundamental difficulty of the inference problem in the semi-supervised setting and is independent of specific procedures. This minimax rate also reveals that  the unlabelled data is most effective when the signal strength $\|\beta\|_2$ is large.

We establish the limiting distribution of the CHIVE estimator  and construct data-driven confidence intervals for $\beta^{\intercal}\Sigma\beta$ based on this estimator. The limiting distribution is normal and its variance is scaled to the proportion of the labelled data, which is unique to the semi-supervised setting. A larger amount of unlabelled data leads to a smaller proportion of the labelled data and hence a smaller asymptotic variance, which leads to a shorter confidence interval for  $\beta^{\intercal}\Sigma\beta$. The effect of the unlabelled data is also demonstrated in the numerical studies. Specifically, in comparison with the estimators based only on the labelled data, the RMSE for estimation and the length of confidence intervals can be reduced by as much as 70\%. See details in Section \ref{sec: sim}.

The improvement in semi-supervised inference for $\|\beta\|_2^2$ is similar to that for $\beta^{\intercal}\Sigma\beta$ at a high level but different in technical details. Specifically, the estimation accuracy is significantly improved in the strong signal regime, and the improvement is limited if the signal strength $\|\beta\|_2^2$ is not large enough. Construction of confidence intervals for $\|\beta\|_2^2$ also gets easier in the sense that the condition for sample size and model complexity is weakened by making use of the unlabelled data.

The inference results obtained in this paper are applied to (i) signal detection and global testing, (ii) prediction accuracy evaluation, and (iii) confidence ball construction. For signal detection, we control the type I error and characterize the type II error by establishing the power function under a local alternative. The results can be easily extended to the general global testing problem. For evaluation of out-of-sample prediction accuracy of a given sparse estimator of $\beta$,  both the point and interval estimators are developed. We establish the estimation error bound for the point estimator of the prediction accuracy and control the length of the corresponding confidence interval.  A confidence ball for the regression vector $\beta$ with controlled radius is also constructed. We stress that these  procedures are data-driven and do not require a priori knowledge of the design covariance matrix $\Sigma$ or the noise level $\sigma$. See more details in Section \ref{sec: stat applications}. 




\subsection{Related Work}
\label{sec: related work}

Estimation and inference for quadratic functionals have been studied in the literature in a range of settings. In particular, minimax and adaptive estimation of quadratic functionals plays an important role in  nonparametric inference and has been well studied in density estimation, nonparametric regression, and white noise with drift model. See, for example, \citet{Bickel1988,donoho1990minimax,efromovich1996optimal,laurent2000adaptive,cai2005nonquadratic,cai2006optimal,collier2015minimax}. 

The most related works to the current paper are \citet{verzelen2016adaptive} and \citet{guo2016optimal}, 
which considered estimation of $\beta^{\intercal}\Sigma\beta/\sigma^2$ and $\|\beta\|_2^2$, respectively, in high-dimensional linear regression. The main difference between the current paper and these two related works are two-fold: (a) \citet{verzelen2016adaptive} and \citet{guo2016optimal} only considered the supervised setting instead of the semi-supervised setting. As demonstrated in both theoretical and numerical justifications, a careful integration of the unlabelled data proves useful in improving the estimation accuracy and reducing the length of constructed confidence intervals.  (b) The focus of \citet{verzelen2016adaptive} and \citet{guo2016optimal} is about point estimation while the current paper studies the more challenging problem of uncertainty quantification and also related hypothesis testing, in addition to point estimation. As is well known, uncertainty quantification in high dimensions is significantly different from and more involved than point estimation \citep{nickl2013confidence,cai2017confidence}. 

Another related paper, \citet{janson2015eigenprism}, studied the construction of confidence intervals for $\|\beta\|_2^2$ in the setting of $\Sigma={\rm I}$, moderate dimension where $n/p\rightarrow \xi \in (0,1)$ and no sparsity assumption on $\beta$. 
The inference problem  considered in the current paper is significantly different from the setting considered in \citet{janson2015eigenprism}, mainly due to the complicated geometry induced by the sparsity structure and the unknown design covariance matrix $\Sigma$.  Other works related to quadratic functional inference include construction of confidence intervals for the $\ell_{2}$ loss of the estimator considered in \citet{cai2016accuracy}. 
In addition, \citet{javanmard2017flexible,zhu2017projection} considered hypothesis testing for high-dimensional linear regression.  As another significant difference, the current paper studies how to efficiently integrate the labelled and unlabelled data in the general semi-supervised setting while all the aforementioned works  solely focused on the supervised regression. 
 
The statistical applications studied in this paper have also been considered separately in the literature. Signal detection was studied in \citet{ingster2010detection,arias2011global}  under the linear model \eqref{eq: high dim linear model} in a special setting where the design covariance matrix $\Sigma$ is equal to or closed to the identity matrix. In this setting, \citet{ingster2010detection,arias2011global} established optimal signal detection method and theory.  The results established in the present paper enable the study of signal detection under a general setting where the design covariance matrix $\Sigma$ is unknown. 
The confidence ball construction for the whole regression vector was considered in \citet{nickl2013confidence} in the case of known $\sigma$ and the optimal size and possibility of adaptive confidence balls was also established. The  results obtained in the current paper lead to a confidence ball construction for $\beta$ in the case of unknown $\sigma$. A problem related to prediction accuracy is inference for the estimation accuracy, which was considered in \citet{cai2016accuracy,janson2015eigenprism}. However, inference for the prediction accuracy and that for the estimation accuracy are different problems.

\subsection{Organization of the Paper}

The rest of the paper is organized as follows. In Section \ref{sec: est}, we introduce in detail the CHIVE estimator and establish its minimax rate optimality in the semi-supervised setting. Section \ref{sec: ci} focuses on quantifying the uncertainty of the CHIVE estimator and construction of the confidence intervals for $\beta^{\intercal}\Sigma\beta$. In Section \ref{sec: semi-infer general}, we extend the methodology to the semi-supervised inference for $\|\beta\|_2^2$. We apply in Section \ref{sec: stat applications} the developed procedures to tackle three important problems, signal detection and global testing, prediction accuracy evaluation and confidence ball construction. Simulation results are given in Section \ref{sec: sim} to illustrate the numerical improvement through incorporating the unlabelled data. An analysis of a yeast data set is presented in Section \ref{sec: real data}. A discussion is provided in Section \ref{sec: discussion}. The proofs and the additional simulation results are presented in the appendix.

\section{Semi-supervised Estimation of $\beta^{\intercal}\Sigma\beta$}
\label{sec: est}

In this section, we first introduce the calibration methodology for estimating the explained variance in the general semi-supervised framework and then establish the minimax convergence rate of estimating $\beta^{\intercal}\Sigma\beta$. A significant statistical gain is obtained by carefully integrating the unlabelled data and the proposed estimator is shown to achieve the optimal rate in the semi-supervised setting.
The supervised setting and the setting with known design covariance matrix are then discussed as special cases. We begin with the notation that will be used in the rest of the paper. 

For a matrix $A$, $A_{i\cdot}$,  $A_{\cdot j}$, and $A_{i,j}$ denote respectively the $i$-th row,  $j$-th column, and  $(i,j)$ entry of the matrix $A$. The spectral norm of $A$ is $\|A\|_{2}=\sup_{\|x\|_2 = 1} \|Ax\|_2$ and the matrix $\ell_1$ norm is $\|A\|_{L_1}=\sup_{1\leq j\leq p}\sum_{i=1}^{p}|A_{ij}|$. 
For a symmetric matrix $A$, $\lambda_{\min}\left(A\right)$ and $\lambda_{\max}\left(A\right)$  denote respectively the smallest and largest eigenvalue of $A$. 
 For a set $S$, $\left|S\right|$ denotes the cardinality of $S$. 
For a vector $x\in \R^{p}$, 
$\suppx$ denotes the support of $x$ and
the $\ell_q$ norm of $x$ is defined as $\|x\|_{q}=\left(\sum_{i=1}^{p}|x_i|^q\right)^{\frac{1}{q}}$ for $q \geq 0$ with $\|x\|_0=|\suppx|$ and $\|x\|_{\infty}=\max_{1\leq j \leq p}|x_j|$. 
For $a\in \R$, $a_{+}=\max\left\{a,0\right\}$. 
 We use $c$ and $C$ to denote generic positive constants that may vary from place to place. For a sequence of random variables $X_n$ indexed by $n$, we use $X_n \cip X$ and $X_{n} \cid X$ to represent that $X_n$ converges to $X$ in probability and in distribution, respectively. For a sequence of random variables $X_n$ and numbers $a_n$, we define $X_{n}=o_{p}(a_n)$ if $X_n/a_n$ converges to zero in probability. For two positive sequences $a_n$ and $b_n$,  $a_n \lesssim b_n$ means $a_n \leq C b_n$ for all $n$ and $a_n \gtrsim b_n $ if $b_n\lesssim  a_n$ and $a_n \asymp b_n $ if $a_n \lesssim b_n$ and $b_n \lesssim a_n$, and $a_n \ll b_n$ if $\limsup_{n\rightarrow\infty} \frac{a_n}{b_n}=0$ and $a_n \gg b_n$ if $b_n \ll a_n$. We define the signal-to-noise ratio ($\SNR$) in the context of model \eqref{eq: high dim linear model} as
$ \SNR=\frac{1}{\sigma} \sqrt{\beta^{\intercal}\Sigma\beta}.$

\subsection{Calibration of Plug-in Estimators}
\label{sec: calibration}
In semi-supervised learning, we observe the labelled data $(X_{1\cdot},y_1),\cdots, (X_{n\cdot},y_n)$ and the unlabelled data $\W_{n+1\cdot},,\cdots, \W_{n+N\cdot}$, where $X_{1\cdot},\cdots,X_{n\cdot},\W_{n+1\cdot},,\cdots,\W_{n+N\cdot}$ are i.i.d realizations of $p$-dimensional covariates. 
We use $\widehat{\beta}$ and $\widehat{\Sigma}$ to denote some estimators of $\beta$ and $\Sigma$, which will be specified later. A preliminary estimator of the quadratic functional $\Q=\beta^{\intercal}\Sigma\beta$ is the plug-in estimator $\widehat{\beta}^{\intercal} \widehat{\Sigma}\widehat{\beta}$, which
 has the following error decomposition, 
\begin{equation}
\widehat{\beta}^{\intercal} \widehat{\Sigma}\widehat{\beta}-\beta^{\intercal}\Sigma \beta=2\widehat{\beta}^{\intercal}\widehat{\Sigma}(\widehat{\beta}-\beta) -(\widehat{\beta}-\beta)^{\intercal}\widehat{\Sigma}(\widehat{\beta}-\beta)+\beta^{\intercal}(\widehat{\Sigma}-\Sigma)\beta.
\label{eq: key decomposition}
\end{equation}
Since the first term $2\widehat{\beta}^{\intercal}\widehat{\Sigma}(\widehat{\beta}-\beta)$ on the right hand side can be estimated in a data-dependent way, the corresponding estimation error of the preliminary estimator $\widehat{\beta}^{\intercal} \widehat{\Sigma}\widehat{\beta}$ can be further reduced. We estimate the term $2\widehat{\beta}^{\intercal}\widehat{\Sigma}(\widehat{\beta}-\beta)$ by $-2\widehat{\beta}^{\intercal}\frac{1}{n}\sum_{i=1}^{n}X_{i\cdot}(y_i-X_{i\cdot}\widehat{\beta})$ and propose the following calibrated estimator, 
\begin{equation}
{\widehat{\Q}}(\widehat{\beta},\widehat{\Sigma})=\widehat{\beta}^{\intercal} \widehat{\Sigma}\widehat{\beta}+2\widehat{\beta}^{\intercal}\frac{1}{n}\sum_{i=1}^{n}X_{i\cdot}(y_i-X_{i\cdot}\widehat{\beta}).
\label{eq: EST}
\end{equation}
This estimator is referred to as  the Calibrated High-dimensional Inference for Variance Explained (CHIVE) estimator.
The calibration step in \eqref{eq: EST} is essentially to improve the plug-in estimator $\widehat{\beta}^{\intercal}\widehat{\Sigma}\widehat{\beta}$ through re-balancing the bias and variance.

 The CHIVE estimator requires three inputs, the initial estimators $\widehat{\beta}$ and $\widehat{\Sigma}$ and the data $(X, y)$. With this machinery, we have the flexibility of choosing the initial estimators $\widehat{\beta}$ (and also $\widehat{\sigma}^2$) and $\widehat{\Sigma}$ based on the observed data. We begin with the estimator for $\beta$ and $\sigma^2$ and then move on to the estimator for $\Sigma$.
Throughout the paper, we assume that the estimators $\widehat \beta$ and $\widehat \sigma^2$ satisfy the following conditions. 
\begin{enumerate}
	{ \item[(B1)] With probability larger than $1-\gamma(n)$ where $\gamma(n)\rightarrow 0$, 
	 the estimator $\widehat{\beta}$ satisfies 
	{\small
	\begin{equation*}
	\max\{{\frac{1}{{n}}\sum_{i=1}^{n}[X_{i\cdot}^{\intercal}(\widehat{\beta}-\beta)]^2}, \|\widehat{\beta}-\beta\|^2_2\} \lesssim {\frac{k \log p}{n}}\sigma,\; 	\|(\widehat{\beta}-\beta)_{S^{c}}\|_1 \leq C_0 \|(\widehat{\beta}-\beta)_{S}\|_1
	\end{equation*}}
where $S={\rm supp}(\beta)$ and $C_0>0$ is some positive constant. 
	\item[(B2)] $\widehat{\sigma}^2$ is a consistent estimator of $\sigma^2$, that is, $\left|{\widehat{\sigma}^2}/{\sigma^2}-1\right|\cip 0$.
	}
\end{enumerate}
One of the key assumptions for the general penalized estimators satisfying {\rm (B1)} and  {\rm (B2)} is the following restricted eigenvalue condition on the population covariance matrix $\Sigma$, 
\begin{equation*}
\kappa(k,C_0,\Sigma)=\min_{S \in \{1,\cdots,p\}, \, |S|\leq k} \; \min_{v\neq 0, \|v_{S^{c}}\|_1\leq C_0 \|v_{S}\|_1 } \frac{\|\Sigmahalf v\|_2}{\|v_{S}\|_2}\geq c,
\label{eq: RE}
\end{equation*}
for some positive constant $c>0$. 
This population version restricted eigenvalue condition implies the sample version restricted eigenvalue condition introduced in \citet{bickel2009simultaneous}, under the assumption that the covariates $X_{i\cdot}$ are in a certain broad family of sub-gaussian random vectors and the sparsity $k$ satisfies $k\lesssim n/\log p$;
See \citet{zhou2009restricted,raskutti2010restricted} for the exact statement. \\
\noindent {\bf Estimators satisfying {\rm (B1)} and {\rm (B2)}.} The scaled lasso estimator $\{\widehat{\beta},\widehat{\sigma}\}$ defined by\begin{equation}
\{\widehat{\beta},\widehat{\sigma}\}=\arg\min_{\beta \in \R^{p},\sigma
	\in \R^{+}}\frac{\|y-X\beta\|_2^2}{2n\sigma}+\frac{\sigma}{2}+ \sqrt{\frac{2.01\log p}{{n}}} \sum_{j=1}^{p} \frac{\|X_{\cdot j}\|_2}{\sqrt{n}} |\beta_j| 
\label{eq: scaled Lassoa complete}
\end{equation} 
has been shown in \citet{sun2012scaled} to satisfy (B1) and (B2) under regularity conditions.
See also Lemma 1 in \citet{guo2016optimal} for more details. 
Since the square root lasso estimator \citep{belloni2011square} is numerically the same with the scaled Lasso estimator, the square root lasso estimators of $\beta$ and $\sigma$ also satisfy {\rm (B1)} and {\rm (B2)}.
In addition, with a prior knowledge of $\sigma$, the Lasso estimator of $\beta$ and other variants are also shown to satisfy the above condition ${\rm (B1)}$; see \citet{candes2007dantzig,zhang2010nearly,ye2010rate} for more details. 

Now, we turn to the estimators of $\Sigma$. This is exactly the place where we make use of the unlabelled data. 
Specifically, we pool the information contained in both the labelled and unlabelled data and estimate $\Sigma$ by $\widehat{\Sigma}^{S}=\frac{1}{n+N}\sum_{i=1}^{n+N}X_{i\cdot}X_{i\cdot}^{\intercal}$. Then we use $\widehat{\beta}$ and $\widehat{\Sigma}^{S}$ as inputs and utilize the calibration idea introduced in \eqref{eq: EST}, 
\begin{equation}
{\widehat{\Q}}(\widehat{\beta},\widehat{\Sigma}^{S})=\widehat{\beta}^{\intercal} \widehat{\Sigma}^{S}\widehat{\beta}+2\widehat{\beta}^{\intercal}\frac{1}{n}\sum_{i=1}^{n}X_{i\cdot}(y_i-X_{i\cdot}^{\intercal}\widehat{\beta}).
\label{eq: EST semi}
\end{equation}
When there is no confusion, we use ${\widehat{\Q}}$ to denote the estimator proposed in \eqref{eq: EST semi}.
We introduce the following regularity conditions and then establish 
 the convergence rate of the proposed estimator in \eqref{eq: EST semi} in Theorem \ref{thm: est bound semi}. 

\begin{enumerate}
\item[(A1)] The regression vector $\beta$ is assumed to be $k$-sparse; The errors $\{\epsilon_i\}_{1\leq i\leq n}$ are independent of $\{X_{i\cdot}\}_{1\leq i\leq n+N}$ and follow i.i.d sub-gaussian random variable with mean zero and variance $\sigma^2$; The rows $X_{i\cdot}$ are i.i.d. $p$-dimensional random vectors and can be expressed in the form of $X_{i\cdot}=\Sigmahalf Z_{i\cdot}$ where $Z_{i\cdot}\in \RR^{p}$ is a subgaussian random vector of mean $0$ and identity covariance matrix and $\Sigma$ has a bounded restricted largest eigenvalue $\rho_{\rm max}(k,\Sigma)$, which is defined as 
$
\rho_{\rm max}(k,\Sigma)=\max_{\|v\|_2=1,\, \|v\|_0\leq k} v^{\intercal}\Sigma v.
$

	    \item[(A2)] {{$\sqrt{\E\left({\beta^{\intercal}X_{1\cdot} X_{1\cdot}^{\intercal}\beta}-\beta^{\intercal}\Sigma\beta\right)^2}\geq c_0 \beta^{\intercal}\Sigma\beta$.} for some positive constant $c_0>0.$} 
\end{enumerate}

%
%
%

Assumption (A1) requires that the restricted largest eigenvalue $\rho_{\rm max}(k,\Sigma)$ is upper bounded,  where the ``restricted" here means that the maximum in the definition of $\rho_{\rm max}(k,\Sigma)$ is taken with respect to $k$-sparse vectors.  Note that the restricted (smallest) eigenvalue condition is not required for the theoretical analysis of the proposed estimator $\widehat{\Q}$ as long as the estimator $\widehat{\beta}$ of $\beta$ satisfies the condition ${\rm (B1)}$. 
Define $U=X_{i\cdot}^{\intercal}\beta/\sqrt{\beta^{\intercal}\Sigma\beta}$, where $\E(U)=0$ and $\E(U^2)=1$. Assumption (A2) is placed on this random variable $U$ such that ${\rm Var}(U^2)$ is not vanishing. This assumption is imposed such that ${\rm Var}(U^2)$ can be well estimated and this type of assumption has been introduced in covariance matrix estimation literature \citep{cai2011adaptive} for the same purpose.

\begin{Theorem} Suppose that Condition {\rm (A1)} holds and $k\leq c n/\log p$ for some constant $c>0$. For any estimator $\widehat{\beta}$ satisfying  Condition  {\rm (B1)}, with probability at least $\hprob$, the estimator $\widehat{\Q}={\widehat{\Q}}(\widehat{\beta},\widehat{\Sigma}^{S})$ defined in \eqref{eq: EST semi} satisfies
	\begin{equation}
	\left|{\widehat{\Q}}-\Q\right|\lesssim t \frac{\sigma\|\Sigmahalf \beta\|_2}{\sqrt{n}}+t \frac{\beta^{\intercal}\Sigma \beta}{\sqrt{N+n}}+(1+\frac{\|\Sigmahalf \beta\|_2}{\sigma} \frac{N}{n+N})\frac{k \log p}{n}\sigma^2.
	\label{eq: convergence rate semi}
	\end{equation}
Under the additional assumptions $k \ll {\sqrt{n}}/{\log p}$ and $\SNR\gg {k \log p}/{\sqrt{n}},$
	\begin{equation}
	\frac{\sqrt{n}\left({\widehat{\Q}}-\Q\right)}{\sqrt{4\sigma^2\beta^{\intercal}\Sigma\beta+\rho\E\left(\beta^{\intercal}X_{1\cdot} X_{1\cdot}^{\intercal}\beta-\beta^{\intercal}\Sigma\beta\right)^2}}\cid N(0,1)
	\label{eq: distribution semi}
	\end{equation}
where $\rho=\lim_{n\rightarrow \infty} \frac{n}{N+n}.$	
	\label{thm: est bound semi}
\end{Theorem}
As a remark, the probability $\hprob$ holds for the finite sample $n$ and finite dimension $p$ and also any non-negative constant $t\geq 0$. However, the established result is more interesting over the regime $\min\{p,n\}\rightarrow \infty$ and $t\rightarrow \infty$ as in this scenario, the corresponding probability $\hprob$ approaches 1.
Since $\Q\geq 0$, the convergence rate \eqref{eq: convergence rate semi} also holds for $\widehat{\Q}_{+}$, the positive part of $\widehat{\Q}$. To keep the notation simpler, we only present the results for $\widehat{\Q}$ in this paper. 

The rate of convergence in \eqref{eq: convergence rate semi} reveals the effect of the unlabelled data.  The sample size of the unlabelled data, $N$, appears only in the second term  $t \frac{\beta^{\intercal}\Sigma \beta}{\sqrt{N+n}}$. An interesting observation is that the usefulness of the unlabelled data varies across different signal strengths. If the signal is strong in the sense that ${\rm SNR}\gtrsim \max\{1, {k \log p}/{\sqrt{n}}\}$, in which case the term $t \frac{\beta^{\intercal}\Sigma \beta}{\sqrt{N+n}}$ is  dominant in \eqref{eq: convergence rate semi}, then the additional unlabelled data reduces the estimation error significantly; if the signal is weak in the sense that ${\rm SNR}\ll \max\{1, {k \log p}/{\sqrt{n}}\}$, then the impact of the additional unlabelled data is limited.

To demonstrate the effect of calibration, we note that an upper bound for the term $\widehat{\beta}^{\intercal}\widehat{\Sigma}(\widehat{\beta}-\beta)$ in \eqref{eq: key decomposition} is at the order of magnitude $\sigma\|\Sigmahalf \beta\|_2{\sqrt{{k \log p}/{n}}}$ while the remaining error after the calibration step is $t \frac{\sigma\|\Sigmahalf \beta\|_2}{\sqrt{n}}+(1+\frac{\|\Sigmahalf \beta\|_2}{\sigma} \frac{N}{n+N})\frac{k \log p}{n}\sigma^2$, as shown in \eqref{eq: convergence rate semi}. By comparing these upper bounds, we note that the calibration step is useful in reducing the upper bound for the rate of convergence. This reduction of estimation error has also been numerically demonstrated in Section \ref{sec: comparison}. 
The terms $t \frac{\beta^{\intercal}\Sigma \beta}{\sqrt{N+n}}+\frac{k \log p}{n}\sigma^2$ in \eqref{eq: convergence rate semi} capture the convergence rate of the last two terms in \eqref{eq: key decomposition}.

The distributional result in \eqref{eq: distribution semi} is established under the additional assumptions $k \ll {\sqrt{n}}/{\log p}$ and $\SNR\gg {k \log p}/{\sqrt{n}}$. These additional assumptions are imposed to ensure that the variance component $t \frac{\sigma\|\Sigmahalf \beta\|_2}{\sqrt{n}}+t \frac{\beta^{\intercal}\Sigma \beta}{\sqrt{N+n}}$, captured by the normal limiting distribution after re-scaling, dominates the bias component $(1+\frac{\|\Sigmahalf \beta\|_2}{\sigma} \frac{N}{n+N})\frac{k \log p}{n}\sigma^2.$ 
Since the bias term is hard to characterize, we impose these sufficient conditions such that the variance term is the dominating term. The normal limiting distribution in \eqref{eq: distribution semi} can be  used in Section \ref{sec: ci} to construct confidence intervals for $\beta^{\intercal}\Sigma\beta$. 

Another interesting phenomenon is that the limiting distribution established in \eqref{eq: distribution semi} depends on the proportion of the labelled data, which is unique in the semi-supervised inference problem. If the amount of unlabelled data dominates that of labelled data (that is, $\rho=0$), then the limiting distribution in \eqref{eq: distribution semi} is simplified to 
$\frac{\sqrt{n}\left({\widehat{\Q}}-\Q\right)}{\sqrt{4\sigma^2\beta^{\intercal}\Sigma\beta}}\cid N(0,1).
$ Theorem \ref{thm: est bound semi} demonstrates that the CHIVE estimator integrating the unlabelled data improves the rate of convergence in estimating the explained variance. The lower bound given in the next subsection shows that CHIVE is optimal in terms of the rate of convergence.

\subsection{Optimal Estimation in the Semi-supervised Setting}
\label{sec: semi optimality}
In this section, we further investigate the fundamental limit for estimating $\Q=\beta^{\intercal}\Sigma\beta$ in the general semi-supervised setting over the following specific parameter space, 
{\footnotesize	
$$
	\Theta\left(k,M\right)=\left\{\theta=\left(\beta,\Sigma,\sigma\right):
	\|\beta\|_0\leq k,\; M/2\leq \|\beta\|_2\leq M, \;\frac{1}{\Md} \leq \lambda_{\min}\left(\Sigma\right) \leq \lambda_{\max}\left(\Sigma\right) \leq \Md,\; \sigma \leq \Mv
	\right\},
$$}	
\noindent where $\Md\geq 1$ and $\Mv>0$ are positive constants. Here $k$ quantifies the sparsity of $\beta$ and $M$ quantifies the signal strength of the true signal $\beta$ in terms of its $\ell_2$ norm. Both $k$ and $M$ are allowed to grow with $n$ and $p$. The other conditions ${1}/{\Md} \leq \lambda_{\min}\left(\Sigma \right) \leq \lambda_{\max}\left(\Sigma\right) \leq \Md$ and $\sigma \leq \Mv$ are regularity conditions. 
The following theorem establishes the minimax lower bounds for estimating $\Q$ over the parameter space $\Theta(k,M)$.
\begin{Theorem}
	\label{thm: lower bound}
	Suppose $k\leq c \min\left\{{n/\log p}, p^{\nu}\right\}$ for some constants $c>0$ and $0\leq \nu<{1\over 2}$. Then
	\begin{equation}
	\inf_{\widetilde{\rm Q}}\sup_{\theta\in \Theta\left(k,M\right)} \PP\left(\left|\widetilde{\rm Q}-\Q \right|\gtrsim {M^2\over \sqrt{N+n}}+ \min\left\{{M\over \sqrt{n}} + {k \log p\over n}, M^2\right\} \right) \geq \frac{1}{4}.
	\label{eq: lower bound for semi-supervised}
	\end{equation}
\end{Theorem}
One interesting observation of the above theorem is that only the first term in the lower bound is involved with the amount of the unlabelled data. 
Theorems \ref{thm: est bound semi}  and \ref{thm: lower bound} together show that the estimator proposed in Section \ref{sec: calibration} is  minimax rate optimal under regularity conditions. 
\begin{Corollary} 
Suppose that Condition {\rm (A1)} holds and $k\leq c \min\left\{{n/\log p}, p^{\nu}\right\}$ for some constants $c>0$ and $0\leq \nu<{1\over 2}$. For any estimator $\widehat{\beta}$ satisfying  Condition  {\rm (B1)},
the estimator $\widehat{\Q}$ defined in \eqref{eq: EST semi} is minimax rate optimal over $\Theta(k,M)$ where $\sqrt{k \log p/n}\lesssim M\leq C$ for some constant $C>0$, that is, 
\begin{equation}
	\sup_{\theta \in \Theta(k,M)}\PP\left(\left|{\widehat{\Q}}-\Q\right|\gtrsim t {M^2\over \sqrt{n+N}}+ {M\over \sqrt{n}} + {k \log p\over n}\right)\leq \lprobsup
\label{eq: optimal statement 1}
	\end{equation}

\label{cor: optimal convergence}
\end{Corollary}
The CHIVE estimator attains the optimal convergence rate when the $\ell_2$ norm of $\beta$ is relatively strong, that is, $M$ is bounded away from zero by $\sqrt{{k \log p}/{n}}$. As shown in Theorem \ref{thm: lower bound}, for the case where $M\ll \sqrt{{k \log p}/{n}}$, the lower bound of estimating $\beta^{\intercal}\Sigma\beta$  is $M^2$. This optimal convergence rate can be achieved by a trivial estimator $0$.

In Corollary \ref{cor: optimal convergence}, the lower bound \eqref{eq: lower bound for semi-supervised} is only matched for the regime where $ M\leq C$ for some constant $C>0$. For theoretical interest, we will modify the proposed estimator $\widehat{\Q}$ defined in \eqref{eq: EST semi} such that the modified version achieves the lower bound \eqref{eq: lower bound for semi-supervised} over the regime $M\gtrsim \sqrt{k \log p/n}$. We randomly split the data $\left(y,X\right)$ into two subsamples $\left(\ya,\xa\right)$ with sample size $ \na $ and $\left(y^{(2)},X^{(2)}\right)$ with sample size $ \nb$, where $\na \asymp \nb.$
Let $\widehat{\beta}$ denote an estimator which is produced by the first sub-sample $\left(\ya,\xa\right)$ and satisfies Condition $({\rm A1})$. One example of such an estimator is the scaled Lasso estimator \eqref{eq: scaled Lassoa complete} applied to the subsample $\left(\ya,\xa\right)$. 
We propose the following estimator of $\Q$,
\begin{equation}
\widehat{\Q}(\widehat{\beta},\widehat{\Sigma}^{(2)})=\widehat{\beta}^{\intercal} \widehat{\Sigma}^{(2)}\widehat{\beta}+2\widehat{\beta}^{\intercal}\frac{1}{\nb}\sum_{i=\na+1}^{n}X_{i\cdot}^{\intercal}(y_i-X_{i\cdot}\widehat{\beta}),
\label{eq: EST semi optimal}
\end{equation}
where $\widehat{\Sigma}^{\left(2\right)}=\frac{1}{n+N-\na}\sum_{i=\na+1}^{n+N}X_{i\cdot} X_{i\cdot}^{\intercal}$. The following theorem establishes the convergence rate of $\widehat{\Q}(\widehat{\beta},\widehat{\Sigma}^{(2)})$ and shows that this estimator achieves the optimal convergence rate of estimating $\Q$ for $M\gtrsim \sqrt{k \log p/n}$. 
\begin{Theorem} Suppose that Condition {\rm (B1)} holds and $k\leq c n/\log p$ for some constant $c>0$. Let $\widehat{\beta}$ be an estimator depending on the first half sample $\left(\ya,\xa\right)$ and satisfying Condition $({\rm A1})$.  Then with probability larger than $\hprob$, the estimator $\widehat{\Q}(\widehat{\beta},\widehat{\Sigma}^{(2)})$ defined in \eqref{eq: EST semi optimal} satisfies
	\begin{equation}
	\left|\widehat{\Q}(\widehat{\beta},\widehat{\Sigma}^{(2)})-\Q\right|\lesssim (t+1)\frac{\sigma \|\Sigmahalf \beta\|_2}{\sqrt{n}}+t\frac{\beta^{\intercal}\Sigma\beta}{\sqrt{N+n}}+\frac{k\log p}{n}\sigma^2.
	\label{eq: optimal upper}
	\end{equation}
Hence, the estimator $\widehat{\Q}(\widehat{\beta},\widehat{\Sigma}^{(2)})$ defined in \eqref{eq: EST semi optimal} achieves the optimal estimation rate 
over $\Theta(k,M)$ in the sense of \eqref{eq: optimal statement 1} over the regime $k\leq c \min\left\{{n/\log p}, p^{\nu}\right\}$ for some constants $c>0$ and $0\leq \nu<{1\over 2}$ and $M\gtrsim \sqrt{k \log p/n}$.	
	\label{thm: est bound semi optimal}
\end{Theorem}
\subsection{Two Special Cases}
\label{sec: special cases}

We now turn to two important special cases, the inference in the supervised setting and the setting with known design covariance matrix. 



\subsubsection{Case I: Supervised Inference}

In the supervised setting without any additional unlabelled data, $\Sigma$ is estimated by $\widehat{\Sigma}^{L}=\frac{1}{n}\sum_{i=1}^{n}X_{i\cdot} X_{i\cdot}^{\intercal}$. The following corollary establishes the convergence rate of the estimator $\widehat{\Q}={\widehat{\Q}}(\widehat{\beta},\widehat{\Sigma}^{L})$, which is a special case of the estimator \eqref{eq: EST semi} with $N=0$. 
\begin{Corollary} Suppose that Condition {\rm (A1)} holds and $k\leq c n/\log p$ for some constant $c>0$. For any estimator $\widehat{\beta}$ satisfying  {\rm (B1)}, with probability larger than $\hprobsup$,  ${\widehat{\Q}}(\widehat{\beta},\widehat{\Sigma}^{L})$ proposed in \eqref{eq: EST} with $\widehat{\Sigma}^{L}=\frac{1}{n}\sum_{i=1}^{n}X_{i\cdot} X_{i\cdot}^{\intercal}$ satisfies
	\begin{equation}
	\left|{\widehat{\Q}}(\widehat{\beta},\widehat{\Sigma}^{L})-\Q\right|\lesssim t \frac{\sigma\|\Sigmahalf\beta\|_2+\beta^{\intercal}\Sigma\beta}{\sqrt{n}}+ \frac{k\log p}{n}\sigma^2.
	\label{eq: convergence rate}
	\end{equation}
Under the additional assumption ${\rm (A2)}$ and
 $\SNR\gg \min\left\{{k \log p}/{\sqrt{n}},\left({k \log p}/{\sqrt{n}}\right)^{1/2}\right\}$, 
\begin{equation}
\frac{\sqrt{n}\left({\widehat{\Q}}(\widehat{\beta},\widehat{\Sigma}^{L})-\Q\right)}{\sqrt{4\sigma^2\beta^{\intercal}\Sigma\beta+\E\left(\beta^{\intercal}X_{1\cdot} X_{1\cdot}^{\intercal}\beta-\beta^{\intercal}\Sigma\beta\right)^2}}\cid N(0,1)
\label{eq: distribution}
\end{equation}
\label{cor: est bound}
\end{Corollary}

Corollary \ref{cor: est bound} basically follows from Theorem  \ref{thm: est bound semi} with $N=0$ except for some technical difference.   By comparing Corollary \ref{cor: est bound} with Theorems \ref{thm: est bound semi} and \ref{thm: est bound semi optimal}, we observe that the unlabelled data leads to a faster convergence rate by reducing $\beta^{\intercal}\Sigma\beta/\sqrt{n}$ in \eqref{eq: convergence rate} to $\beta^{\intercal}\Sigma\beta/\sqrt{N+n}$ in \eqref{eq: convergence rate semi} and \eqref{eq: optimal upper}; the unlabelled data does not affect other terms in the convergence rate. The effect of the unlabelled data is also revealed in the limiting distribution in \eqref{eq: distribution}, where the exact variance level is reduced from $[4\sigma^2\beta^{\intercal}\Sigma\beta+\E\left(\beta^{\intercal}X_{1\cdot} X_{1\cdot}^{\intercal}\beta-\beta^{\intercal}\Sigma\beta\right)^2]/n$ in \eqref{eq: distribution} to $[4\sigma^2\beta^{\intercal}\Sigma\beta+\rho\E\left(\beta^{\intercal}X_{1\cdot} X_{1\cdot}^{\intercal}\beta-\beta^{\intercal}\Sigma\beta\right)^2]/n$ in \eqref{eq: distribution semi} for $\rho=\lim_{n\rightarrow \infty} \frac{n}{N+n}\in [0,1].$ The following corollary further establishes the minimax rate for estimating $\beta^{\intercal}\Sigma\beta$ in the supervised setting.

\begin{Corollary} Suppose that Condition {\rm (A1)} holds and $k\leq c \min\left\{{n/\log p}, p^{\nu}\right\}$ for some constants $c>0$ and $0\leq \nu<{1\over 2}$. For any estimator $\widehat{\beta}$ satisfying  Condition  {\rm (B1)}, the estimator $\widehat{\Q}={\widehat{\Q}}(\widehat{\beta},\widehat{\Sigma}^{L})$ defined in \eqref{eq: EST} with $\widehat{\Sigma}^{L}=\frac{1}{n}\sum_{i=1}^{n}X_{i\cdot} X_{i\cdot}^{\intercal}$ achieves the optimal estimation rate over $\Theta(k,M)$ for $M\gtrsim \sqrt{k \log p/n}$,  
that is, ${\widehat{\Q}}(\widehat{\beta},\widehat{\Sigma}^{L})$ satisfies
	\begin{equation}
	\sup_{\theta \in \Theta(k,M)}\PP\left(\left|{\widehat{\Q}}(\widehat{\beta},\widehat{\Sigma}^{L})-\Q\right|\gtrsim t {M^2\over \sqrt{n}}+ {M\over \sqrt{n}} + {k \log p\over n}\right)\leq \lprob.
\label{eq: optimal special 1}
	\end{equation}
\label{cor: optimal convergence special 1}
\end{Corollary}

\begin{Remark}\rm
In the supervised setting,  \citep{guo2016optimal} established that the optimal rate of estimating $\|\beta\|_2^2$ over $\Theta(k,M)$ for $M\gtrsim \sqrt{k \log p/n}$ is $M/{\sqrt{n}}+ (M+1){k \log p}/{n}$. In contrast to \eqref{eq: optimal special 1}, we can see that neither of these two problems is easier than the other, where there is an additional term $M^2/\sqrt{n}$ in \eqref{eq: optimal special 1} and an additional term $M{k \log p}/{n}$ in the optimal convergence rate of estimating $\|\beta\|_2^2$.
\end{Remark}
Inference for $\beta^{\intercal}\Sigma\beta$ in the supervised setting is closely connected to \citet{sun2012scaled, verzelen2016adaptive}, where \citet{sun2012scaled} studied the inference problem for $\sigma^2$ and \citet{verzelen2016adaptive} studied the estimation of $\beta^{\intercal}\Sigma\beta/\sigma^2$. In particular, \citet{sun2012scaled}  proposed the scaled lasso estimator $\widehat{\sigma}^2$ in \eqref{eq: scaled Lassoa complete} to estimate $\sigma^2$ and  \citet{verzelen2016adaptive} proposed to estimate $\beta^{\intercal}\Sigma\beta$ by $\left(\frac{1}{n}\|y\|_2^2-\widehat{\sigma}^2\right)_{+}$ as an intermediate step of estimating $\beta^{\intercal}\Sigma\beta/\sigma^2$.  For the estimator ${\widehat{\Q}}(\widehat{\beta},\widehat{\Sigma}^{L})$ defined in \eqref{eq: EST}, if $\widehat{\beta}$ is taken as the scaled Lasso estimator,  then ${\widehat{\Q}}(\widehat{\beta},\widehat{\Sigma}^{L})$ is reduced to being the same as the estimator proposed in \citet{verzelen2016adaptive}, where the equivalence is shown by the following expression, 
\begin{equation}
\widehat{\beta}^{\intercal} \widehat{\Sigma}^{L}\widehat{\beta}+2\widehat{\beta}^{\intercal}\frac{1}{n}\sum_{i=1}^{n}X_{i\cdot}(y_i-X_{i\cdot}\widehat{\beta})=\frac{1}{n}\left(\|y\|_2^2-\|y-X\widehat{\beta}\|_2^2\right)=\frac{1}{n}\|y\|_2^2-\widehat{\sigma}^2.
\label{eq: connection}
\end{equation}
As a remark, in the supervised setting, the calibration idea in \eqref{eq: EST} provides a completely new perspective on estimation of $\beta^{\intercal}\Sigma\beta$, where instead of using the expression $\Q=\E(y_i^2)-\sigma^2$ and estimating $\sigma^2$ first, we estimate $\Q$ directly by calibrating the plug-in estimator. This new perspective establishes a general machinery taking reasonable good initial estimators of $\beta$ and $\Sigma$ as inputs. As shown in \eqref{eq: EST semi}, the flexibility of the calibrated estimator has proven useful in efficiently pooling additional information on $\Sigma$ while the estimation method introduced in \citet{verzelen2016adaptive} cannot be directly extended to integrating the unlabelled data in the semi-supervised setting. 

In the numerical studies, we have demonstrated that the effect of including unlabelled data is of great practical significance, where in the case of dense $\Sigma$, the RMSE of the new proposed CHIVE estimator is 60\% to 70\% smaller than the estimators in \eqref{eq: connection} without using the unlabelled data. See Table \ref{tab: semi-helpful} in Section \ref{sec: sim} for details. 

Additionally, \citet{verzelen2016adaptive} focused on the estimation problem instead of confidence interval construction and hypothesis testing problems. In terms of technical details on estimation optimality, the results in \citet{verzelen2016adaptive} allowed for a more general regime $k\geq \sqrt{p}$ than Corollary \ref{cor: optimal convergence special 1} but did not handle the optimality in the semi-supervised setting and did not allow the signal strength $M$ to grow with $n,p$.


\subsection{Case II: Known $\Sigma$}
The general semi-supervised results also shed light on another interesting setting where the design covariance $\Sigma$ is known. In the semi-supervised setting, the unlabelled data is used for estimating $\Sigma$, so the case of known $\Sigma$ is an extreme case of the semi-supervised setting with $N$ taken as infinity. The estimator \eqref{eq: EST semi optimal} can be modified as
$
\widehat{\Q}(\widehat{\beta},{\Sigma}, Z^{(2)})=\widehat{\beta}^{\intercal}\Sigma\widehat{\beta}+2\widehat{\beta}^{\intercal}\frac{1}{\nb}\sum_{i=\na+1}^{n}X_{i\cdot}^{\intercal}(y_i-X_{i\cdot}\widehat{\beta}).$ Similarly, the estimator proposed in \eqref{eq: EST semi} is changed to 
$
\widehat{\Q}(\widehat{\beta},{\Sigma})=\widehat{\beta}^{\intercal} {\Sigma}\widehat{\beta}+2\widehat{\beta}^{\intercal}\frac{1}{n}\sum_{i=1}^{n}X_{i\cdot}(y_i-X_{i\cdot}^{\intercal}\widehat{\beta}).
$
\begin{Corollary}
Suppose that Condition {\rm (A1)} holds and $k\leq c n/\log p$ for some constant $c>0$. 
\begin{enumerate}
\item For any estimator $\widehat{\beta}$ depending on the first half sample $\left(\ya,\xa\right)$ and satisfying Condition $({\rm B1})$, then with probability larger than $\hprobsup$, 	\begin{equation}
	\left|\widehat{\Q}(\widehat{\beta},{\Sigma}, Z^{(2)})-\Q\right|\lesssim (t+1) \frac{\sigma\|\Sigmahalf \beta\|_2}{\sqrt{n}}+\frac{k\log p}{n}\sigma^2.
	\label{eq: known optimal bound}
	\end{equation}
\item For any estimator $\widehat{\beta}$ satisfying Condition $({\rm B1})$, then with probability larger than $\hprobsup$,	\begin{equation}
	\left|\widehat{\Q}(\widehat{\beta},{\Sigma})-\Q\right|\lesssim t\frac{\|\Sigmahalf \beta\|_2}{\sqrt{n}}+(\frac{\|\Sigmahalf \beta\|_2}{\sigma}+1)\frac{k\log p}{n}\sigma^2.
	\label{eq: known bound}
	\end{equation}
\end{enumerate}
\label{cor: upper bound special 2}
\end{Corollary}
Through comparing \eqref{eq: known optimal bound} with \eqref{eq: optimal upper} and \eqref{eq: known bound} with \eqref{eq: convergence rate semi}, the uncertainty of estimating the design covariance matrix leads to the additional term $\beta^{\intercal}\Sigma\beta/\sqrt{N+n}$. By applying Theorem \ref{thm: lower bound}, it can be shown that the upper bound in \eqref{eq: known optimal bound} leads to  the optimal convergence rate ${M}/{\sqrt{n}}+{k\log p}/{n}$. The term $M^2/\sqrt{N+n}$ disappears due to the known design covariance matrix $\Sigma$.

\section{Semi-supervised Confidence Intervals for $\beta^{\intercal}\Sigma\beta$} 
\label{sec: ci}

In this section, we quantify the uncertainty of the CHIVE estimator proposed in Section \ref{sec: est} and then construct confidence intervals for $\beta^{\intercal}\Sigma\beta$ in the semi-supervised setting. 

\subsection{Confidence Interval Construction}
\label{sec: ci construction}



The main next step of confidence interval construction for $\Q$ is to consistently estimate the standard error $\sqrt{4\sigma^2\beta^{\intercal}\Sigma\beta+\rho\E\left(\beta^{\intercal}X_{1\cdot} X_{1\cdot}^{\intercal}\beta-\beta^{\intercal}\Sigma\beta\right)^2}/\sqrt{n}$ of the limiting distribution established in \eqref{eq: distribution semi}. Specifically, we estimate $4\sigma^2\beta^{\intercal}\Sigma\beta$ by $\widehat{\phi}_1$, $\rho$ by $\widehat{\rho}=n/(N+n)$ and $\E\left(\beta^{\intercal}X_{1\cdot} X_{1\cdot}^{\intercal}\beta-\beta^{\intercal}\Sigma\beta\right)^2$ by $\widehat{\phi}_2$, where 
$\widehat{\phi}_1={\widehat{\sigma}^2\widehat{\beta}^{\intercal}\widehat{\Sigma}^{S}\widehat{\beta}} \quad \text{and} \quad \widehat{\phi}_2={\frac{1}{n+N}\sum_{ i=1}^{n+N} \left(\widehat{\beta}^{\intercal} X_{i\cdot} X_{i\cdot}^{\intercal}\widehat{\beta}-\widehat{\beta}^{\intercal}\widehat{\Sigma}^{S}\widehat{\beta}\right)^2},$ 
with $\widehat{\Sigma}^{S}$ defined in \eqref{eq: EST semi}.
Then we propose the following confidence interval centered at $\widehat{\Q}$,  
\begin{equation}
{\rm CI}(Z)=[(\widehat{\Q}-z_{\alpha/2}\widehat{\phi})_{+},\; \widehat{\Q}+z_{\alpha/2}\widehat{\phi}], \;\text{where}\; \widehat{\phi}=\sqrt{({4 \widehat{\phi}_1+\widehat{\rho}\widehat{\phi}_2})/{n}},
\label{eq: CI}
\end{equation}
where $z_{\alpha/2}$ is the upper $\alpha/2$ quantile of standard normal distribution. 
The following theorem establishes the coverage and precision properties of ${\rm CI}(Z)$, where the length of the interval ${\rm CI}(Z)=\left(L(Z), U(Z)\right)$ is defined as $\LL({\rm CI}(Z))=U(Z)-L(Z)$. 


\begin{Theorem}
	\label{thm: explained variance}
Suppose that Conditions {\rm (A1)} and {\rm (A2)} hold, $k\ll \min \{n/(\log(N+n)\log p),{\sqrt{n}}/{\log p}\}$ and $\SNR\gg {k \log p}/{\sqrt{n}}$. For $\widehat{\beta}$ and $\widehat{\sigma}^2$ satisfying Conditions {\rm (B1)} and {\rm (B2)}, respectively, the confidence interval given in \eqref{eq: CI} satisfies,
\begin{equation}
\liminf_{n\rightarrow\infty}\PP\left(\beta^{\intercal}\Sigma\beta \in {\rm CI}(Z)\right)\geq 1-\alpha
\label{eq: coverage CI2}
\end{equation}
\begin{equation}
\lim_{n\rightarrow\infty}\PP\left(\LL({\rm CI}(Z))\geq (1+\delta_0)\sqrt{{4\sigma^2\beta^{\intercal}\Sigma\beta}/{n}+{\E\left(\beta^{\intercal}X_{1\cdot} X_{1\cdot}^{\intercal}\beta-\beta^{\intercal}\Sigma\beta\right)^2}/(N+n)}\right)=0
\label{eq: precision CI2}
\end{equation}
for any positive constant $\delta_0>0$.
\end{Theorem}
The effect of the unlabelled data on the length of confidence interval is carefully characterized in \eqref{eq: precision CI2}, where the unlabelled data shrinks part of the length of confidence interval, ${\E\left(\beta^{\intercal}X_{1\cdot} X_{1\cdot}^{\intercal}\beta-\beta^{\intercal}\Sigma\beta\right)^2}/(N+n).$ This term corresponds to the uncertainty of estimating $\beta^{\intercal}\Sigma\beta$ in the oracle setting of known $\beta$. The most effective regime of integrating the unlabelled data is when the ratio $\frac{\E\left(\beta^{\intercal}X_{1\cdot} X_{1\cdot}^{\intercal}\beta-\beta^{\intercal}\Sigma\beta\right)^2}{\sigma^2\beta^{\intercal}\Sigma\beta}$ is not vanishing to zero. Otherwise, the dominating term in the length of \eqref{eq: precision CI2} is ${4\sigma^2\beta^{\intercal}\Sigma\beta}/{n}$ and the additional unlabelled data is not helpful in this regime.
In the numerical studies, we investigate how much shorter confidence intervals can be after integrating the unlabelled data. The lengths of CIs in the semi-supervised setting  can be reduced to being as short as 30\% to 40\%  of those in the supervised setting. See Table \ref{tab: semi-helpful} for details.

The upper bound for CI length established in \eqref{eq: precision CI2} is further upper bounded by ${\sigma\|\Sigmahalf\beta\|_2}/{\sqrt{n}}+{\beta^{\intercal}\Sigma\beta}/{\sqrt{N+n}}$, which matches the optimal convergence rate of estimation $M/\sqrt{n}+M^2/\sqrt{N+n}$ over the parameter space $\Theta(k,M)$ for $k\ll {\sqrt{n}}/{\log p}$ and $M\gg {k \log p}/{\sqrt{n}}$. 

As shown in Theorem \ref{thm: explained variance}, the validity of the proposed confidence interval \eqref{eq: CI} requires the condition that $\SNR$ is bounded away from zero by ${k \log p}/{\sqrt{n}}$. Although ${k \log p}/{\sqrt{n}}$ converges to zero over the extreme sparse regime $k\ll {\sqrt{n}}/{\log p}$, it reveals the difficulty of constructing stable confidence intervals for $\beta^{\intercal}\Sigma\beta$ when $\SNR$ is at a local neighborhood of zero. The next section will address the inference problem when $\SNR$ is at a local neighborhood of $0$. 

\vspace{-5mm}
\subsection{Inference for Weak Signals}
\label{sec: randomization}

As discussed in the introduction, uncertainty quantification of $\Q=\beta^{\intercal}\Sigma\beta$ is closely connected to other important statistical problems, including (1) signal detection and global testing; (2) prediction accuracy evaluation and (3) confidence ball construction.  These applications provide a strong motivation for studying the inference problem for the explained variance under the settings of weak signals (that is, $\SNR\lesssim {k\log p}/{\sqrt{n}}$). The main goal of this section is to discuss extensions of the proposed procedure to conduct statistical inference uniformly over different levels of signal strength, measured by $\SNR$. 

To begin with, we recall the reasoning for the non-uniformity assumption $\SNR\gg {k\log p}/{\sqrt{n}}$. This assumption is imposed such that the variance component of the CHIVE estimator dominates the bias component and in this case an asymptotic limiting distribution for the variance component is used to construct confidence interval for the explained variance. 
Specifically, we discuss two possible solutions to remove this stringent assumption, a) to enlarge the confidence interval by an upper bound for the bias in Section \ref{sec: sol 1}; b) to increase the variance level by randomized calibration in Section \ref{sec: sol 2}.
\subsubsection{Bound the bias term}
\label{sec: sol 1}
One way to construct confidence intervals uniformly over all $\SNR$ is to enlarge the estimated variance level defined in \eqref{eq: CI} to 
\begin{equation}
\widehat{\phi}^{E}=\widehat{\phi}^{E}(y,X,\tau_0)=\sqrt{\frac{1}{n}4\widehat{\sigma}^2\left(\widehat{\beta}^{\intercal}\widehat{\Sigma}^{S}\widehat{\beta}+\tau_0^2\right)+{\frac{1}{(n+N)^2}\sum_{i=1}^{n+N} \left(\widehat{\beta}^{\intercal} X_{i\cdot} X_{i\cdot}^{\intercal}\widehat{\beta}-\widehat{\beta}^{\intercal}\widehat{\Sigma}^{S}\widehat{\beta}\right)^2}},
\label{eq: var level enlarge}
\end{equation}
for some positive constant $\tau_0>0$. 
Then we construct the confidence interval as
\begin{equation}
{\rm CI}^{E}(Z)=[(\widehat{\Q}-z_{\alpha/2}\widehat{\phi}^{E})_{+},\quad \widehat{\Q}+z_{\alpha/2}\widehat{\phi}^{E}], 
\label{eq: CI enlarged}
\end{equation}
where $z_{\alpha/2}$ is the upper $\alpha/2$ quantile of standard normal distribution. The reason for adding the term $\frac{1}{n}4\widehat{\sigma}^2\tau_0^2$ in the width \eqref{eq: var level enlarge} is that this additional term is an upper bound for the bias term in the regime $k \ll \sqrt{n}/\log p$. The following corollary establishes the coverage and the precision property of the enlarged confidence interval, ${\rm CI}^{E}(Z)$.  
 \begin{Corollary}
Suppose that Conditions {\rm (A1)} and {\rm (A2)} hold, $k\ll \min \{n/(\log(N+n)\log p),{\sqrt{n}}/{\log p}\}$ and $\tau_0>0$ is a positive constant. For $\widehat{\beta}$ and $\widehat{\sigma}^2$ satisfying Conditions {\rm (B1)} and {\rm (B2)}, respectively, then the confidence interval defined in \eqref{eq: CI enlarged} satisfies, 
\begin{equation}
\liminf_{n\rightarrow\infty}\PP\left(\beta^{\intercal}\Sigma\beta \in {\rm CI}^{E}(Z)\right)\geq 1-\alpha
\label{eq: coverage CI enlarge}
\end{equation}
\begin{equation}
\lim_{n\rightarrow\infty}\PP\left(\LL({\rm CI}^{E}(Z))\geq (1+\delta_0)\sqrt{{4\sigma^2\left(\beta^{\intercal}\Sigma\beta+\tau_0^2\right)}/{n}+{\E\left(\beta^{\intercal}X_{1\cdot} X_{1\cdot}^{\intercal}\beta-\beta^{\intercal}\Sigma\beta\right)^2}/{(N+n)}}\right)=0
\label{eq: precision CI enlarge}
\end{equation}
for any positive constant $\delta_0>0$.
\label{cor: explained variance enlarge}
\end{Corollary}
In contrast to the length of confidence interval in \eqref{eq: precision CI2}, the length in \eqref{eq: precision CI enlarge} is enlarged by the exact amount ${4\sigma^2\tau_0^2}/{n}$. In contrast to Theorem \ref{thm: explained variance}, the inference is uniform over all levels of $\SNR$ at the expense of a slightly longer confidence interval.
\subsubsection{Randomized Calibration}
\label{sec: sol 2}

 The construction in \eqref{eq: CI enlarged} still uses the CHIVE estimator as the center and enlarges the constructed confidence interval. 
We introduce a randomized version of the CHIVE estimator as the new center, where the main intuition is to increase the variance level through randomization such that the variance of this randomized estimator dominates its bias level. We generate random variables $u_i \stackrel{iid}{\sim} N(0,\tau_0^2)$ for $1\leq i\leq n$, independent of the observed data $Z$
and  propose the following randomized calibrated estimator, 
\begin{equation}
{\widehat{\Q}^{R}}={\widehat{\Q}}^{R}(\widehat{\beta},\widehat{\Sigma}^{S},u)=\widehat{\beta}^{\intercal}\widehat{\Sigma}^{S}\widehat{\beta}+2\frac{1}{n}\sum_{i=1}^{n}({X}_{i\cdot}^{\intercal}\widehat{\beta}+u_i)({y}_i-{X}_{i\cdot}^{\intercal}\widehat{\beta}).
\label{eq: randomized EST}
\end{equation}
When there is no confusion, we use ${\widehat{\Q}^{R}}$ to denote the estimator proposed in \eqref{eq: randomized EST}.
In contrast to \eqref{eq: EST semi}, the calibration step in \eqref{eq: randomized EST} is involved with an additional term $2\frac{1}{n}\sum_{i=1}^{n}u_i({y}_i-{X}_{i\cdot}^{\intercal}\widehat{\beta})$. If $u_i$ is zero instead of being generated as normal random variables in \eqref{eq: randomized EST}, the estimator ${\widehat{\Q}}^{R}(\widehat{\beta},\widehat{\Sigma}^{S},0)$ is reduced to being exactly the same as ${\widehat{\Q}}(\widehat{\beta},\widehat{\Sigma}^{S})$ defined in \eqref{eq: EST semi}.
Since $u_i$ in \eqref{eq: randomized EST} is randomly generated normal random variables, this additional term approximately follows a normal distribution with mean zero and variance $4\sigma^2\tau_0^2/n$. Even in  the presence of weak signals, this additional term further enlarges the variance level of the calibrated estimator such that the bias level of the calibrated estimator is  dominated by the corresponding variance level. 
The following theorem establishes the limiting distribution of the estimator $\widehat{\Q}^{R}$ after randomized calibration.
\begin{Theorem} Suppose that Condition {\rm (A1)} holds, $k \ll \sqrt{n}/\log p$ and $\tau_0>0$ is a positive constant. For any estimator $\widehat{\beta}$ satisfying  Condition  {\rm (B1)}, then
\begin{equation}
\sqrt{n}\frac{\widehat{\Q}^{R}- \Q}{\sqrt{4\sigma^2\left(\beta^{\intercal}\Sigma\beta+\tau_0^2\right)+\rho\E\left(\beta^{\intercal}X_{1\cdot} X_{1\cdot}^{\intercal}\beta-\beta^{\intercal}\Sigma\beta\right)^2}}\cid N\left(0,1\right)
\label{eq: limiting distribution boosting}
\end{equation}
where $\rho=\lim \frac{n}{n+N}$.
\label{thm: randomization} 
\end{Theorem}
In comparison to the limiting distribution \eqref{eq: distribution semi} in Theorem \ref{thm: est bound semi}, Theorem \ref{thm: randomization} requires no condition on $\SNR$ to establish the asymptotic limiting distribution while the variance level of the established normal distribution is enlarged by the amount ${4\sigma^2 \tau_0^2}/{n}$.
This additional variance term is a side effect of the randomized calibration. However, it enables a uniform inference procedure over all levels of $\SNR$. 
Then we propose the following confidence interval, 
${\rm CI}^{R}(Z)=\left[(\widehat{\Q}^{R}-z_{\alpha/2}\widehat{\phi}^{E})_{+},\quad \widehat{\Q}^{R}+z_{\alpha/2}\widehat{\phi}^{E}\right],$ where $\widehat{\phi}^{E}$ is defined in \eqref{eq: var level enlarge}. 
This confidence interval has the same length as that of \eqref{eq: CI enlarged} but different centers. The proposed estimator $\widehat{\Q}^{R}$ enjoys the advantage of having an asymptotic normal distribution but it suffers from the disadvantage as all randomized procedure, where the output is random even given the same data set. The following corollary characterizes the coverage and precision properties of ${\rm CI}^{R}(Z)$.



\begin{Corollary}
Under the same conditions as Corollary \ref{cor: explained variance enlarge}, the coverage property in \eqref{eq: coverage CI enlarge} and precision property in \eqref{eq: precision CI enlarge} hold for the confidence interval ${\rm CI}^{R}(Z)$.
\label{cor: explained variance ran}
\end{Corollary}
Algorithm \ref{alg: SIVE-CRAN} summarizes the uncertainty quantification methods for $\beta^{\intercal}\Sigma\beta$. 
\begin{algorithm}[h]
    \SetKwInOut{Input}{Input}
    \SetKwInOut{Output}{Output}
    \Input{Labelled data $\{y_i,X_{i\cdot}\}_{1\leq i\leq n}$ and unlabelled data $\{\W_{i\cdot}\}_{n+1\leq i\leq n+N}$; $\tau_0>0$ }
    \Output{Point estimator $\widehat{\Q}=\widehat{\Q}(y,X)$, $\widehat{\Q}^{R}=\widehat{\Q}^{R}(y,X,\tau_0)$ and variance estimator $\widehat{\phi}^{E}=\widehat{\phi}^{E}(y,X,\tau_0)$}
    Initialization: Construct point estimator $\widehat{\beta}$ and $\widehat{\sigma}^2$ satisfying ${\rm (B1)}$ and ${\rm (B2)}$; Estimate $\Sigma$ by $\widehat{\Sigma}^{S}$ defined in \eqref{eq: EST semi};\\
   Calibration: Estimate $\Q$ by the CHIVE estimator $\widehat{\Q}$ in \eqref{eq: EST} or its randomized version $\widehat{\Q}^{R}$ in \eqref{eq: randomized EST}.\\ 
Uncertainty Quantification: Quantify the error of the proposed estimator by $\widehat{\phi}^{E}$ defined in \eqref{eq: var level enlarge}.
 \caption{Semi-supervised Uncertainty Quantification for $\beta^{\intercal}\Sigma\beta$}
 \label{alg: SIVE-CRAN}
\end{algorithm}

We conclude this section with some additional comments. Compared to point estimation, construction of confidence intervals  for the explained variance is a more challenging problem, mainly due to the fact that one needs to characterize the uncertainty of the proposed estimator. Specifically, accurate estimation of $\Q$ can be conducted uniformly over all levels of $\SNR$ while construction of confidence intervals uniformly over all levels of $\SNR$ requires much more  efforts. Another interesting observation is that inference for explained variance is different from that for linear functional \citep{zhang2014confidence,van2014asymptotically,javanmard2013hypothesis,javanmard2014confidence,cai2017confidence}, where the valid inference results for the latter do not depend on the magnitude of $\SNR$. 

\section{Related Semi-supervised Inference Problem}
\label{sec: semi-infer general}
The improvement due to integrating the unlabelled data is not just limited to the inference problem for $\beta^{\intercal}\Sigma\beta$, but can also be obtained in the semi-supervised inference for $\|\beta\|_2^2$. This unweighted quadratic functional is different from $\beta^{\intercal}\Sigma\beta$ as the covariance matrix $\Sigma$ does not appear in the expression. Hence, it is even unclear whether the unlabelled data can be of any help. We introduce in this section a procedure integrating the unlabelled data and also carefully quantify the improvement with making use of the additional unlabelled data in the semi-supervised setting.

The estimation of $\|\beta\|_2^2$ in the supervised setting was studied in \citet{guo2016optimal}, where the error decomposition of the plug-in estimator $\|\widehat{\beta}\|_2^2$ was established as
$
\|\widehat{\beta}\|_2^2-\|\beta\|_2^2=2\widehat{\beta}^{\intercal}(\widehat{\beta}-\beta) -(\widehat{\beta}-\beta)^{\intercal}(\widehat{\beta}-\beta).
$
In \citet{guo2016optimal}, the bias term $2\widehat{\beta}^{\intercal}(\widehat{\beta}-\beta)$ in the decomposition was estimated and hence the plug-in estimator $\|\widehat{\beta}\|_2^2$ was corrected. 

We illustrate here how the additional unlabelled data facilitates the bias-correction step. We randomly split the labelled data $\left(y,X\right)$ into two subsamples $\left(\ya,\xa\right)$ with sample size $ \na $ and $\left(y^{(2)},X^{(2)}\right)$ with sample size $ \nb$, where $\na \asymp \nb.$
Let $\widehat{\beta}$ denote an estimator of $\beta$ produced by the first sub-sample $\left(\ya,\xa\right)$ satisfying Condition $({\rm B1})$, where one example is the scaled Lasso estimator \eqref{eq: scaled Lassoa complete} applied to  $\left(\ya,\xa\right)$.  Then we construct a projection direction $\widehat{u}\in \R^{p}$ and propose the estimator $\|\widehat{\beta}\|_2^2$ as
\begin{equation}
\widehat{\|\beta\|_2^2}=\|\widehat{\beta}\|_2^2+2\widehat{u}^{\intercal}\frac{1}{\nb}\sum_{i=\na+1}^{n}X_{i\cdot}\left(y_i-X_{i\cdot}^{\intercal}\widehat{\beta}\right).
\label{eq: semi extend}
\end{equation}
The unlabelled data is particularly useful in estimating the projection direction $\widehat{u}\in \R^{p}$. The projection direction $\widehat{u}$ is constructed as 
$
\widehat{u}=\widehat{\Omega}\widehat{\beta}=\sum_{l \in {\rm supp}(\widehat{\beta})}\widehat{\Omega}_{\cdot l}\widehat{\beta}_l
$
where $\widehat{\Omega}_{\cdot l}$ is the CLIME estimator defined as 
\begin{equation}
\begin{aligned}
\widehat{\Omega}_{\cdot l}=\argmin \|m\|_1 \quad
\text{subject to}\;\; \|\widetilde{\Sigma}m-e_{l}\|_\infty\leq \lambda_{S}
\end{aligned}
\label{eq: CLIME estimator}
\end{equation}
with $\widetilde{\Sigma}=\frac{1}{N+\na}(\sum_{i=1}^{\na}X_{i\cdot} X_{i\cdot}^{\intercal}+\sum_{i=n+1}^{n+N}X_{i\cdot} X_{i\cdot}^{\intercal})$ and $\lambda_{S}\asymp \sqrt{{\log p}/{(\na+N)}}$. 
The additional unlabelled data plays a role in constructing the sample covariance matrix $\widetilde{\Sigma}$ in \eqref{eq: CLIME estimator} and hence constructing the projection direction $\widehat{u}$.
The specific way of including the unlabelled data to improve the estimation accuracy of $\|\beta\|_2^2$ is different from that of $\beta^{\intercal}\Sigma\beta$, where the additional unlabelled data is used to estimate $\Sigma$ directly in estimating $\beta^{\intercal}\Sigma\beta$ while the additional unlabelled is used to estimate $\Sigma^{-1}$ in estimating $\|\beta\|_2^2$. However, the high level idea is the same, that is, making use of the flexibility of calibrated estimator and properly incorporating the information about $\Sigma$ contained in the unlabelled data. 

Precision matrix estimation has been studied in the literature; see \citet{cai2011constrained} and the reference in the paper.  We restrict the attention to $\widehat{\Omega}$ satisfying the following condition.
\begin{enumerate}
\item[(B3)] The estimator $\widehat{\Omega}$ satisfies $
\PP\left(\|\widehat{\Omega}-\Omega\|_2\gtrsim C_{\Omega} s \sqrt{{\log p}/{(N+n)}}\right)\geq 1-\gamma_1(N+n)$
where $\gamma_1(N+n)\rightarrow 0$, $s=\max_{1\leq l\leq p} \|\Omega_{l\cdot}\|_0$ and $C_{\Omega}$ is a constant depending on $\|\Omega\|_{L_1}$.
\end{enumerate}
The CLIME estimator $\widehat{\Omega}=\begin{pmatrix} \widehat{\Omega}_{\cdot 1}&\widehat{\Omega}_{\cdot 2}&\cdots \widehat{\Omega}_{\cdot p}\end{pmatrix}$ with $\widehat{\Omega}_{l\cdot}$ constructed in \eqref{eq: CLIME estimator} is shown to satisfy the condition {\rm (B3)} under certain regularity conditions. See the exact statement in \citet{cai2011constrained}.  We show in the following theorem that, with a sufficiently large amount of unlabelled data, the inference results for the semi-supervised setting are distinguished from those in the supervised data. 
\begin{Theorem}
Suppose that Condition {\rm (A1)} holds, $k\leq c n/\log p$ for some constant $c>0$ and $c_0 \leq \lambda_{\min}\left(\Omega\right) \leq \lambda_{\max}\left(\Omega\right) \leq C_0$ for some positive constants $C_0\geq c_0>0$. Suppose that  $\widehat{\beta}$ satisfies  {\rm (B1)} and $\widehat{\Omega}$ satisfies  {\rm (B3)}. Under the sample size condition $N+n\gg C_{\Omega}^2 k\left(s\log p\right)^2,$ 
then with probability larger than $\hprobsup-\gamma_1(N+n)$,  
\begin{equation}
\left|\widehat{\|\beta\|_2^2}-\|\beta\|_2^2\right|\lesssim \sigma\frac{\|\beta\|_2}{\sqrt{n}}+k\frac{\log p}{n}\sigma^2.
\label{eq: special case 2}
\end{equation}
In addition, if $\frac{1}{\sigma}\|\beta\|_2\gg {k \log p}/{\sqrt{n}}$ and $\epsilon_i$ are i.i.d Gaussian random variables, then 
\begin{equation}
\sqrt{{n}/(\sigma^2 {\rm V})}\left(\widehat{\|\beta\|_2^2}-\|\beta\|_2^2\right)\cid N(0,1), \quad \text{with}\quad {\rm V}={4}\sum_{i=n_1+1}^{n}(\widehat{u}^{\intercal}X_{i\cdot})^2/\nb^2
\label{eq: limiting unweighted}
\end{equation}
\label{cor: l2 norm}
\end{Theorem}
 The limiting distribution in \eqref{eq: limiting unweighted} leads to the confidence interval construction 
\begin{equation*}
{\rm CI}_{\|\beta\|_2^2}=\left(\widehat{\|\beta\|_2^2}-z_{\alpha/2}\widehat{\sigma} \sqrt{\rm V},\widehat{\|\beta\|_2^2}+z_{\alpha/2}\widehat{\sigma} \sqrt{\rm V}\right)
\end{equation*}
where $\widehat{\|\beta\|_2^2}$ is defined in \eqref{eq: semi extend}, ${\rm V}$ is defined as \eqref{eq: limiting unweighted} and $\widehat{u}=\sum_{l \in {\rm supp}(\widehat{\beta})}\widehat{\Omega}_{\cdot l}\widehat{\beta}_l$.

A few remarks are in order for the semi-supervised inference for $\|\beta\|_2^2$. The results established in \citet{guo2016optimal} showed that the optimal rate for estimating $\|\beta\|_2^2$ in the supervised setting is ${\sigma {\|\beta\|_2}\over {\sqrt{n}}}+(1+\|\beta\|_2) k\frac{\log p}{n}\sigma^2$. In contrast, the term $\|\beta\|_2\cdot k\frac{\log p}{n}\sigma$ disappears in the rate of convergence (37) by efficiently incorporating the unlabelled data.  The improvement varies across different signal strengths, where the reduction in RMSE is limited if the signal strength $\|\beta\|_2$ is small but is significant if $\|\beta\|_2$ is large.
While integrating the unlabelled data is useful in reducing the RMSE for estimating both $\beta^{\intercal}\Sigma\beta$ and $\|\beta\|_2^2$, it is interesting to observe that the improvement by incorporating the unlabelled data is different, where for estimating $\beta^{\intercal}\Sigma\beta$, part of the variance component is reduced but for estimating $\|\beta\|_2^2$, the bias component is reduced by $\|\beta\|_2\cdot k\frac{\log p}{n}\sigma$. More interestingly, when the size of the unlabelled data is large enough and the spectrum of $\Sigma$ is bounded away from zero and infinity, the rate of estimating $\|\beta\|_2^2$ in \eqref{eq: special case 2} coincides with that of estimating $\beta^{\intercal}\Sigma\beta$ in \eqref{eq: known optimal bound}.

Theorem \ref{cor: l2 norm} requires the additional sample size condition for the unlabelled data, $N+n\gg C_{\Omega}^2 k\left(s\log p\right)^2.$ The general results for any $N\geq 0$  are given in Section \ref{sec: unweighted add} in the supplementary material.
 
The additional unlabelled data is not just useful in improving the estimation accuracy, but is also useful in confidence interval construction.  The specific effect is different from that   for $\beta^{\intercal}\Sigma\beta$; the confidence interval for $\beta^{\intercal}\Sigma\beta$  is shortened as in \eqref{eq: precision CI2} while the length of confidence interval ${\rm CI}_{\|\beta\|_2^2}$ is not shortened in terms of order of magnitude. However, the additional unlabelled data significantly weakens the model complexity and sample size condition for establishing the limiting distribution, where the sufficient condition for the supervised setting is $\frac{1}{\sigma}\|\beta\|_2\gg {k \log p}/{\sqrt{n}}$ and $k \ll {\sqrt{n}}/{\log p}$. Corollary \ref{cor: l2 norm} has shown that the condition $k \ll {\sqrt{n}}/{\log p}$ is not needed if there is sufficient amount of unlabelled data.


\section{Statistical Applications}
\label{sec: stat applications}
In this section, we apply the inference procedure related to the CHIVE estimator  to tackle several important statistical problems. 
\subsection{Application 1: Signal Detection and Global Testing}
\label{sec: detection}
Signal detection is of great importance in statistics and related scientific applications and the detection problem in high-dimensional linear regression was studied in \citet{arias2011global,ingster2010detection}. The inference procedure stated in Algorithm \ref{alg: SIVE-CRAN} has profound implications on signal detection and the general global testing in high-dimensional linear regression.
We consider the global hypothesis testing problem $
H_0: (\beta-\beta^{\rm null})^{\intercal}\Sigma (\beta-\beta^{\rm null})=0$ v.s. $H_1: (\beta-\beta^{\rm null})^{\intercal}\Sigma (\beta-\beta^{\rm null})>0.$ , which includes signal detection as a special case with $\beta^{\rm null}=0$. 
We apply Algorithm \ref{alg: SIVE-CRAN} with a given $\tau_0>0$ and obtain the point estimator $\widehat{\Q}^{R}(y-X\beta^{\rm null},X,\tau_0)$ and its standard error estimator $\widehat{\phi}^{E}(y-X\beta^{\rm null},X,\tau_0)$. Then we propose the detection procedure, with Type I error controlled at $\alpha\in(0,1)$ as
$D(\tau_0)={\bf{1}}\left(\widehat{\Q}^{R}(y-X\beta^{\rm null},X,\tau_0)\geq{\widehat{\phi}^{E}(y-X\beta^{\rm null},X,\tau_0)}{z_{\alpha}}\right). $
Define the null parameter space 
$
\mathcal{H}_0=\left\{\theta=\left(\beta^{\rm null},\Sigma,\sigma\right):\frac{1}{\Md} \leq \lambda_{\min}\left(\Sigma\right) \leq \lambda_{\max}\left(\Sigma\right) \leq \Md,\; \sigma \leq \Mv\right\}
$
and the local alternative parameter space as
{\footnotesize
\begin{equation*}
\mathcal{H}_1\left(\Delta\right)=\left\{\theta=\left(\beta,\Sigma,\sigma\right): (\beta-\beta^{\rm null})^{\intercal}\Sigma (\beta-\beta^{\rm null})= \frac{\Delta}{\sqrt{n}},\;\frac{1}{\Md} \leq \lambda_{\min}\left(\Sigma\right) \leq \lambda_{\max}\left(\Sigma\right) \leq \Md,\; \sigma \leq \Mv\right\}.
\label{eq: local alternative}
\end{equation*}}
The following corollary establishes that $D(\tau_0)$ controls the type I error asymptotically and also establishes the asymptotic power function of the proposed test.
\begin{Corollary}
Suppose that Conditions {\rm (A1)} and  {\rm (A2)} hold, $\tau_0>0$ is a positive constant and the vector $\delta=\beta-\beta^{\rm null}$ satisfies the conditions that $\|\delta\|_0 \ll \min \{n/(\log(N+n)\log p),{\sqrt{n}}/{\log p}\}$ and $\sqrt{\E\left({\delta^{\intercal}X_{1\cdot} X_{1\cdot}^{\intercal}\delta}-\delta^{\intercal}\Sigma\delta\right)^2}\geq c_0\delta^{\intercal}\Sigma\delta$ for some positive constant $c_0$. Then for any $\theta\in \mathcal{H}_0$, the type I error is controlled, $
\limsup_{n\rightarrow \infty}{\mathbb P}_{\theta}\left(D(\tau_0)=1\right)\leq \alpha.$
For $\rho>0$ and any $\theta\in \mathcal{H}_1(\Delta)$ with some positive constant $\Delta>0$, then
\begin{equation}
\small
\lim_{n\rightarrow \infty}{\mathbb P}_{\theta}\left(D(\tau_0)=1\right)=1-\Phi^{-1}\left(z_{\alpha}-\frac{\Delta}{{\sqrt{4\sigma^2\left(\delta^{\intercal}\Sigma\delta+\tau_0^2\right)+\rho\E\left(\delta^{\intercal}X_{1\cdot} X_{1\cdot}^{\intercal}\delta-\delta^{\intercal}\Sigma\delta\right)^2}}}\right).
\label{eq: power}
\end{equation}
\label{cor: signal detection}
\end{Corollary}
The assumptions of Corollary \ref{cor: signal detection} are the same as those of Corollary \ref{cor: explained variance ran} from the perspective that the conditions imposed on $\beta$ in Corollary \ref{cor: explained variance ran} are now  imposed on the difference vector $\delta=\beta-\beta^{\rm null}$. One sufficient condition for the difference vector $\delta$ being sparse is that both the true signal $\beta$ and the null hypothesis $\beta^{\rm null}$ are sparse. Corollary \ref{cor: signal detection} shows that for any positive constant $\tau_0$, $D(\tau_0)$ controls the type I error asymptotically. The asymptotic power of the proposed test is established in \eqref{eq: power}, where the additional unlabelled data proves useful in improving the power. See Section \ref{sec: detection-sim} for the improvement in the numerical studies. For the finite sample performance, we have investigated how to choose the randomization level $\tau_0$ in the simulation section. See Section \ref{sec: sim2} for the numerical performance.

\subsection{Application 2: Prediction Accuracy Assessment}
\label{sec: pred loss}
Inference for explained variance has important applications to evaluating the out-of-sample prediction for a given sparse estimator $\check{\beta}$. To keep the notation consistent, we assume $\check{\beta}$ is estimated based on a training data set $(X^{0},y^{0})$ and $(X,y)$ is an independent test data to evaluate its prediction accuracy. We start with computing the residual on the test data set   $y-X\check{\beta}=X(\beta-\check{\beta})+\epsilon.$
The  out-of-sample prediction accuracy is defined as ${{\rm PA}(\check{\beta})}=\E_{x_{\rm new}} \left(x_{\rm new}^{\intercal}(\check{\beta}-\beta)\right)^2=(\check{\beta}-\beta)^{\intercal}\Sigma(\check{\beta}-\beta)$ and it is reduced to the explained variance for the residual model with outcome $r=y-X\check{\beta}$ and covariates $X$. Let $\widehat{\Q}^{R}(r,X,\tau_0)$ and $\widehat{\phi}^{E}(r,X,\tau_0)$ denote the outputs of Algorithm \ref{alg: SIVE-CRAN} with the labeled data $\{(r_i,X_{i\cdot})\}_{1\leq i\leq n}$ and unlabelled data $\{X_{i\cdot}\}_{n+1\leq i\leq n+N}$ as inputs. Then we propose the point estimator of ${{\rm PA}(\check{\beta})}$ as $\widehat{\Q}^{R}(r,X,\tau_0)$ and the interval estimator for ${{\rm PA}(\check{\beta})}$ as
\begin{equation}
{\rm CI}_{{{\rm PA}(\check{\beta})}}=[(\widehat{\Q}^{R}(r,X,\tau_0)-z_{\alpha/2}{\widehat{\phi}^{E}(r,X,\tau_0)})_{+},\widehat{\Q}^{R}(r,X,\tau_0)+z_{\alpha/2}{\widehat{\phi}^{E}(r,X,\tau_0)}]
\label{eq: prediction accuracy CI}
\end{equation}
The following corollary establishes the convergence rate for the point estimator and the coverage and precision properties of the interval estimator.
\begin{Corollary} Suppose that Conditions {\rm (A1)} and {\rm (A2)} hold, $\tau_0>0$ is a positive constant and $c_0 \leq \lambda_{\min}\left(\Omega\right) \leq \lambda_{\max}\left(\Omega\right) \leq C_0$, $\sigma \leq \Mv$ for some positive constants $C_0\geq c_0>0$ and $\Mv>0$. For any sparse estimator satisfying $\|\check{\beta}\|_0 \leq C \|\beta\|_0$ and $C>0$,
\begin{enumerate}
\item If $k \leq c {n}/\log p$ for some positive constant $c>0$, then with probability larger than $\hprob$, 
	\begin{equation}
	\left|\widehat{\Q}^{R}(r,X,\tau_0)-\Q\right|\lesssim t\frac{\|\check{\beta}-\beta\|_2+\tau_0}{\sqrt{n}}+t\frac{\|\check{\beta}-\beta\|_2^2}{\sqrt{N+n}}+\left(\|\check{\beta}-\beta\|_2+1\right)\frac{k\log p}{n}
	\label{eq: estimation prediction accuracy}
	\end{equation}
\item	If $k \ll \min \{n/(\log(N+n)\log p),{\sqrt{n}}/{\log p}\}$ and $\sqrt{\E\left({\delta^{\intercal}X_{1\cdot} X_{1\cdot}^{\intercal}\delta}-\delta^{\intercal}\Sigma\delta\right)^2}\geq c_0\delta^{\intercal}\Sigma\delta$ for $\delta=\beta-\check{\beta}$ and some positive constant $c_0$,
then the confidence interval defined in \eqref{eq: prediction accuracy CI} satisfies the coverage property $
\liminf_{n\rightarrow\infty}\PP\left({{\rm PA}(\check{\beta})} \in {\rm CI}_{{{\rm PA}(\check{\beta})}}\right)\geq 1-\alpha$ and
\begin{equation}
\lim_{n\rightarrow\infty}\PP\left(\LL({\rm CI}_{{{\rm PA}(\check{\beta})}})\geq C \left(\frac{\|\check{\beta}-\beta\|_2+\tau_0}{\sqrt{n}}+\frac{\|\check{\beta}-\beta\|_2^2}{\sqrt{N+n}}\right)\right)=0
\label{eq: precision prediction accuracy}
\end{equation}
for some constant $C>0$.
\end{enumerate}
\label{cor: prediction accuracy}
\end{Corollary}
The above corollary has shown that the precision of confidence interval for the prediction accuracy is not just related to the sample sizes $n,N$, the sparsity $k$ and the dimension $p$, but also related to the accuracy of the evaluated estimator $\|\check{\beta}-\beta\|_2$. As characterized in \eqref{eq: estimation prediction accuracy} and \eqref{eq: precision prediction accuracy}, the integration of the unlabelled data is useful in improving the estimation accuracy and confidence interval precision. See Sections \ref{sec: semi-use} and \ref{sec: sim3} for the numerical performance.
\subsection{Application 3: Confidence Ball Construction}
\label{sec: confidence set}

The prediction accuracy evaluation established in \eqref{eq: prediction accuracy CI} can be used to construct confidence ball for $\beta$. For the setting where $\lambda_{\min}(\Sigma)$ is known, then we have $\lambda_{\min}(\Sigma)\|\check{\beta}-\beta\|_2^2\leq (\check{\beta}-\beta)^{\intercal}\Sigma(\check{\beta}-\beta)$ and construct the confidence ball for $\beta$ as 
\begin{equation}
{\rm CB}(\check{\beta})=\left\{\beta:  \|\beta-\check{\beta}\|_2^2\leq z_{\alpha/2}\frac{1}{\lambda_{\min}(\Sigma)}{\widehat{\phi}^{E}(r,X,\tau_0)}\right\}
\label{eq: CB}
\end{equation}
As shown in \eqref{eq: precision prediction accuracy}, the radius of the confidence ball ${\rm CB}(\check{\beta})$ is upper bounded by $\frac{\|\check{\beta}-\beta\|_2+\tau_0}{\sqrt{n}}+\frac{\|\check{\beta}-\beta\|_2^2}{\sqrt{N+n}}$. To minimize the radius, we need to select the center $\check{\beta}$ for the confidence ball in \eqref{eq: CB} such that $\check{\beta}$ is sparse and $\|\check{\beta}-\beta\|_2$ is small. In the high-dimensional literature, several penalized estimators are shown to satisfy such properties, such as Lasso, scaled Lasso and Dantzig Selector. 

%
\section{Simulation Study} 
\label{sec: sim}
We carry out simulation studies in this section to demonstrate the numerical performance of the CHIVE estimator. Specifically, we illustrate the numerical improvement of pooling over the unlabelled data in Section \ref{sec: semi-use}; we compare the performance of the CHIVE estimator with the plug-in estimator in Section \ref{sec: comparison}. Additional simulation results are postponed to Section \ref{sec: additional sim} in the supplementary material.

We first introduce the general simulation setting up used for this section. We generate the high-dimensional linear regression \eqref{eq: high dim linear model} with the dimension $p=800$ and the labelled data with sample size $n$ and unlabelled data with sample size $N$. For the linear model \eqref{eq: high dim linear model},  the covariates $\{X_{i\cdot}\}_{1\leq i\leq n}$ for the labelled data and also $\{X_{i\cdot}\}_{n+1\leq i\leq n+N}$ for the unlabelled data are generated in i.i.d. fashion to follow multivariate normal distribution with mean zero and covariance matrix $\Sigma\in \R^{800\times 800}$ and the errors $\{\epsilon_i\}_{1\leq i\leq n}$ are generated as i.i.d standard normal distribution. 

\subsection{Effect of Pooling-over Additional Unsupervised Data}
\label{sec: semi-use}
The focus of this section is to illustrate the improvement after integrating the unlabelled data in the semi-supervised setting. We first consider the inference problem for $\beta^{\intercal}\Sigma\beta$ and then the out-of-sample prediction loss evaluation.\\
\noindent \underline{\textbf{Inference for $\beta^{\intercal}\Sigma\beta$}}
We fix the labelled data sample size as $n=400$ and vary the unlabelled data sample size $N$ across $\{2,000, 6,000, 20,000\}$. We consider the following settings for the design covariance matrix $\Sigma$ and high-dimensional regression vector $\beta$, 
\begin{itemize}
\item Across Settings 1,2 and 3, the regression coefficients are generated as $\beta_i=i/10$ for $1\leq i\leq 0$ and $\beta_i=0$ for $i\geq 11$; The covariance matrix $\Sigma$ is generated as follows,
\begin{itemize}
\item Setting 1: $\Sigma_{ij}=0.5^{|i-j|}$;
\item Setting 2: $\Sigma_{ij}=0.35$ for $1\leq i\neq j \leq p$ and $\Sigma_{ii}=1$ for $1\leq i\leq p$;
\item Setting 3: $\Sigma_{ij}=0.7$ for $1\leq i\neq j \leq p$ and $\Sigma_{ii}=1$ for $1\leq i\leq p$.
\end{itemize}
\item Across Settings 4,5 and 6, the regression coefficients are generated as $\beta_i=1.5\cdot 0.8^{i}$ for $1\leq i\leq 800$. The covariance matrix $\Sigma$ is generated as follows,
\begin{itemize}
\item Setting 4: $\Sigma_{ij}=0.5^{|i-j|}$;
\item Setting 5: $\Sigma_{ij}=0.35$ for $1\leq i\neq j \leq p$ and $\Sigma_{ii}=1$ for $1\leq i\leq p$;
\item Setting 6: $\Sigma_{ij}=0.7$ for $1\leq i\neq j \leq p$ and $\Sigma_{ii}=1$ for $1\leq i\leq p$.
\end{itemize}
\end{itemize}
\begin{table}[htp]
\centering
\begin{tabular}{|r|r|rrr|rr|rrr|}
  \hline
  &&\multicolumn{3}{c}{RMSE}\vline&\multicolumn{2}{c}{Coverage}\vline &\multicolumn{3}{c}{Length}\vline\\
  \hline
 Setting& N& Semi-S & S & Ratio & Semi-S & S& Semi-S & S & Ratio \\ 
  \hline
\multirow{3}{*}{1}& 2000 & 0.420 & 0.733 & 57.2\% & 0.950 & 0.942 & 1.598 & 2.769 & 57.7\% \\ 
  &6000 & 0.370 & 0.751 & 49.2\% & 0.950 & 0.949 & 1.388 & 2.791 & 49.7\% \\ 
  &20000 & 0.341 & 0.732 & 46.6\% & 0.933 & 0.940 & 1.291 & 2.777 & 46.5\% \\ 
  \hline
\multirow{3}{*}{2}& 2000& 0.554 & 1.026 & 54.0\% & 0.933 & 0.928 & 2.077 & 3.834 & 54.2\% \\ 
  &6000  & 0.421 & 0.949 & 44.3\%& 0.951 & 0.957 & 1.741 & 3.855 & 45.2\% \\ 
  &20000 & 0.407 & 0.994 & 41.0\% & 0.940 & 0.948 & 1.581 & 3.838 & 41.2\% \\ 
    \hline
\multirow{3}{*}{3}& 2000 & 0.813 & 1.612 & 50.4\%& 0.950 & 0.958 & 3.213 & 6.539 & 49.1\% \\ 
   &6000 & 0.642 & 1.654 & 38.8\%& 0.960 & 0.946 & 2.510 & 6.530 & 38.4\% \\ 
   &20000& 0.559 & 1.597 & 35.0\%& 0.942 & 0.956 & 2.148 & 6.509 & 33.0\% \\ 
     \hline
\multirow{3}{*}{4}& 2000& 0.415 & 0.745 & 55.7\% & 0.938 & 0.939 & 1.591 & 2.740 & 58.1\% \\ 
  &6000  & 0.361 & 0.742 & 48.6\%& 0.932 & 0.935 & 1.383 & 2.742 & 50.4\% \\ 
   &20000 & 0.324 & 0.738 & 43.9\% & 0.955 & 0.949 & 1.290 & 2.748 & 46.9\% \\ 
     \hline
\multirow{3}{*}{5}& 2000& 0.589 & 1.088 & 54.2\% & 0.953 & 0.968 & 2.329 & 4.462 & 52.2\% \\ 
    &6000  & 0.496 & 1.149 & 43.2\% & 0.934 & 0.939 & 1.909 & 4.447 & 42.9\% \\ 
  &20000& 0.465 & 1.181 & 39.4\% & 0.936 & 0.935 & 1.713 & 4.441 & 38.6\% \\ 
    \hline
\multirow{3}{*}{6}& 2000 & 0.924 & 2.013 & 45.9\%& 0.962 & 0.949 & 3.698 & 7.689 & 48.1\% \\ 
  &6000  & 0.724 & 1.914 & 37.8\%& 0.945 & 0.951 & 2.822 & 7.692 & 36.7\% \\ 
  &20000 & 0.632 & 1.894 & 33.3\% & 0.935 & 0.959 & 2.371 & 7.696 & 30.8\% \\ 
   \hline
\end{tabular}
\caption{Inference for $\beta^{\intercal}\Sigma\beta$ with $n=400$ and $N=2000,6000,20000$}
\label{tab: semi-helpful}
\end{table}
Settings 1 to 3 correspond to the exact sparse case while Settings 4 to 6 correspond to the approximate sparse case. Settings 1 and 4 correspond to the case of approximated banded covariance matrix while Settings 2,3,5 and 6 are about denser covariance matrices.  
The simulations are replicated over 1,000 simulations. The Root Mean Squared Error (RMSE) and the coverage and length of confidence intervals are present in Table \ref{tab: semi-helpful}, where the columns under ``Semi-S" correspond to the semi-supervised method and the columns under ``S" correspond to the supervised method. Regarding RMSE, we observe that incorporation of unlabelled data reduces the RMSE significantly. The column under ``Ratio" reports the ratio of RMSE of the semi-supervised method to that of the supervised method and RMSE of the semi-supervised method is reduced to 33\% to 57\% of that of the supervised method, depending on the amount of the unlabelled data and also the structure on $\Sigma$. Since $X_{i\cdot}$ follows multivariate Gaussian, the variance component depending on the unlabelled data is expressed as ${\E\left(\beta^{\intercal}X_{1\cdot} X_{1\cdot}^{\intercal}\beta-\beta^{\intercal}\Sigma\beta\right)^2}/(N+n)=2 (\beta^{\intercal}\Sigma\beta)^2/(N+n)$. From setting 1 to setting 3, the value $\beta^{\intercal}\Sigma\beta$ increases as $\Sigma$ becomes denser and this explains why the effect of using the unlabelled data becomes more significant; the same phenomenon holds for settings 4 to 6. 

In terms of constructed confidence intervals, both confidence intervals constructed in the semi-supervised setting and supervised setting have near $95\%$ coverage while the confidence interval constructed using the unlabelled data have much shorter lengths. Specifically, we use ``Ratio" to measure the ratio of the length of CI in the semi-supervised setting to that in the supervised setting and observe that the length of confidence intervals can be reduced by as much as 70\%.

The unlabelled data is not just useful in inference for $\beta^{\intercal}\Sigma\beta$, but also useful in prediction loss evaluation, which will be illustrated in the following.\\
\noindent \underline{\textbf{Prediction Loss Evaluation}}
We generate $\beta_i=i/5$ for $1\leq i\leq 0$ and $\beta_i=0$ for $i\geq 11$ and $\Sigma_{ij}=0.5^{|i-j|}$. 
We fix the labelled data sample size as $n=400$ and vary the unlabelled data sample size $N$ across $\{2,000, 6,000, 20,000\}$. We use this generated data (both labelled and unlabelled) to evaluate the out-of-sample prediction accuracy $(\widehat{\beta}(\lambda)-\beta)^{\intercal}\Sigma(\widehat{\beta}(\lambda)-\beta)$, where $\widehat{\beta}(\lambda)$ is the Lasso estimator based on an independent training data $\left(X^{(0)},y^{(0)}\right)$ with sample size $300$ with the tuning parameter $\lambda,$
$
\widehat{\beta}\left(\lambda\right)=\arg\min_{\beta \in \R^{p}}\frac{\|y^{(0)}-X^{(0)}\beta\|_2^2}{2 n_0}+  {\lambda} \sum_{j=1}^{p} \frac{\|X^{(0)}_{\cdot j}\|_2}{\sqrt{n_0}} |\beta_j|. 
$
Note that $\left(X^{(0)},y^{(0)}\right)$ is an independent copy of the labelled data $\left(X,y\right)$. Specifically, we consider three estimators $\widehat{\beta}(\lambda_0),\widehat{\beta}(6\lambda_0)$ and $\widehat{\beta}(10\lambda_0)$ with 
$\lambda_{0}=\sqrt{\frac{z_{1-1/(10p)}}{n_0}}$ and use the randomization  level $\tau_0=2$ in terms of estimating this out-of-sample prediction accuracy. 

\begin{table}[htp]
\centering
\resizebox{\columnwidth}{!}{
\begin{tabular}{|r|r|r|rrr|rr|rrr|}
  \hline
 &&&\multicolumn{3}{c}{RMSE}\vline&\multicolumn{2}{c}{Coverage}\vline &\multicolumn{3}{c}{Length}\vline\\
  \hline
Estimator& Loss&N& Semi-S & S & Ratio & Semi-S & S& Semi-S & S & Ratio \\ 
  \hline
\multirow{3}{*}{$\widehat{\beta}(\lambda_0)$}&\multirow{3}{*}{0.145}&2000& 0.269 & 0.279 & 96.3\% & 0.910 & 0.898 & 0.896 & 0.898 & 99.8\% \\ 
   &&6000& 0.270 & 0.281 & 96.3\% & 0.918 & 0.902 & 0.895 & 0.897 & 99.8\% \\ 
   &&10000& 0.262 & 0.273 & 96.0\% & 0.924 & 0.910 & 0.895 & 0.897 & 99.8\% \\ 
  \hline
\multirow{3}{*}{$\widehat{\beta}(6\lambda_0)$} &\multirow{3}{*}{1.818}&2000& 0.294 & 0.363 & 81.2\% & 0.924 & 0.896 & 1.046 & 1.148 & 91.1\% \\ 
   &&6000& 0.308 & 0.373 & 82.6\% & 0.918 & 0.888 & 1.042 & 1.148 & 90.8\% \\ 
   &&10000& 0.299 & 0.368 & 81.1\% & 0.926 & 0.892 & 1.038 & 1.148 & 90.3\%\\ 
  \hline
\multirow{3}{*}{$\widehat{\beta}(10\lambda_0)$}&\multirow{3}{*}{4.679}& 2000& 0.365 & 0.548 & 66.5\% & 0.930 & 0.928 & 1.318 & 1.841 & 71.6\% \\ 
   &&6000& 0.378 & 0.553 & 68.3\% & 0.934 & 0.902 & 1.291 & 1.839 & 70.2\% \\ 
   &&10000& 0.362 & 0.551 & 65.8\% & 0.920 & 0.916 & 1.267 & 1.841 & 68.8\%\\ 
   \hline
\end{tabular}}
\caption{Inference for the out-of-sample prediction accuracy $(\widehat{\beta}-\beta)^{\intercal}\Sigma(\widehat{\beta}-\beta)$.} 
\label{tab: PA semi}
\end{table}

The simulations are replicated over 1,000 simulations and we report the numerical performance of both point and interval estimators of the corresponding prediction accuracy in Table \ref{tab: PA semi}. The observation is consistent with that for $\beta^{\intercal}\Sigma\beta$, where CIs in both semi-supervised and supervised settings have coverage but the semi-supervised estimators are uniformly better than the supervised estimators in terms of both RMSE and the length of CI. As observed in Table \ref{tab: PA semi}, across the three estimators $\widehat{\beta}(\lambda_0),\widehat{\beta}(6\lambda_0)$ and $\widehat{\beta}(10\lambda_0)$, the effect of unlabelled data is different. The effect of unlabelled data for estimating $\widehat{\beta}(\lambda_0)$ is marginal while the effect of unlabelled data $\widehat{\beta}(10 \lambda_0)$ is much more significant, where RMSE and length of CI can be reduced by 30\%. This matches with the theory, where in the simulation setting of Gaussian design, the unlabelled data reduces the term $((\widehat{\beta}-\beta)^{\intercal}\Sigma(\widehat{\beta}-\beta))^2/(N+n)$ and $(\widehat{\beta}-\beta)^{\intercal}\Sigma(\widehat{\beta}-\beta)$ is pretty small ($0.145$) for $\widehat{\beta}=\widehat{\beta}(\lambda_0)$ and is much larger ($4.679$) for $\widehat{\beta}=\widehat{\beta}(10\lambda_0)$.

The semi-supervised data is also useful for improving the power for signal detection. Since the detection power is only improved around 5\%, we defer the detailed results to the supplementary material Section \ref{sec: detection-sim}.


\subsection{Comparison with Other Estimators}
\label{sec: comparison}
In the following, we compare the CHIVE estimator with the plug-in estimator. We fix the size of unlabelled data at $N=2,000$ and vary the labelled data sample size $n$ across $\{200,400,600,800,1,000\}.$
The simulations are replicated over 500 simulations. 
We generate the design covariance matrix as $\Sigma_{ij}=0.5^{|i-j|}$ and the high-dimensional regression vector $\beta$ across the following three settings,
\begin{enumerate}
	\item[a.] {\bf Setting a:} $\beta$ is generated with sparsity $10$ where $\beta_{j}=j/10$ for $1\leq j\leq 10$ and $\beta_j=0$ for $j\geq 11$;
	\item[b.] {\bf Setting b: } $\beta$ is generated with sparsity $50$ where $\beta_{j}=j/50$ for $1\leq j\leq 50$ and $\beta_j=0$ for $j\geq 51$;
	\item[c.] {\bf Setting c: } $\beta$ is generated as approximate sparse vector with $\beta_j=(0.5)^{p-1}$.
\end{enumerate}	  

\begin{figure}[ht]
\centering
\includegraphics[height=5.5cm, width=15cm]{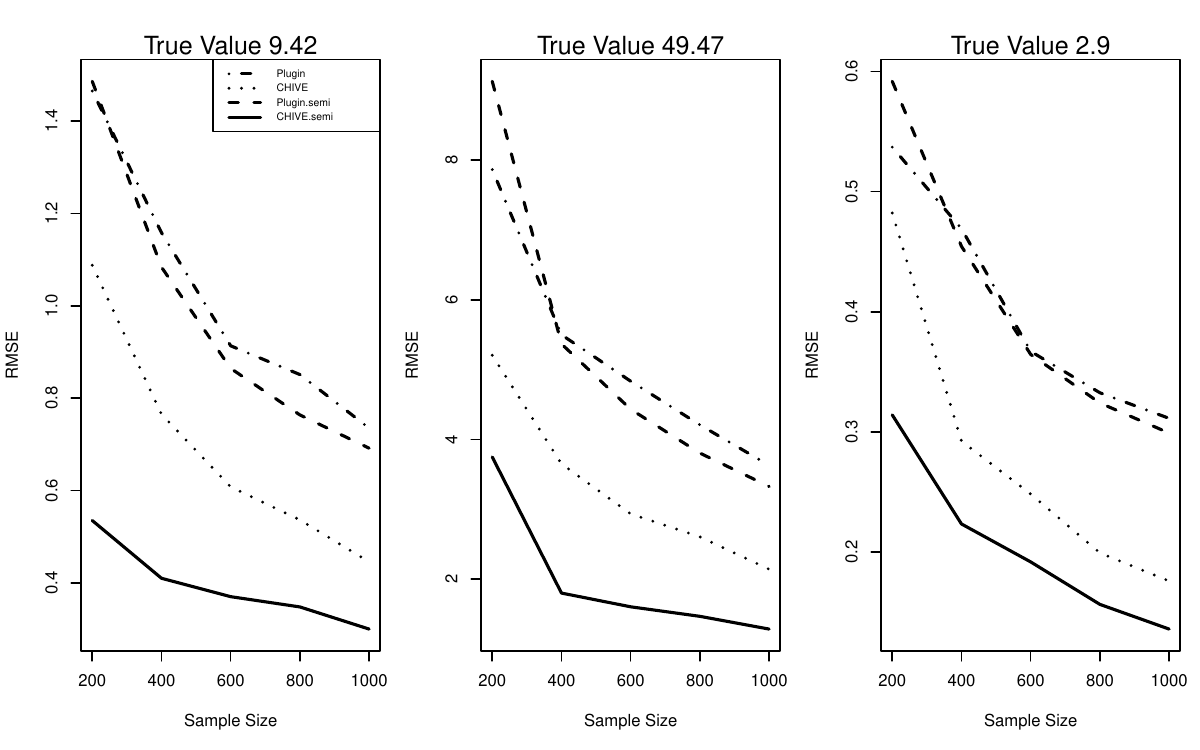}
\caption{\footnotesize  Root Mean Squared Error (RMSE) of different estimators of $\beta^{\intercal}\Sigma\beta$. The x-axis stands for the sample size and y-axis stands for the RMSE of corresponding estimators. The dotted line and the solid line represent the corresponding RMSEs of the CHIVE estimator in the supervised setting and semi-supervised setting, respectively; The dashed line and the dotted-dashed solid line represent the corresponding RMSE of the plug-in estimator in the supervised setting and semi-supervised setting, respectively. From the leftmost to the rightmost, the figures correspond to setting a to c, with the corresponding values for $\beta^{\intercal}\Sigma\beta$ as 9.42, 49.47 and 2.9.}
\label{fig: sim-figure1}
\end{figure}

We compare four different estimators, where ``CHIVE" and ``CHIVE.semi" stand for the CHIVE estimator in the supervised setting and semi-supervised setting, respectively; ``Plugin" and ``Plugin.semi" stand for the plug-in estimator $\widehat{\beta}^{\intercal}\widehat{\Sigma}\widehat{\beta}$ in the supervised setting and semi-supervised setting, respectively.
The numerical comparison has been reported in Figure \ref{fig: sim-figure1}. Across all three settings, it is observed that the proposed CHIVE estimator has achieved uniformly  much better estimation accuracy than the plug-in estimators, in both supervised and semi-supervised settings. This numerical observation demonstrates that the calibration step is useful in improving the estimation accuracy.

 We shall also point out that the unlabelled data is useful only if it is incooporated in a proper way.  ``Plugin.semi" is another estimator also using the unlabelled data to estimate ${\Sigma}$, but it is only slightly better than the ``Plugin" estimator. In contrast, together with the calibration machinery, ``CHIVE.semi" uses the additional data in an efficient way and the corresponding RMSE is significantly reduced in comparison to the ``CHIVE" estimator. 


\section{Real Data Application}
\label{sec: real data}
 In this section, we analyze a yeast data set reported in \citet{bloom2013finding} and study how the genetic variants explain the colony sizes under different growth media. The goal is to estimate the heritability measures of colony sizes under different growth media, which represent the variance of the colony sizes explained by the genetic variants. 
 
 \citet{bloom2013finding}  investigated a large scale genome-wide association study of 46 quantitative traits based on  1,008  {\it Saccharomyces cerevisiae} segregants crossbred from a laboratory strain and a wine strain. These quantitative traits are measures of end-point colony size under 46 different growth media, including Hydrogen Peroxide, Cadmium Chloride, Calcium Chloride, Lactose, Raffinose, Sorbitol, Yeast Nitrogen Base (YNB) and Yeast Peptone Dextrose (YPD). The genetic maker genotypes are coded  as $1$ or $-1$, according to which strain it comes from. A set of 11,623 unique genotype markers of the 1,008 segregants is measured. Since many of these markers are highly correlated and the corresponding codes are only different in serval samples, \citet{bloom2013finding} further selected a  set of $4,410$ markers that are weakly dependent based on the linkage disequilibrium information. All traits are normalized to have unit variance and hence the explained variance is a measure for heritability. 
\citet{bloom2013finding}  showed that the genetic variants are associated with many of such trait values and highlighted the importance of addressing {\em missing heritability}. \citet{bloom2013finding} pointed out one key reason for missing heritability as ``the undiscovered factors could have effects that are too small to be detected with current sample sizes, or even too small to ever be individually detected with statistical significance".  We demonstrate that the CHIVE estimator has exactly addressed  this concern of missing heritability. As reported in Table \ref{tab: heritability inference}, we choose $6$ traits out of the total $46$ traits and observe that the CHIVE estimates are always larger than the corresponding plug-in estimates. This means that the calibration step adds back the missing heritability due to plugging in the Lasso estimator, where the Lasso estimator tends to ignore the genetic markers with small effects. The results for all $46$ traits is reported in Section \ref{sec: add real} in the supplementary material.

We also construct confidence intervals for heritability of all $46$ traits and report part of the results in Table \ref{tab: heritability inference}. Note that a proportion of the outcome variables for different growth media have missing values, with the proportion of missing ranging from 0.2\% to 40.58\%. This forms the semi-supervised type data naturally (note that the unlabelled data is of a smaller size than the labelled data in this specific example). After applying the proposed methods to analyzing the corresponding outcomes, we have the following interesting observations, 1) the heritability measures of the colony sizes under different growth media range from 0.3 to 0.8 and all of the confidence interval estimators do not contain zero. This means the colony sizes under different growth media are strongly genetically heritable; 2) The integration of the unlabelled data has shortened the length of the constructed confidence intervals. For example the length is shorten by around $3\%$ for Sorbitol (with 40.58\% outcome missing), around $2\%$ for Raffinose (with 34.33\% outcome missing) and around $1\%$ for Hydrogen Peroxide (with 23.71\% outcome missing).


\begin{table}[ht]
\resizebox{\columnwidth}{!}{
\begin{tabular}{|c|ccc|ccc|c|}
  \hline
&\multicolumn{3}{c}{Supervised}\vline& \multicolumn{3}{c}{Semi-Supervised}\vline&\\
\hline
 Media& Plug & CHIVE & CI & Plug & CHIVE & CI & Missing \\ 
  \hline\hline
Cadmium & 0.6240 &  0.7682 & [0.7077, 0.8286] &0.6215 & 0.7657 &  [0.7058, 0.8256] & 20.73\% \\ 
Chloride&&(0.0308)&&&(0.0306)&&\\
\hline
  Calcium&  0.1807 & 0.3701 & [0.3068, 0.4333] & 0.1785 & 0.3679  & [0.3050, 0.4308] & 5.85\% \\ 
Chloride&&(0.0323)&&&(0.0321)&&\\
\hline
  Hydrogen& 0.2909  & 0.4835 & [0.4090, 0.5581] & 0.2879 & 0.4806  & [0.4071, 0.5540] & 23.71\% \\ 
Peroxide&&(0.0380) &&& (0.0375)&&\\
\hline
  Raffinose & 0.3168  & 0.5105 &  [0.4300, 0.5909] & 0.3105& 0.5041  & [0.4259, 0.5824] & 34.33\% \\ 
  &&(0.0410)&&& (0.0399)&&\\
\hline  
  Sorbitol & 0.2968  & 0.4893 & [0.4049, 0.5737]  & 0.2864 & 0.4789  & [0.3972, 0.5606] & 40.58\% \\ 
  &&(0.0431) &&& (0.0417)&&\\
\hline  
  YPD & 0.3754  & 0.5960 &  [0.5275,0.6645] & 0.3761 & 0.5966  & [0.5282, 0.6651] & 0.20\% \\ 
  &&(0.0349) &&& (0.0349)&&\\
   \hline
\end{tabular}
}
\caption{\footnotesize Confidence intervals for heritability. The column indexed with `` Media" represents the growth media for the yeast segragents; The three columns under ``Supervised" corresponds to the case of only using the labelled data, where the column indexed with ``Plug" represents the plug-in estimator, indexed with ``CHIVE" represents the CHIVE estimator, and indexed with ``CI" represents the constructed confidence interval; Similarly, the three columns under ``Semi-Supervised" corresponds to analyzing the semi-supervised type data, that is also using the observations with missing outcome variables. The numbers inside the parenthesis represent the standard errors of the proposed CHIVE estimators. The column indexed with ``Missing" represents the proportion of missing outcome for the corresponding media.}
\label{tab: heritability inference}
\end{table}


\section{Discussions}
\label{sec: discussion}
This paper studies statistical inference for the explained variance $\beta^{\intercal}\Sigma\beta$ in the semi-supervised setting, which includes the supervised setting as a special case. By comparing the theoretical as well as the numerical results for the semi-supervised and supervised settings, it is easy to see the significant contributions of the unlabelled data to the inference accuracy. In addition, the constructed confidence interval, using the idea of calibration, has been shown to be useful in tackling other important statistical applications, including signal detection and global testing, prediction accuracy evaluation and confidence ball construction. There remain a few open questions for future research. 

Although the CHIVE estimator has been shown to achieve the optimal rates over the whole sparse regime $k\lesssim {n}/{\log p}$, construction of confidence intervals for $\beta^{\intercal}\Sigma\beta$ is only considered over the ultra-sparse regime $k\ll {\sqrt{n}}/{\log p}$. Since both point and interval estimator do not require the prior knowledge of the exact sparsity level, they are referred to as adaptive estimation and adaptive confidence interval, respectively. However, it remains open whether it is possible to construct adaptive confidence intervals over the moderate sparse regime ${\sqrt{n}}/{\log p}\lesssim k\lesssim {n}/{\log p}$. The possibility of adaptive confidence interval for the general linear functional $\eta^{\intercal}\beta$ for $\eta\in \R^{p}$ has been studied in \citet{cai2017confidence} and the technical tools developed in \citet{cai2017confidence} can be useful to study the adaptive confidence intervals for  $\beta^{\intercal}\Sigma\beta$. 
%
 
Due to the emerging semi-supervised data sets, it is of significant importance to propose procedures incorporating the unlabelled data efficiently and study how the unlabelled data affects the statistical accuracy. This paper has studied both methodological and theoretical perspectives of the semi-supervised statistical inference for the explained variance $\beta^{\intercal}\Sigma\beta$ and the unweighted quadratic functional  $\|\beta\|_2^2$. However, it is largely unknown how these unlabelled data can facilitate the statistical inference problem for other quantities of interests, such as the general linear functional $\eta^{\intercal}\beta$ for some given $\eta\in \R^{p}$ and the variance level $\sigma^2$. These are interesting problems left for future research. 
\section*{Acknowledgments}

The research of Tony Cai was supported in part by NSF Grant DMS-1712735 and NIH grants R01-GM129781 and R01-GM123056. The research of Zijian Guo was supported in part by NSF DMS 1811857. The authors are grateful for the constructive and helpful comments from the Editor, the Associate Editor and three referees.

\bibliographystyle{plainnat}
\bibliography{HDRef}

\input{Supplement}

\end{document}

%% file: Supplement.tex

\newpage
\appendix

\section{Additional Results on Semi-supervised Inference for $\|\beta\|_2^2$}
\label{sec: unweighted add}
The following theorem establishes the rate of convergence of the estimator defined in \eqref{eq: semi extend} for general $N\geq 0$, which is more general than the results presented in Theorem \ref{cor: l2 norm}.
\begin{Theorem} Suppose that Condition {\rm (A1)} holds, $k\leq c n/\log p$ for some constant $c>0$ and $c_0 \leq \lambda_{\min}\left(\Omega\right) \leq \lambda_{\max}\left(\Omega\right) \leq C_0$ for some positive constants $C_0\geq c_0>0$. For any estimator $\widehat{\beta}$ satisfying  {\rm (B1)} and $\widehat{\Omega}$ satisfying  {\rm (B3)}, with probability larger than $\hprobsup-\gamma_1(N+n)$,  then $\widehat{\|\beta\|_2^2}$ proposed in \eqref{eq: semi extend} satisfies 
\begin{equation}
\left|\widehat{\|\beta\|_2^2}-\|\beta\|_2^2\right|\lesssim \sigma\frac{\|\beta\|_2}{\sqrt{n}}\left(1+C_{\Omega}\frac{s\sqrt{k}\log p}{\sqrt{N+n}}\right)+k\frac{\log p}{n}\sigma^2\left(1+C_{\Omega}\frac{s\sqrt{\log p}}{\sqrt{N+n}}\right).
\label{eq: special case 1}
\end{equation}
\label{thm: general thm of l2 norm}
\end{Theorem}
The above theorem illustrates the usefulness of the unlabelled data, where the amount of the unlabelled data $N$ plays a role in \eqref{eq: special case 1}.

\section{Proof}
\label{sec: proof}
In this section, we will prove theorems and corollaries in the main paper and the proofs of lemmas are present in Section \ref{sec: lemma proof}.  The proofs of Theorem \ref{thm: est bound semi} and Corollary \ref{cor: est bound} are present in Section \ref{sec: proof 1}; The proof of Theorem \ref{thm: randomization} is present in Section \ref{sec: proof 2};  The proof of Theorem \ref{thm: explained variance} is present in Section \ref{sec: proof 3}; The proof of Theorem \ref{thm: est bound semi optimal} is present in Section \ref{sec: proof 4}; The proof of Theorem \ref{thm: lower bound} is present in Section \ref{sec: proof 5}; The proof of Corollary \ref{cor: upper bound special 2} is present in Section \ref{sec: proof 6}; Thr proofs of Corollaries \ref{cor: explained variance enlarge} and \ref{cor: explained variance ran} are present in Section \ref{sec: proof 7}; The proofs of Theorem \ref{thm: general thm of l2 norm} and Corollary \ref{cor: l2 norm} are present in Section \ref{sec: proof 8}; The proofs of Corollaries \ref{cor: signal detection} and \ref{cor: prediction accuracy} are present in Sections  \ref{sec: proof 9} and \ref{sec: proof 10}, respectively.

\subsection{Proofs of Theorem \ref{thm: est bound semi} and Corollary \ref{cor: est bound}}
\label{sec: proof 1}

To establish Theorem \ref{thm: est bound semi} and Corollary \ref{cor: est bound}, we first decompose the difference between the calibrated estimator $\widehat{\Q}={\widehat{\Q}}(\widehat{\beta},\widehat{\Sigma}^{S})$ and $\Q={\beta^{\intercal}\Sigma\beta}$, 
\begin{equation}
\small
\begin{aligned}
&\widehat{\Q}-\Q=\frac{2}{n}\widehat{\beta}^{\intercal}X^{\intercal}\epsilon+\beta^{\intercal}\left(\widehat{\Sigma}^{S}-\Sigma\right)\beta-(\widehat{\beta}-\beta)^{\intercal}\widehat{\Sigma}^{S}(\widehat{\beta}-\beta)+2\widehat{\beta}^{\intercal}(\widehat{\Sigma}^{S}-\SigmaLa)(\widehat{\beta}-\beta)\\
=&\frac{2}{n}{\beta}^{\intercal}X^{\intercal}\epsilon+\beta^{\intercal}\left(\widehat{\Sigma}^{S}-\Sigma\right)\beta+\frac{2}{n}(\widehat{\beta}-\beta)^{\intercal}X^{\intercal}\epsilon-(\widehat{\beta}-\beta)^{\intercal}\widehat{\Sigma}^{S}(\widehat{\beta}-\beta)+2\widehat{\beta}^{\intercal}(\widehat{\Sigma}^{S}-\SigmaLa)(\widehat{\beta}-\beta).\\
\end{aligned}
\label{eq: decomposition of the diff}
\end{equation}
Lemma \ref{lem: key lemma 1} characterizes the convergence rates of the last three terms in \eqref{eq: decomposition of the diff}. The proof can be found in Section \ref{sec: lemma 1 proof} in the appendix. 
\begin{Lemma}
\label{lem: key lemma 1}
Suppose that Condition {\rm (A1)} holds and $k\leq c n/\log p$ for some constant $c>0$.  For any estimator $\widehat{\beta}$ satisfying  Condition  {\rm (B1)},  then with probability larger than $1-\gamma(n)-cp^{-c}-\exp(-c{N})-e^{-ct^2}$,
\begin{equation}
\left|\frac{1}{n}(\widehat{\beta}-\beta)^{\intercal}X^{\intercal}\epsilon\right| \leq \|\widehat{\beta}-\beta\|_1\left\|\frac{1}{n}X^{\intercal}\epsilon\right\|_{\infty}\lesssim \frac{k \log p}{n}\sigma^2;
\label{eq: error bound 1}
\end{equation}
\begin{equation}
\left|(\widehat{\beta}-\beta)^{\intercal}\widehat{\Sigma}^{S}(\widehat{\beta}-\beta)\right| = \frac{1}{N+n}\sum_{i=1}^{N+n}\left(X_{i\cdot}^{\intercal}(\widehat{\beta}-\beta)\right)^2\lesssim \frac{k \log p}{n}\sigma^2. 
\label{eq: error bound 2}
\end{equation}
\begin{equation}
\left|\widehat{\beta}^{\intercal}(\widehat{\Sigma}^{S}-\SigmaLa)(\widehat{\beta}-\beta)\right|\lesssim k \frac{\log p}{n}\sigma^2 +  \|\Sigmahalf \beta\|_2 \sigma \left(t\frac{\sqrt{N}}{n+N}\sqrt{\frac{k \log p}{n}}+\frac{N}{n+N}\frac{k \log p}{n}\right)
\label{eq: error bound 3}
\end{equation}
\end{Lemma}
\noindent For the first two terms in \eqref{eq: decomposition of the diff}, the following lemma establishes their convergence rate and also the limiting distribution. The proof can be found in Section \ref{sec: lemma 2 proof} in the appendix. 
\begin{Lemma}
\label{lem: key lemma 2}
Suppose that Condition {\rm (A1)} holds and $k\leq c n/\log p$ for some constant $c>0$.  Then with probability larger than $1-e^{-ct^2}$,
\begin{equation}
\left|\frac{2}{n}{\beta}^{\intercal}X^{\intercal}\epsilon\right|\lesssim t \frac{\|\Sigmahalf \beta\|_2 }{\sqrt{n}}\sigma, \quad \left|\beta^{\intercal}\left(\widehat{\Sigma}^{S}-\Sigma\right)\beta\right|\lesssim t \frac{\|\Sigmahalf \beta\|_2^2}{\sqrt{N+n}}
\label{eq: key distribution 1}
\end{equation}
In addition, we establish the limiting distribution
\begin{equation}
\sqrt{n}\frac{\frac{2}{n}{\beta}^{\intercal}X^{\intercal}\epsilon+\beta^{\intercal}\left(\widehat{\Sigma}^{S}-\Sigma\right)\beta}{\sqrt{4\sigma^2\beta^{\intercal}\Sigma\beta+\rho\E\left(\beta^{\intercal}X_{1\cdot} X_{1\cdot}^{\intercal}\beta-\beta^{\intercal}\Sigma\beta\right)^2}}\cid N\left(0, 1\right)
\label{eq: key distribution 2}
\end{equation}
\end{Lemma}
\noindent \underline{\em Proof of Theorem \ref{thm: est bound semi}.} In proving Theorem \ref{thm: est bound semi}, the convergence rate in \eqref{eq: convergence rate semi} follows from the decomposition \eqref{eq: decomposition of the diff}, Lemma \ref{lem: key lemma 1}, \eqref{eq: key distribution 1} in Lemma \ref{lem: key lemma 2} and the fact that $\frac{\sqrt{N}}{n+N}\sqrt{\frac{k \log p}{n}}\ll \frac{1}{\sqrt{n}}. $ Under the additional assumptions $k \ll {\sqrt{n}}/{\log p}$ and $\SNR=\frac{1}{\sigma}\|\Sigmahalf \beta\|_2\gg {k \log p}/{\sqrt{n}}$, it follows from Lemma \ref{lem: key lemma 1} that
\begin{equation*}
\frac{\sqrt{n}\left(\frac{2}{n}(\widehat{\beta}-\beta)^{\intercal}X^{\intercal}\epsilon-(\widehat{\beta}-\beta)^{\intercal}\widehat{\Sigma}(\widehat{\beta}-\beta)+2\widehat{\beta}^{\intercal}\left(\widehat{\Sigma}-\SigmaLa\right)(\widehat{\beta}-\beta)\right)}{{\sqrt{4\sigma^2\beta^{\intercal}\Sigma\beta+\rho\E\left(\beta^{\intercal}X_{1\cdot} X_{1\cdot}^{\intercal}\beta-\beta^{\intercal}\Sigma\beta\right)^2}}}\cip 0.
\end{equation*}
Combined with \eqref{eq: key distribution 2} in Lemma \ref{lem: key lemma 2}, we establish the limiting distribution \eqref{eq: distribution semi}. 

\noindent \underline{\em Proof of Corollary \ref{cor: est bound}.} The proof of Corollary \ref{cor: est bound} is similar to that of Theorem \ref{thm: est bound semi}. The main change is that $\widehat{\Sigma}^{S}$ in \eqref{eq: decomposition of the diff} is replaced by $\widehat{\Sigma}^{L}=\frac{1}{n}\sum_{i=1}^{n} X_{i\cdot} X_{i\cdot}^{\intercal}$ and hence the last term $2\widehat{\beta}^{\intercal}(\widehat{\Sigma}^{S}-\SigmaLa)(\widehat{\beta}-\beta)$ in the decomposition \eqref{eq: decomposition of the diff} becomes zero in this case.  Hence, the convergence rate in \eqref{eq: convergence rate} follows from the decomposition \eqref{eq: decomposition of the diff} and Lemma \ref{lem: key lemma 1} and \eqref{eq: key distribution 1} in Lemma \ref{lem: key lemma 2}. Under the additional assumptions $\SNR=\frac{1}{\sigma}\|\Sigmahalf \beta\|_2\gg \min\left\{{k \log p}/{\sqrt{n}},\left({k \log p}/{\sqrt{n}}\right)^{1/2}\right\}$ and the condition ${\rm (A2)}$
 it follows from Lemma \ref{lem: key lemma 1} that
$\frac{\sqrt{n}\left(\frac{2}{n}(\widehat{\beta}-\beta)^{\intercal}X^{\intercal}\epsilon-(\widehat{\beta}-\beta)^{\intercal}\widehat{\Sigma}(\widehat{\beta}-\beta)\right)}{\sqrt{4\sigma^2\beta^{\intercal}\Sigma\beta+\E\left(\beta^{\intercal}X_{1\cdot} X_{1\cdot}^{\intercal}\beta-\beta^{\intercal}\Sigma\beta\right)^2}}\cip 0.$
Combined with \eqref{eq: key distribution 2} in Lemma \ref{lem: key lemma 2}, we establish the limiting distribution \eqref{eq: distribution}. 

\subsection{Proof of Theorem \ref{thm: randomization}}
\label{sec: proof 2}
Following from \eqref{eq: decomposition of the diff}, we establish the error decomposition of ${\widehat{\Q}^{R}}-\Q$, 
\begin{equation}
\begin{aligned}
{\widehat{\Q}^{R}}-\Q&=\frac{2}{n}{\beta}^{\intercal}X^{\intercal}\epsilon+\frac{2}{n}u^{\intercal}\epsilon+\beta^{\intercal}\left(\widehat{\Sigma}^{S}-\Sigma\right)\beta+\frac{2}{n}u^{\intercal}X^{\intercal}\left(\beta-\widehat{\beta}\right)\\
&+\frac{2}{n}(\widehat{\beta}-\beta)^{\intercal}X^{\intercal}\epsilon-(\widehat{\beta}-\beta)^{\intercal}\widehat{\Sigma}^{S}(\widehat{\beta}-\beta)+2\widehat{\beta}^{\intercal}(\widehat{\Sigma}^{S}-\SigmaLa)(\widehat{\beta}-\beta).
\label{eq: decomp randomize}
\end{aligned}
\end{equation}
The theorem follows from Lemma \ref{lem: key lemma 1}, the decomposition \eqref{eq: decomp randomize} and the following lemma, whose proof is postponed to Section \ref{sec: lemma3 proof} in the appendix.
\begin{Lemma}
\label{lem: key lemma random}
Under the same assumptions as Theorem \ref{thm: randomization}, we have 
\begin{equation}
\sqrt{n}\frac{\frac{2}{n}{\beta}^{\intercal}X^{\intercal}\epsilon+\frac{2}{n}u^{\intercal}\epsilon+\beta^{\intercal}\left(\widehat{\Sigma}^{S}-\Sigma\right)\beta+\frac{2}{n}u^{\intercal}X^{\intercal}\left(\beta-\widehat{\beta}\right)}{\sqrt{4\sigma^2\left(\beta^{\intercal}\Sigma\beta+\tau_0^2\right)+\rho\E\left(\beta^{\intercal}X_{1\cdot} X_{1\cdot}^{\intercal}\beta-\beta^{\intercal}\Sigma\beta\right)^2}}\cid N\left(0, 1\right)
\label{eq: key distribution 2 random}
\end{equation}
\end{Lemma}

\subsection{Proof of Theorem \ref{thm: explained variance}}
\label{sec: proof 3}
Define $\phi_1={\sigma^2}{\beta}^{\intercal}{\Sigma}{\beta}$ and $\phi_2=\E\left(\beta^{\intercal}X_{1\cdot} X_{1\cdot}^{\intercal}\beta-\beta^{\intercal}\Sigma\beta\right)^2$. Recall the definitions $$\widehat{\phi}_1={\widehat{\sigma}^2\widehat{\beta}^{\intercal}\widehat{\Sigma}^{S}\widehat{\beta}} \quad \text{and} \quad \widehat{\phi}_2={\frac{1}{n+N}\sum_{i=1}^{n+N} \left(\widehat{\beta}^{\intercal} X_{i\cdot} X_{i\cdot}^{\intercal}\widehat{\beta}-\widehat{\beta}^{\intercal}\widehat{\Sigma}^{S}\widehat{\beta}\right)^2}.$$ 
The coverage property \eqref{eq: coverage CI2} follows from the following observation,
\begin{equation}
\PP\left(\beta^{\intercal}\Sigma\beta \in {\rm CI}(Z)\right)
=\PP\left(-z_{\frac{\alpha}{2}}\leq \frac{{\sqrt{n}}\left(\widehat{\Q}-\beta^{\intercal}\Sigma\beta\right)}{\sqrt{4{\phi}_1+{\rho}{\phi}_2}}\cdot\sqrt{\frac{{4{\phi}_1+{\rho}{\phi}_2}}{{4\widehat{\phi}_1+\widehat{\rho}\widehat{\phi}_2}}}\leq z_{\frac{\alpha}{2}}\right)
\label{eq: sample est coverage}
\end{equation}
The precision property of the constructed confidence intervals require the following lemma and the proof can be found in Section \ref{sec: lemma4 proof} in the appendix.
\begin{Lemma}
	\label{lem: concentration bound}
Under the same assumptions as Theorem \ref{thm: explained variance}, then
	\begin{equation}
	\left|\frac{\widehat{\phi}_1-\phi_1}{\phi_1}\right|\cip 0 
	\label{eq: consistent var 1}
		\end{equation}
	\begin{equation}
	\left|\frac{\widehat{\phi}_2-\phi_2}{4\phi_1+\rho \phi_2}\right|\cip 0 \;\;\text{for}\;\; \rho> 0
	\label{eq: consistent var 2}
	\end{equation}
	\begin{equation}
	\frac{\frac{1}{\sqrt{n}}\sqrt{4\widehat{\phi}_1+\widehat{\rho}\widehat{\phi}_2}}{\sqrt{{4\phi_1}/{n}+{\phi_2}/{(N+n)}}}\cip 1
	\label{eq: consistent var 3}
	\end{equation}
\end{Lemma}
To establish the coverage property \eqref{eq: coverage CI2}, we consider the following two cases, 
\begin{enumerate}
\item For the case $\rho=0$, we have $\widehat{\rho}\widehat{\phi}_2\geq \rho\phi_2$ and hence
\begin{equation}
\left|\frac{{\sqrt{n}}\left(\widehat{\Q}-\beta^{\intercal}\Sigma\beta\right)}{\sqrt{4{\phi}_1+{\rho}{\phi}_2}}\cdot\sqrt{\frac{{4{\phi}_1+{\rho}{\phi}_2}}{{4\widehat{\phi}_1+\widehat{\rho}\widehat{\phi}_2}}}\right|\leq \left|\frac{{\sqrt{n}}\left(\widehat{\Q}-\beta^{\intercal}\Sigma\beta\right)}{\sqrt{4{\phi}_1+{\rho}{\phi}_2}}\right|\cdot\sqrt{\frac{{{\phi}_1}}{{\widehat{\phi}_1}}}.
\label{eq: sample limit 1}
\end{equation}
Together with \eqref{eq: consistent var 1}, we establish  the coverage property \eqref{eq: coverage CI2}.
\item For the case $\rho>0$, by Lemma \ref{lem: concentration bound}, we have $\frac{{4{\phi}_1+{\rho}{\phi}_2}}{{4\widehat{\phi}_1+\widehat{\rho}\widehat{\phi}_2}}\cip 1$ and hence  
\begin{equation}
\frac{{\sqrt{n}}\left(\widehat{\Q}-\beta^{\intercal}\Sigma\beta\right)}{\sqrt{4{\phi}_1+{\rho}{\phi}_2}}\cdot\sqrt{\frac{{4{\phi}_1+{\rho}{\phi}_2}}{{4\widehat{\phi}_1+\widehat{\rho}\widehat{\phi}_2}}}\cid N(0,1),
\label{eq: sample limit 2}
\end{equation}
which leads to the coverage property \eqref{eq: coverage CI2}.
\end{enumerate}
The precision property \eqref{eq: precision CI2} follows from \eqref{eq: consistent var 3}.
\subsection{Proof of Theorem \ref{thm: est bound semi optimal}} 
\label{sec: proof 4}
The error $\widehat{\Q}(\widehat{\beta},\widehat{\Sigma}^{(2)}, Z^{(2)})-\Q$ is decomposed as follows, \begin{equation}
\frac{2}{\nb}\widehat{\beta}^{\intercal}(\xb)^{\intercal}\eb+\widehat{\beta}^{\intercal}\left(\widehat{\Sigma}^{(2)}-\Sigma\right)\widehat{\beta}\\
-(\widehat{\beta}-\beta)^{\intercal}{\Sigma}(\widehat{\beta}-\beta)+2\widehat{\beta}^{\intercal}\left(\Sigma-\frac{1}{n_2}(\xb)^{\intercal}\xb\right)(\widehat{\beta}-\beta)
\label{eq: key decomposition optimal}
\end{equation}
The following Lemma controls the terms involved in the above decomposition and the corresponding proof is present in Section \ref{sec: lemma5 proof}.
\begin{Lemma}
	With probability larger than $1-p^{-c_1}-e^{-c_1 t^2}-\gamma(n)$, 
	\begin{equation}
	(\widehat{\beta}-\beta)^{\intercal}{\Sigma}(\widehat{\beta}-\beta)\lesssim {\frac{k \log p}{n}}\sigma^2,
	\label{eq: error bound 1 optimal}
	\end{equation}
	\begin{equation}
\left|\widehat{\beta}^{\intercal}\left(\Sigma-\frac{1}{n_2}(\xb)^{\intercal}\xb\right)(\widehat{\beta}-\beta)\right|\lesssim \frac{\|\Sigmahalf\widehat{\beta}\|_2}{\sqrt{n}}\|\Sigmahalf(\widehat{\beta}-\beta)\|_2
  \label{eq: error bound 2 optimal}
	\end{equation}
\begin{equation}
\left|\frac{2}{\nb}\widehat{\beta}^{\intercal}\left(\xb\right)^{\intercal}\eb\right|\lesssim t \sigma\frac{\|\Sigmahalf\widehat{\beta}\|_2 }{\sqrt{n}}, \quad \left|\widehat{\beta}^{\intercal}\left(\widehat{\Sigma}^{(2)}-\Sigma\right)\widehat{\beta}\right|\lesssim t \frac{\|\Sigmahalf\widehat{\beta}\|_2^2}{\sqrt{N+n}}
\label{eq: error bound 3 optimal}
\end{equation}
	\label{lem: key lemma optimal}
\end{Lemma}
\noindent The proof of \eqref{eq: optimal upper} follows from the error decomposition \eqref{eq: key decomposition optimal}, the separate error bounds in Lemma \ref{lem: key lemma optimal} and the fact $k\lesssim {n}/{\log p}$ and the upper bounds $\left|\|\Sigmahalf\widehat{\beta}\|_2 -\|\Sigmahalf{\beta}\|_2\right|\leq \|\Sigmahalf(\widehat{\beta}-\beta)\|_2$ and $\left|\|\Sigmahalf\widehat{\beta}\|_2^2 -\|\Sigmahalf{\beta}\|_2^2\right|\leq \|\Sigmahalf(\widehat{\beta}-\beta)\|_2^2+2\|\Sigmahalf{\beta}\|_2\|\Sigmahalf(\widehat{\beta}-\beta)\|_2.$

\subsection{Proof of Theorem \ref{thm: lower bound}}
\label{sec: proof 5}
We start with introducing two definitions.
Define the $\chi^2$ distance between two distributions $f_1(z)$ and $f_0(z)$ as $
\chi^2(f_{1},f_{0})=\int \frac{\left(f_1(z)-f_0(z)\right)^2}{f_0(z)} dz=\int \frac{f^2_{1}(z)}{f_{0}(z)} dz-1$
and the total variation distance as
$\TV(f_{1},f_{0})=\int \left|f_1(z)-f_0(z)\right| dz.$
It is well known that 
\begin{equation}
\TV(f_{1},f_{0})\leq \sqrt{\chi^2(f_{1},f_{0})}.
\label{eq: relation between chisq and TV}
\end{equation}
Part of the lower bound in Theorem \ref{thm: lower bound} follows from the lower bounds established in \citet{guo2016optimal}, where, using the the current paper's terminology,  equation (29) of Theorem 3 in \citet{guo2016optimal} is expressed as \begin{equation}
	\inf_{\widetilde{\|\beta\|_2^2}}\sup_{\theta\in \Theta\left(k,M\right)} \PP\left(\left|\widetilde{\|\beta\|_2^2}-\|\beta\|_2^2 \right|\gtrsim \min\left\{M/{\sqrt{n}}+ {k \log p}/{n}, M^2\right\} \right) \geq \frac{1}{4}.
	\label{eq: lower bound for quad}
\end{equation}
The constructed least favorable null and alternative hypotheses in the proof of \eqref{eq: lower bound for quad} belong to the subspace $\theta\in \Theta\left(k,M\right)\cap\left\{\Sigma={\rm I}\right\}$. For $\Sigma={\rm I}$, $\Q=\beta^{\intercal}\Sigma\beta$ is reduced to $\|\beta\|_2^2$ and \eqref{eq: lower bound for quad} implies the following lower bound, 
\begin{equation}
	\inf_{\widetilde{\Q}}\sup_{\theta\in \Theta\left(k,M\right)} \PP\left(\left|\widetilde{\Q}-\Q \right|\gtrsim \min\left\{M/{\sqrt{n}}+ {k \log p}/{n}, M^2\right\} \right) \geq \frac{1}{4}.
	\label{eq: lower bound part a}
\end{equation}
It remains to establish the additional term of the lower bound $M^2/\sqrt{N+n}$, whose proof is based on the following version of Le Cam's Lemma (stated as Lemma 4 in \citet{guo2016optimal}; See also \citet{lecam1973convergence,yu1997assouad,ren2015asymptotic}).
\begin{Lemma} Let ${\rm T}\left(\thetab\right)$ denote a functional on $\thetab$. Suppose that $\theta_0,\theta_1 \in \Theta$, $\HH_0=\left\{\thetab_0\right\}$ and $\HH_1=\left\{\thetab_1\right\}$ and $d=\left|{\rm T}\left(\thetab_1\right)-{\rm T}\left(\thetab_0\right)\right|$. Then we have
\begin{equation}
\inf_{\widehat{\rm T}}\sup_{\thetab\in \HH_0\cup \HH_1} \PP_{\thetab}\left(\left|\widehat{\rm T}-{\rm T}\left(\thetab\right)\right|\geq \frac{d}{2}\right)\geq \frac{1-\TV\left(f_{\theta_1},f_{\thetab_0}\right)}{2}.
\end{equation}
\label{lem: lower bound lemma}
\end{Lemma}
To establish the lower bound  $M^2/\sqrt{N+n}$, we need to perturb the design covariance matrix and introduce the following null and alternative parameter spaces,
\begin{equation}
\begin{aligned}
\HH_0&=\left\{\theta_0=\left(\beta,{\rm I},\sigma_0\right)\right\} \\
\HH_1&=\left\{\theta_1=\left(\beta,{\rm I}+\frac{c}{\sqrt{N+n}\|\beta\|_2^2}\beta\beta^{\intercal},\sigma_0\right)\right\},\\
\end{aligned}
\label{eq: perturbing design covariance matrix}
\end{equation}
where $\beta\in \R^{p}$ satisfies $\|\beta\|_0\leq k$ and $\|\beta\|_2= M$ and $c=\min\left\{\sqrt{\log \left(1+\left(\frac{1}{4}-\frac{\alpha}{2}\right)^2\right)},M_1-1\right\}$. Note that $\HH_0,\HH_1 \in \Theta(k,M)$. Since the conditional distribution $f(y|X)$ is the same under both $\theta_0$ and $\theta_1$, then we have the decompositions $f_{\theta_0}(y,X)=f(y|X) f_{\theta_0}(X)$ and $f_{\theta_1}(y,X)=f(y|X) f_{\theta_1}(X)$ and hence \begin{equation}
\begin{aligned}
&\int\int\left|f_{\theta_0}(y,X)-f_{\theta_1}(y,X)\right| dXdy=\int \int f(y|X) \left|f_{{\theta_1}}(X)-f_{{\theta_0}}(X)\right| dX dy\\
&=\int \left(\int f(y|X) dy\right) \left|f_{{\theta_1}}(X)-f_{{\theta_0}}(X)\right| dX= L_1(f_{{\theta_1}}(X),f_{{\theta_0}}(X)).
\end{aligned}
\end{equation}
Hence, it is sufficient to control the $L_1$ or $\chi^2$ distance between $f_{{\theta_1}}(X)$ and $f_{{\theta_0}}(X)$. To control the distance, we introduce the following Lemma, which was established in \citet{cai2012optimal,ren2015asymptotic} and stated as Lemma 3 in \citet{cai2016asupplement}.
\begin{Lemma}
Let $g_i$ be the density function of $N(0,\Sigma_i)$ for $i=0,1,2$, respectively.
Then
$$\int \frac{g_1 g_2}{g_0}=\left({\rm det}\left({\rm I}-\Sigma_0^{-1}\left(\Sigma_1-\Sigma_0\right)\Sigma_0^{-1}\left(\Sigma_2-\Sigma_0\right)\right)\right)^{-\frac{1}{2}}.$$
\label{lem: workhorse Gaussian}
\end{Lemma}
\noindent Note that $\chi^2(f_{\theta_1}(X),f_{\theta_0}(X))+1=\prod_{i=1}^{n} \int \frac{f_{\theta_1}^2(X_{i\cdot})}{f_{\theta_0}(X_{i\cdot})}$. By applying Lemma \ref{lem: workhorse Gaussian} with  $\Sigma_0={\rm I}$ and $\Sigma_1=\Sigma_2={\rm I}+\frac{c_0}{\sqrt{N+n}\|\beta\|_2^2}\beta\beta^{\intercal}$, we have 
$$\chi^2(f_{\theta_1}(X),f_{\theta_0}(X))+1=\left({\rm det}\left({\rm I}-\frac{c_0^2}{{(N+n)}\|\beta\|_2^2}\beta\beta^{\intercal}\right)\right)^{-\frac{N+n}{2}}=\left(1-\frac{c_0^2}{N+n}\right)^{-\frac{N+n}{2}}.$$
For a sufficient small $c$ such that $\frac{c^2}{N+n}<\frac{\log 2}{2}$, we have $\left(1-\frac{c^2}{N+n}\right)^{-\frac{N+n}{2}}\leq \exp\left(c^2\right)\leq 1+\left(\frac{1}{4}-\frac{\alpha}{2}\right)^2,$ where the first inequality follows from the inequality $\frac{1}{1-x}\leq \exp(2x)$ for $x\in[0,\frac{\log 2}{2})$ and the second inequality follows from the definition of $c$. By \eqref{eq: relation between chisq and TV}, we have $L_1(f_{{\theta_1}}(X),f_{{\theta_0}}(X))\leq \frac{1}{4}-\frac{\alpha}{2}$. To apply Lemma \ref{lem: lower bound lemma}, we consider the functional $\T(\theta)=\beta^{\intercal}\Sigma\beta$  and calculate 
\begin{equation*}
\left|\Q(\theta_1)-\Q(\theta_0)\right|=\left|\beta^{\intercal}\beta-\beta^{\intercal}\left({\rm I}+\frac{c}{\sqrt{N+n}\|\beta\|_2^2}\beta\beta^{\intercal}\right)\beta\right| = c \frac{\|\beta\|_2^2}{\sqrt{N+n}}=c\frac{M^2}{\sqrt{N+n}}.
\end{equation*} 
By applying Lemma \ref{lem: lower bound lemma}, we establish 
\begin{equation}
	\inf_{\widetilde{\Q}}\sup_{\theta\in \Theta\left(k,M\right)} \PP\left(\left|\widetilde{\Q}-\Q \right|\geq \frac{c}{2}\frac{M^2}{\sqrt{N+n}}\right) \geq \frac{1}{4}+\frac{\alpha}{2}.
	\label{eq: lower bound part b}
\end{equation}
Combining \eqref{eq: lower bound part a} and \eqref{eq: lower bound part b}, we establish the theorem. 

\subsection{Proof of Corollary \ref{cor: upper bound special 2}}
\label{sec: proof 6}
To establish \eqref{eq: known optimal bound}, we decompose the error $\widehat{\Q}(\widehat{\beta},{\Sigma}, Z^{(2)})-\Q$ as follows, 
 \begin{equation*}
\frac{2}{\nb}\widehat{\beta}^{\intercal}(\xb)^{\intercal}\eb
-(\widehat{\beta}-\beta)^{\intercal}{\Sigma}(\widehat{\beta}-\beta)+2\widehat{\beta}^{\intercal}\left(\Sigma-\frac{1}{n_2}(\xb)^{\intercal}\xb\right)(\widehat{\beta}-\beta).
\end{equation*}
Then \eqref{eq: known optimal bound} follows from the above decomposition and Lemma \ref{lem: key lemma optimal}.
To establish \eqref{eq: known bound}, the error $\widehat{\Q}(\widehat{\beta},{\Sigma}, Z)-\Q$ is decomposed as
\begin{equation}
	\begin{aligned}
\frac{2}{n}{\beta}^{\intercal}X^{\intercal}\epsilon+\frac{2}{n}(\widehat{\beta}-\beta)^{\intercal}X^{\intercal}\epsilon-(\widehat{\beta}-\beta)^{\intercal}{\Sigma}(\widehat{\beta}-\beta)+2\widehat{\beta}^{\intercal}({\Sigma}-\SigmaLa)(\widehat{\beta}-\beta)\\
\end{aligned}
\label{eq: decomposition known}
\end{equation}
By  \eqref{eq: error bound 1}, \eqref{eq: key distribution 1} and \eqref{eq: error bound 1 optimal}, we have
\begin{equation}
\left|\frac{2}{n}{\beta}^{\intercal}X^{\intercal}\epsilon+\frac{2}{n}(\widehat{\beta}-\beta)^{\intercal}X^{\intercal}\epsilon-(\widehat{\beta}-\beta)^{\intercal}{\Sigma}(\widehat{\beta}-\beta)\right|\lesssim t \frac{\|\Sigmahalf\beta\|_2}{\sqrt{n}}\sigma+ \frac{k \log p}{n}\sigma^2.
\label{eq: easy bound}
\end{equation}
By the similar argument as the proof of \eqref{eq: error bound 3}, we establish $\left|2\widehat{\beta}^{\intercal}({\Sigma}-\SigmaLa)(\widehat{\beta}-\beta)\right|\lesssim \frac{k \log p}{n}\sigma^2+\|\Sigmahalf\beta\|_2 \frac{k \log p}{n}\sigma.$ Together with \eqref{eq: decomposition known} and \eqref{eq: easy bound}, we establish \eqref{eq: known bound}.

\subsection{Proof of Corollaries \ref{cor: explained variance enlarge} and \ref{cor: explained variance ran}}
\label{sec: proof 7}
Corollary \ref{cor: explained variance ran} follows from Theorem \ref{thm: randomization} and the consistency of the standard deviation estimator $\widehat{\phi}^{R}$. Define $\phi_3={\sigma^2}\left({\beta}^{\intercal}{\Sigma}{\beta}+\tau_0^2\right)$ and $\widehat{\phi}_3=\widehat{\sigma}^2\left(\widehat{\beta}^{\intercal}\widehat{\Sigma}^{S}\widehat{\beta}+\tau_0^2\right)$. Using the same proof of Lemma \ref{lem: concentration bound}, we can establish the following lemma.
\begin{Lemma}
Under the same assumptions as Corollary \ref{cor: explained variance ran}, then
\begin{equation}
	\left|\frac{\widehat{\phi}_2-\phi_2}{4\phi_3+\rho \phi_2}\right|\cip 0 \;\;\text{for}\;\; \rho> 0 \;\text{and}\; \left|\frac{\widehat{\phi}_3-\phi_3}{\phi_3}\right|\cip 0 
	\label{eq: consistent var 2 ran}
	\end{equation}
	\begin{equation}
	\frac{\frac{1}{\sqrt{n}}\sqrt{4\widehat{\phi}_3+\widehat{\rho}\widehat{\phi}_2}}{\sqrt{{4\phi_3}/{n}+{\phi_2}/{(N+n)}}}\cip 1
	\label{eq: consistent var 3 ran}
	\end{equation}
\label{lem: consistency var}
\end{Lemma}
\noindent Applying the same argument as \eqref{eq: sample limit 1} and \eqref{eq: sample limit 2}, we establish \eqref{eq: coverage CI enlarge} for ${\rm CI}^{R}$; By \eqref{eq: consistent var 3 ran}, we establish \eqref{eq: precision CI enlarge} for ${\rm CI}^{R}$. 
The proof of corollary \ref{cor: explained variance enlarge} follows from the same argument as the proof of Theorem \ref{thm: explained variance} and the fact that
$$(1+\frac{\|\Sigmahalf \beta\|_2}{\sigma} \frac{N}{n+N})\frac{k \log p}{n}\sigma^2 \ll \sqrt{\frac{1}{n}4{\sigma}^2\left({\beta}^{\intercal}{\Sigma}{\beta}+\tau_0^2\right)}\;\; \text{if} \;\; k\ll \frac{\sqrt{n}}{\log p}.$$
Together with Lemma \ref{lem: consistency var}, the bias term $(1+\frac{\|\Sigmahalf \beta\|_2}{\sigma} \frac{N}{n+N})\frac{k \log p}{n}\sigma^2$ in \eqref{eq: convergence rate semi} is upper bounded by the width of the enlarged confidence interval.
\subsection{Proof of Theorem \ref{thm: general thm of l2 norm} and Corollary \ref{cor: l2 norm}}
\label{sec: proof 8}
The proofs rely on the error decomposition of the proposed estimator $\widehat{\|\beta\|_2^2}$ 
\begin{equation}
\widehat{\|\beta\|_2^2}-\|\beta\|_2^2=\frac{2}{\nb}\widehat{\beta}^{\intercal}\widehat{\Omega}X_{i\cdot}^{\intercal}\epsilon_i+2\widehat{\beta}^{\intercal}\left(\widehat{\Omega}\frac{1}{\nb}\sum_{i=\na+1}^{n}X_{i\cdot} X_{i\cdot}^{\intercal}-{\rm I}\right)\left(\beta-\widehat{\beta}\right)-(\widehat{\beta}-\beta)^{\intercal}(\widehat{\beta}-\beta),
\label{eq: l2 norm decomp}
\end{equation}
and the decomposition of the second term on the right hand side of \eqref{eq: l2 norm decomp},
\begin{equation}
\begin{aligned}
&2\widehat{\beta}^{\intercal}\left(\widehat{\Omega}\frac{1}{\nb}\sum_{i=\na+1}^{n}X_{i\cdot} X_{i\cdot}^{\intercal}-{\rm I}\right)\left(\beta-\widehat{\beta}\right)\\
&=2\widehat{\beta}^{\intercal}\left(\widehat{\Omega}-\Omega\right)\frac{1}{\nb}\sum_{i=\na+1}^{n}X_{i\cdot} X_{i\cdot}^{\intercal}\left(\beta-\widehat{\beta}\right)+2\widehat{\beta}^{\intercal}\Omega\left(\frac{1}{\nb}\sum_{i=\na+1}^{n}X_{i\cdot} X_{i\cdot}^{\intercal}-\Sigma\right)\left(\beta-\widehat{\beta}\right).
\end{aligned}
\label{eq: key term decomp}
\end{equation}
To establish the rate of convergence for estimating $\|\beta\|_2^2$, we introduce the following lemma, whose proof is deferred to Section \ref{sec: lemma unweighted}.
\begin{Lemma}
Under the assumption of Theorem \ref{thm: general thm of l2 norm}, then with probability larger than $1-p^{-c}-\gamma(n)-e^{-ct^2}$,
\begin{equation}
\left|\frac{2}{\nb}\sum_{i=n_1+1}^{n}\widehat{\beta}^{\intercal}\widehat{\Omega}X_{i\cdot}^{\intercal}\epsilon_i\right|\lesssim \sigma \sqrt{\frac{1}{\nb}\widehat{\beta}^{\intercal}\widehat{\Omega}\Sigma\widehat{\Omega}\widehat{\beta}}\lesssim  \|\widehat{\Omega}\widehat{\beta}\|_2 \frac{\sigma}{\sqrt{\nb}}
\label{eq: error 1 unweighted}
\end{equation}
\begin{equation}
\left|2\widehat{\beta}^{\intercal}\left(\widehat{\Omega}-\Omega\right)\frac{1}{\nb}\sum_{i=\na+1}^{n}X_{i\cdot} X_{i\cdot}^{\intercal}\left(\beta-\widehat{\beta}\right)\right|\lesssim \|\left(\widehat{\Omega}-\Omega\right)\widehat{\beta}\|_2 \|\Sigmahalf(\widehat{\beta}-\beta)\|_2
\label{eq: error 2 unweighted}
\end{equation}
\begin{equation}
\left|2\widehat{\beta}^{\intercal}\Omega\left(\frac{1}{\nb}\sum_{i=\na+1}^{n}X_{i\cdot} X_{i\cdot}^{\intercal}-\Sigma\right)\left(\beta-\widehat{\beta}\right)\right|\lesssim \frac{1}{\sqrt{\nb}} \|\Omega \widehat{\beta}\|_2  \|\Sigmahalf(\widehat{\beta}-\beta)\|_2
\label{eq: error 3 unweighted}
\end{equation}
\label{lem: error bound unweighted}
\end{Lemma}
By the error decomposition \eqref{eq: l2 norm decomp}, \eqref{eq: key term decomp} and Lemma \ref{lem: error bound unweighted}, we have \begin{equation}
\left|\widehat{\|\beta\|_2^2}-\|\beta\|_2^2\right|\lesssim \sigma \frac{\|\beta\|_2}{\sqrt{n}}+k\frac{\log p}{n}\sigma^2+{\|(\widehat{\Omega}-\Omega)\widehat{\beta}\|_2}\sqrt{\frac{k \log p}{n}}\sigma.
\label{eq: general bound for l2 norm}
\end{equation}
The rate of convergence in \eqref{eq: special case 1} follows from \eqref{eq: general bound for l2 norm}, Condition ${\rm (B3)}$ and the following inequality,
$$\|(\widehat{\Omega}-\Omega)\widehat{\beta}\|_2\leq \|(\widehat{\Omega}-\Omega)\|_2\left(\|\beta\|_2+\|\beta-\widehat{\beta}\|_2\right).$$
The rate of convergence in \eqref{eq: special case 2} follows from \eqref{eq: special case 1} and \eqref{eq: condition semi}.
Note that 
\begin{equation}
\frac{\frac{2}{\sqrt{\nb}}\sum_{i=n_1+1}^{n}\widehat{\beta}^{\intercal}\widehat{\Omega}X_{i\cdot}^{\intercal}\epsilon_i}{\sqrt{\frac{4\sigma^2}{\nb}\widehat{u}^{\intercal}\sum_{i=n_1+1}^{n}X_{i\cdot}^{\intercal}X_{i\cdot}\widehat{u}}}\cid N(0,1)
\label{eq: limitingdist}
\end{equation}
and 
\begin{equation}
\frac{\sqrt{n_2} k\frac{\log p}{n}\sigma^2}{\sqrt{\frac{4\sigma^2}{\nb}\widehat{u}^{\intercal}\sum_{i=n_1+1}^{n}X_{i\cdot}^{\intercal}X_{i\cdot}\widehat{u}}}\asymp \frac{{k\log p}/{\sqrt{n}}}{\|\widehat{\Omega}\widehat{\beta}\|_2}
\end{equation}
Since 
\begin{equation}
\|\widehat{\Omega}\widehat{\beta}-\Omega\beta\|_2\leq \|(\widehat{\Omega}-\Omega)\|_2\left(\|\beta\|_2+\|\beta-\widehat{\beta}\|_2\right)+\|\Omega(\widehat{\beta}-\beta)\|_2,
\end{equation}
we have 
\begin{equation}
\begin{aligned}
\|\widehat{\Omega}\widehat{\beta}\|_2&\geq \|\Omega\beta\|_2-\|\widehat{\Omega}\widehat{\beta}-\Omega\beta\|_2\\
&\geq \|\Omega\beta\|_2-\|(\widehat{\Omega}-\Omega)\|_2\left(\|\beta\|_2+\|\beta-\widehat{\beta}\|_2\right)-\|\Omega(\widehat{\beta}-\beta)\|_2
\end{aligned}
\end{equation}
Under condition ${\rm (B3)}$ and the sample size condition \eqref{eq: condition semi}, we have $\|\widehat{\Omega}\widehat{\beta}\|_2\gtrsim \|\beta\|_2-\sqrt{\frac{k \log p}{n}}\sigma$. Under the assumption $\frac{1}{\sigma}\|\beta\|_2\gg {k \log p}/{\sqrt{n}}$, we have $\frac{\sqrt{n_2} k\frac{\log p}{n}\sigma^2}{\sqrt{\frac{4\sigma^2}{\nb}\widehat{u}^{\intercal}\sum_{i=n_1+1}^{n}X_{i\cdot}^{\intercal}X_{i\cdot}\widehat{u}}}\cip 0$. Combined with \eqref{eq: limitingdist}, we establish \eqref{eq: limiting unweighted}.
\subsection{Proof of Corollary \ref{cor: signal detection}}
\label{sec: proof 9}

By applying Corollary \ref{cor: explained variance ran} to the following linear model,
\begin{equation}
y-X\beta^{\rm null}=X(\beta-\beta^{\rm null})+\epsilon,
\label{eq: connection 1}
\end{equation} 
we establish the following limiting distribution,
\begin{equation}
\sqrt{n}\frac{\widehat{\Q}^{R}(y-X\beta^{\rm null},X,\tau_0)-(\beta-\beta^{\rm null})^{\intercal}\Sigma (\beta-\beta^{\rm null})}{\rm SE}\rightarrow N(0,1),
\label{eq: limiting distribution}
\end{equation}
where ${\rm SE}={\sqrt{4\sigma^2\left(\delta^{\intercal}\Sigma\delta+\tau_0^2\right)+\rho\E\left(\delta^{\intercal}X_{1\cdot} X_{1\cdot}^{\intercal}\delta-\delta^{\intercal}\Sigma\delta\right)^2}}$ with $\delta=\beta-\beta^{\rm null}$. 
Hence $\PP\left(D(\tau_0)=1\right)$ can be expressed as
{\footnotesize
\begin{equation*}
\PP\left(\frac{\widehat{\Q}^{R}(y-X\beta^{\rm null},X,\tau_0)-(\beta-\beta^{\rm null})^{\intercal}\Sigma (\beta-\beta^{\rm null})}{{\rm SE}}\geq\frac{{\widehat{\phi}^{E}(y-X\beta^{\rm null},X,\tau_0)}{z_{\alpha}}-(\beta-\beta^{\rm null})^{\intercal}\Sigma (\beta-\beta^{\rm null})}{\rm SE}\right)
\end{equation*}
}
Note that $\lim_{n\rightarrow \infty}\frac{{\widehat{\phi}^{E}(y-X\beta^{\rm null},X,\tau_0)}}{{\rm SE}/\sqrt{n}}\geq 1$, where the equality holds as long as $\rho>0$.
By the limiting distribution \eqref{eq: limiting distribution}, we show that 
\begin{equation}
\lim_{n\rightarrow \infty} \PP\left(D(\tau_0)=1\right)\leq \Phi^{-1}\left({z_{\alpha}}-\frac{\sqrt{n}(\beta-\beta^{\rm null})^{\intercal}\Sigma (\beta-\beta^{\rm null})}{\rm SE}\right),
\label{eq: general power}
\end{equation}
where the equality holds as long as $\rho>0$. By applying \eqref{eq: general power} with $(\beta-\beta^{\rm null})^{\intercal}\Sigma (\beta-\beta^{\rm null})=0$, we controls the type I error; For the case $\rho>0$, by applying \eqref{eq: general power} with 
$(\beta-\beta^{\rm null})^{\intercal}\Sigma (\beta-\beta^{\rm null})= \frac{\Delta}{\sqrt{n}}$, we establish 
\eqref{eq: power}.
\subsection{Proof of Corollary \ref{cor: prediction accuracy}}
\label{sec: proof 10}

The estimation bound \eqref{eq: estimation prediction accuracy} follows from the argument of Theorem \ref{thm: est bound semi} and the decomposition of \eqref{eq: decomp randomize}. Note that the additional randomization term can be controlled as in \eqref{eq: key distribution 2 random}.

The proof of the coverage and precision properties follows from the application of Corollary \ref{cor: explained variance ran} to the following linear model,
\begin{equation}
y-X\check{\beta}=X(\beta-\check{\beta})+\epsilon.
\label{eq: connection 2}
\end{equation} 
Note that the precision property also relies on the following observation, 
\begin{equation*}
\E\left(\delta^{\intercal}X_{1\cdot} X_{1\cdot}^{\intercal}\delta-\delta^{\intercal}\Sigma\delta\right)^2\leq 4 \|X_{1\cdot}\|_{\psi_2}^2\|\delta\|_2^4,
\end{equation*}
which follows from Lemma \ref{lem: product of sub-gaussian} and the definition of sub-exponential random variable.
\section{Proof of Lemmas}
\label{sec: lemma proof}

To establish the technical lemmas, we introduce the following definitions. For a random variable $U$, its sub-gaussian norm is defined as
$\|U\|_{\psi_2}=\sup_{q\geq 1} \frac{1}{\sqrt{q}} \left(\E |U|^{q}\right)^{\frac{1}{q}},$ and its sub-exponential norm is defined as
$\|U\|_{\psi_1}=\sup_{q\geq 1} \frac{1}{q} \left(\E |U|^{q}\right)^{\frac{1}{q}}$. For a random vector $U\in \R^{p}$, its sub-gaussian norm is defined as $\|U\|_{\psi_2}=\sup_{v\in S^{p-1}} \|\langle v, U\rangle\|_{\psi_2}$ and sub-exponential norm is defined as $\|U\|_{\psi_1}=\sup_{v\in S^{p-1}} \|\langle v, U\rangle\|_{\psi_1},$ where $S^{p-1}$ is the unit sphere in $\R^{p}$. The following lemma shows that the product of two sub-gaussian variables is a sub-exponential variable, whose proof is present in Section \ref{sec: proof of lemmas}.
\begin{Lemma}
Suppose that $U$ and $V$ are sub-gaussian random variables, then 
\begin{equation}
\|UV\|_{\psi_1}\leq 2\|U\|_{\psi_2}\|V\|_{\psi_2} \quad \text{and} \quad \|UV-\E UV\|_{\psi_1}\leq 4\|U\|_{\psi_2}\|V\|_{\psi_2} 
\end{equation}
\label{lem: product of sub-gaussian}
\end{Lemma}
\noindent We introduce the following events to facilitate the proofs,
\begin{equation}
\begin{aligned}
G_1&=\left\{\max\left\{\|\widehat{\beta}-\beta\|_2^2,\frac{1}{{n}}\sum_{i=1}^{n}\left(X_{i\cdot}^{\intercal}(\widehat{\beta}-\beta)\right)^2\right\}\lesssim \frac{k \log p}{n}\sigma^2\right\},\;
G_2=\left\{\|\widehat{\beta}-\beta\|_1\lesssim k\sqrt{\frac{\log p}{n}}\sigma\right\},\\
G_3&=\left\{\|\frac{1}{n}\sum_{i=1}^{n}X_{i\cdot}\epsilon_i\|_{\infty}\lesssim C \sqrt{\frac{\log p}{n}}\sigma\right\},\;
G_4=\left\{\frac{1}{{N}}\sum_{i=n+1}^{n+N}\left(X_{i\cdot}^{\intercal}(\widehat{\beta}-\beta)\right)^2\lesssim \frac{k \log p}{n}\sigma^2\right\},
\label{eq: high prob event 1}
\end{aligned}
\end{equation}
and
\begin{equation}
\begin{aligned}
G_5(w,t)&=\left\{\left|\frac{1}{n}{w}^{\intercal}X^{\intercal}\epsilon\right|\lesssim t \frac{\|\Sigmahalf w\|_2 }{\sqrt{n}}\sigma\right\},\\
G_6(w,v,t)&=\left\{\left|{w^{\intercal}\left(\frac{1}{m}\sum_{i=1}^{m}X_{i}X_i^{\intercal}\right)v}-w^{\intercal}\Sigma v\right|\lesssim t\frac{\|\Sigmahalf w\|_2\|\Sigmahalf v\|_2}{\sqrt{m}}\right\}
\end{aligned}
\label{eq: high prob event 2}
\end{equation}
for $w,v\in \R^{p}$.  
Define $G=\cap_{i=1}^{4}G_i$. The following Lemma demonstrates that the above events happen with high probability and the corresponding proof is present in Section \ref{sec: proof of lemmas}.
\begin{Lemma} For any estimator $\widehat{\beta}$ satisfying ${\rm (B1)}$, then \begin{equation}
\PP(G)\geq 1-\gamma(n)-cp^{-c}-\exp(-c{N}).
\label{eq: con 1}
\end{equation}
For given $w,v\in \R^{p}$ and $t>0$, then 
\begin{equation}
\PP(G_5(w,t))\geq 1-2\exp(-ct^2) \quad \text{and}\quad
\PP(G_6(w,v,t))\geq 1-2\exp(-ct^2).
\label{eq: con 2}
\end{equation}
\label{lem: concentration}
\end{Lemma}
\begin{Lemma}
Suppose that the condition holds for $\Sigma$, then we have 
\begin{equation}
\max_{\|v_{S^{c}}\|_1\leq C_0 \|v_{S}\|_1 } {\|\Sigma^{\frac{1}{2}} v\|_2}\leq \rho_{\rm max}(k,\Sigma)(2+C_0){\|v\|_2}.
\end{equation}
\label{lem: max RE}
\end{Lemma}
The proof of the above lemmas is present in next subsection. 
\subsection{Proof of Lemmas \ref{lem: product of sub-gaussian}, \ref{lem: concentration} and \ref{lem: max RE}}
\label{sec: proof of lemmas}
\underline{\bf Proof of Lemma \ref{lem: product of sub-gaussian}}
The proof for $\|UV\|_{\psi_1}$  follows from the following inequality
\begin{equation*}
\|UV\|_{\psi_1}=\sup_{q\geq 1}\frac{1}{q} \left(\E |UV|^{q}\right)^{\frac{1}{q}}\leq 2\frac{1}{\sqrt{2q}} \left(\E |U|^{2q}\right)^{\frac{1}{2q}} \frac{1}{\sqrt{2q}}\left(\E |V|^{2q}\right)^{\frac{1}{2q}}\leq 2\|U\|_{\psi_2}\|V\|_{\psi_2},
\end{equation*}
where the first inequality is by Cauchy-Schwarz and the second inequality follows from the definition of sub-gaussian norm. The proof of the centered part $\|UV-\E UV\|_{\psi_1}$ follows from the upper bound for $\|UV\|_{\psi_1}$ and the remark 5.18 in \citet{vershynin2010introduction}.\\
\underline{\bf Proof of Lemma \ref{lem: concentration}}
The control of the events $G_1$ and $G_2$ follows from the definition of ${\rm (B1)}$ and the following inequality,
\begin{equation*}
\|\widehat{\beta}-\beta\|_1\leq (1+C_0)\|(\widehat{\beta}-\beta)_{S}\|_1\leq (1+C_0)\sqrt{k}\|(\widehat{\beta}-\beta)_{S}\|_2.
\end{equation*}
In the following, we first establish \eqref{eq: con 2} and then come back to the control of events $G_3$ and $G_4$.  
By Lemma \ref{lem: product of sub-gaussian}, $w^{\intercal}\left(X_{i}X_i^{\intercal}-\Sigma\right)v=w^{\intercal}\left(\Sigmahalf Z_{i}Z_i^{\intercal}\Sigmahalf-\Sigma\right)v=(\Sigmahalf {w})^{\intercal}\left(Z_{i}Z_i^{\intercal}-{\rm I}\right)\Sigmahalf v$ is centered random variable with sub-exponential norm 
\begin{equation*}
\|w^{\intercal}\left(X_{i}X_i^{\intercal}-\Sigma\right)v\|_{\psi_1}\leq 2\|\Sigmahalf w\|_2\|\Sigmahalf v\|_2\|Z_{i\cdot}\|_{\psi_2}^2= K_1 \|\Sigmahalf w\|_2\|\Sigmahalf v\|_2
\label{eq: sub-exponential 1}
\end{equation*}
where $K_1= 2\|Z_{i\cdot}\|_{\psi_2}^2$. 
Similarly, ${w}^{\intercal}X_{i\cdot}\epsilon_i=(\Sigmahalf {w})^{\intercal} Z_{i\cdot}\epsilon_i$ is centered sub-exponential random variable with sub-exponential norm $\|{w}^{\intercal}X_{i\cdot}\epsilon_i\|_{\psi_1}\leq \|\Sigmahalf w\|_2\|Z_{i\cdot}\|_{\psi_2}\|\epsilon_i\|_{\psi_2}\leq K_2\|\Sigmahalf w\|_2\sigma$ where $K_2=\|Z_{i\cdot}\|_{\psi_2}\|\epsilon_i/\sigma\|_{\psi_2}$. By applying Corollary 5.17 in \citet{vershynin2010introduction}, we have for $t\leq \sqrt{n}$,
\begin{equation*}
\PP\left(\left|\frac{1}{n}\sum_{i=1}^{n} {w}^{\intercal}X_{i\cdot}\epsilon_i\right|\geq \frac{t}{\sqrt{n}}\cdot K_2\|\Sigmahalf w\|_2\sigma\right)\leq 2\exp(-c t^2)
\end{equation*}
and for $t\leq \sqrt{m}$,
\begin{equation*}
\PP\left(\left|{w^{\intercal}\left(\frac{1}{m}\sum_{i=1}^{m}X_{i}X_i^{\intercal}\right)v}-w^{\intercal}\Sigma v\right|\geq \frac{t}{\sqrt{m}}\cdot{K_1 \|\Sigmahalf w\|_2\|\Sigmahalf v\|_2}\right)\leq 2\exp(-c t^2)
\end{equation*}
Then \eqref{eq: con 2} follows from the above two concentration inequality. Note that, on the event $\cap_{i=1}^{p}G_5(e_i,\sqrt{\log p})$, the event $G_3$ holds and hence we have $\PP(G_3)\geq 1-2p\exp(-c (\sqrt{\log p})^2)$; on the event $G_6(\widehat{\beta}-\beta,\widehat{\beta}-\beta,{t})$, $$\frac{1}{{N}}\sum_{i=n+1}^{n+N}\left(X_{i\cdot}^{\intercal}(\widehat{\beta}-\beta)\right)^2\leq \left(1+\frac{t}{\sqrt{N}}\right)(\widehat{\beta}-\beta)^{\intercal}\Sigma(\widehat{\beta}-\beta)\lesssim \left(1+\frac{t}{\sqrt{N}}\right) \frac{k \log p}{n}\sigma^2.$$ 
By taking $t=\sqrt{N}$, we have $\PP(G_4)\geq 1-2\exp(-c{N})$.\\
\underline{\bf Proof of Lemma \ref{lem: max RE}}
For a given set $S$ and vector $v$, we divide $S^{c}$ into disjoint sets, $S^{c}=\cup_{j=1}^{M} T_{j}$ where $|T_1|=|T_2|=\cdots=|T_{M-1}|=k$ and $|T_{M}|\leq k$ and also 
\begin{equation}
\min_{i\in T_{j}}|v_{i}|\geq \max_{l \in T_{j+1}} |v_{l}|
\label{eq: ordering}
\end{equation}
For any given unit vector $\delta \in \R^{p}$, we have 
\begin{equation}
\left|\langle \delta, \Sigma^{\frac{1}{2}} v \rangle\right|\leq \left|\langle \delta, \Sigma^{\frac{1}{2}} v_{S} \rangle\right|+\sum_{j=1}^{M}\left|\langle \delta, \Sigma^{\frac{1}{2}} v_{T_{j}} \rangle\right| \leq  \rho_{\rm max}(k,\Sigma) \left(\|v_{S}\|_2+\sum_{j=1}^{M}\| v_{T_{j}}\|_2\right).
\label{eq: key decomp}
\end{equation}
Due to the property \eqref{eq: ordering}, we have 
$
\| v_{T_{j}}\|_2 \leq \frac{1}{\sqrt{k}}\|v_{T_{j-1}}\|_1$ for $j\geq 2
$
and hence 
{\small
\begin{equation*}
\sum_{j=1}^{M}\| v_{T_{j}}\|_2\leq \| v_{T_{1}}\|_2+ \frac{1}{\sqrt{k}}\sum_{j=1}^{M-1}\|v_{T_{j-1}}\|_1\leq \| v_{T_{1}}\|_2+\frac{1}{\sqrt{k}} C_0 \|v_{S}\|_1 \leq \| v_{T_{1}}\|_2+C_0 \|v_{S}\|_2 \leq (1+C_0)\|v\|_2.
\end{equation*}
}
Then we have 
$
\|v_{S}\|_2+\sum_{j=1}^{M}\| v_{T_{j}}\|_2\leq (2+C_0)\|v\|_2.
$
By \eqref{eq: key decomp}, we have 
$
\left|\langle \delta, \Sigma^{\frac{1}{2}} v \rangle\right|\leq \rho_{\rm max}(k,\Sigma)(2+C_0)\|v\|_2.
$
Taking the maximum over the unit vector $\delta \in \R^{p}$, we have 
$
\|\Sigma^{\frac{1}{2}}v\|_2\leq \rho_{\rm max}(k,\Sigma)(2+C_0)\|v\|_2.
$

\subsection{Proof of Lemma \ref{lem: key lemma 1}}
\label{sec: lemma 1 proof}
The first inequality in \eqref{eq: error bound 1} follows from the Holder's inequality while the second inequality holds under the event $G_2\cap G_3$. On the event $G_1\cap G_4$, the second error bound \eqref{eq: error bound 2} follows from the following decomposition $$\left|(\widehat{\beta}-\beta)^{\intercal}\widehat{\Sigma}^{S}(\widehat{\beta}-\beta)\right|=\frac{n}{N+n}\frac{1}{n}\sum_{i=1}^{n}\left(X_{i\cdot}^{\intercal}(\widehat{\beta}-\beta)\right)^2+\frac{N}{N+n}\frac{1}{N}\sum_{i=n}^{n+N}\left(X_{i\cdot}^{\intercal}(\widehat{\beta}-\beta)\right)^2.$$
Together with \eqref{eq: con 1}, we establish \eqref{eq: error bound 2}.
To establish \eqref{eq: error bound 3}, we start with the following decomposition,
\begin{equation}
\begin{aligned}
&\widehat{\beta}^{\intercal}(\widehat{\Sigma}^{S}-\SigmaLa)(\widehat{\beta}-\beta)=(\widehat{\beta}-\beta)^{\intercal}(\widehat{\Sigma}^{S}-\SigmaLa)(\widehat{\beta}-\beta)\\
&+\frac{N}{N+n}\left(
{\beta}^{\intercal}(\frac{1}{N}\sum_{i=n+1}^{n+N}X_{i\cdot}X_{i\cdot}^{\intercal}-\Sigma)(\widehat{\beta}-\beta)-{\beta}^{\intercal}(\frac{1}{n}\sum_{i=1}^{n}X_{i\cdot}X_{i\cdot}^{\intercal}-\Sigma)(\widehat{\beta}-\beta)\right)
\end{aligned}
\label{eq: tech bound}
\end{equation}
In the following, we are going to bound the terms separately in the above decomposition.
On the event $G_1$, we have $(\widehat{\beta}-\beta)^{\intercal}\SigmaLa  (\widehat{\beta}-\beta)\lesssim \frac{k \log p}{n}\sigma^2$; By \eqref{eq: error bound 2}, we have $(\widehat{\beta}-\beta)^{\intercal}\widehat{\Sigma}^{S}(\widehat{\beta}-\beta)\lesssim \frac{k \log p}{n}\sigma^2$ and hence 
\begin{equation}
(\widehat{\beta}-\beta)^{\intercal}(\widehat{\Sigma}^{S}-\SigmaLa)(\widehat{\beta}-\beta)\lesssim \frac{k \log p}{n}\sigma^2.
\label{eq: tech bound 1}
\end{equation}
On the event $G_6(\beta,\widehat{\beta}-\beta,t)$, we have \begin{equation}
\left|{\beta}^{\intercal}(\frac{1}{N}\sum_{i=n+1}^{n+N}X_{i\cdot}X_{i\cdot}^{\intercal}-\Sigma)(\widehat{\beta}-\beta)\right|\lesssim \frac{t}{\sqrt{N}}\|\Sigmahalf {\beta}\|_2 \|\Sigmahalf (\widehat{\beta}-{\beta})\|_2\lesssim\frac{t}{\sqrt{N}}\|\Sigmahalf {\beta}\|_2 \sqrt{\frac{k \log p}{n}}\sigma,
\label{eq: tech bound 2}
\end{equation}
where the last inequality follows from Lemma \ref{lem: max RE} and condition {\rm (B1)}.
On the event $G_6({\beta},e_i,\sqrt{\log p})$, we have $\left|{\beta}^{\intercal}(\frac{1}{n}\sum_{i=1}^{n}X_{i\cdot}X_{i\cdot}^{\intercal}-\Sigma)e_i\right|\lesssim \sqrt{{\log p}/{n}}\|\Sigmahalf \beta\|_2$ and hence on the even $\cap_{i=1}^{p}G_6({\beta},e_i,\sqrt{\log p})$, we have $\|{\beta}^{\intercal}(\frac{1}{n}\sum_{i=1}^{n}X_{i\cdot}X_{i\cdot}^{\intercal}-\Sigma)\|_{\infty}\lesssim\sqrt{{\log p}/{n}}\|\Sigmahalf \beta\|_2.$ By Holder's inequality, on the event $G_2\cap\left(\cap_{i=1}^{p}G_6({\beta},e_i,\sqrt{\log p})\right)$, we have 
\begin{equation}
\left|{\beta}^{\intercal}(\frac{1}{n}\sum_{i=1}^{n}X_{i\cdot}X_{i\cdot}^{\intercal}-\Sigma)(\widehat{\beta}-\beta)\right|\lesssim \|\Sigmahalf \beta\|_2 {\frac{k\log p}{n}}\sigma
\label{eq: tech bound 3}
\end{equation}
By applying \eqref{eq: tech bound 1}, \eqref{eq: tech bound 2} and \eqref{eq: tech bound 3} to the decomposition \eqref{eq: tech bound}, we establish that  with probability larger than $1-p^{-c}-\gamma(n)-e^{-ct^2}$,
$$\left|\widehat{\beta}^{\intercal}(\widehat{\Sigma}^{S}-\SigmaLa)(\widehat{\beta}-\beta)\right|\lesssim k \frac{\log p}{n}\sigma^2 +  \|\Sigmahalf \beta\|_2 \sigma \left(t\frac{\sqrt{N}}{n+N}\sqrt{\frac{k \log p}{n}}+\frac{N}{n+N}\frac{k \log p}{n}\right).$$ 

\subsection{Proof of Lemma \ref{lem: key lemma 2}}
\label{sec: lemma 2 proof}
On the event $G_5(\beta,t)\cap G_6(\beta,\beta,t)$, the inequality \eqref{eq: key distribution 1} holds. The probability control of \eqref{eq: key distribution 1} follows from \eqref{eq: con 2} with taking $w=v=\beta$. Let $\rho_n$ denote $n/(N+n)$ and hence $\rho_n\rightarrow \rho$ To establish \eqref{eq: key distribution 2}, we start with the decomposition,
\begin{equation}
\begin{aligned}
\sqrt{n}\left(\frac{2}{n}{\beta}^{\intercal}X^{\intercal}\epsilon+\beta^{\intercal}\left(\widehat{\Sigma}^{S}-\Sigma\right)\beta\right)&=\frac{1}{\sqrt{n}}\sum_{i=1}^{n}\left(2{\beta}^{\intercal}X_{i\cdot}\epsilon_i+\rho_{n}\beta^{\intercal}\left(X_{i\cdot}X_{i\cdot}^{\intercal}-\Sigma\right)\beta\right)\\
&+\sqrt{\rho_n(1-\rho_n)}\frac{1}{\sqrt{N}}\sum_{i=n+1}^{N+n}\beta^{\intercal}\left(X_{i\cdot}X_{i\cdot}^{\intercal}-\Sigma\right)\beta
\label{eq: distribution decompostion}
\end{aligned}
\end{equation}
Note that
$
\E\left(2{\beta}^{\intercal}X_{i\cdot}\epsilon_i+\rho_{n}\beta^{\intercal}\left(X_{i\cdot}X_{i\cdot}^{\intercal}-\Sigma\right)\beta\right)^2= 4\sigma^2\beta^{\intercal}\Sigma\beta+\rho_n^2\E\left(\beta^{\intercal}X_{1\cdot} X_{1\cdot}^{\intercal}\beta-\beta^{\intercal}\Sigma\beta\right)^2,$ then we have 
{\small $$
\frac{\sqrt{n}\left(\frac{2}{n}{\beta}^{\intercal}X^{\intercal}\epsilon+\beta^{\intercal}\left(\widehat{\Sigma}^{S}-\Sigma\right)\beta\right)}{\sqrt{4\sigma^2\beta^{\intercal}\Sigma\beta+\rho_n^2\E\left(\beta^{\intercal}X_{1\cdot} X_{1\cdot}^{\intercal}\beta-\beta^{\intercal}\Sigma\beta\right)^2}}\cid N(0,1), \;\; \frac{{\sqrt{4\sigma^2\beta^{\intercal}\Sigma\beta+\rho_n^2\E\left(\beta^{\intercal}X_{1\cdot} X_{1\cdot}^{\intercal}\beta-\beta^{\intercal}\Sigma\beta\right)^2}}}{{\sqrt{4\sigma^2\beta^{\intercal}\Sigma\beta+\rho^2\E\left(\beta^{\intercal}X_{1\cdot} X_{1\cdot}^{\intercal}\beta-\beta^{\intercal}\Sigma\beta\right)^2}}}\rightarrow 1$$}
$$\frac{\sqrt{\rho_n(1-\rho_n)}\frac{1}{\sqrt{N}}\sum_{i=n+1}^{N+n}\beta^{\intercal}\left(X_{i\cdot}X_{i\cdot}^{\intercal}-\Sigma\right)\beta}{\sqrt{\rho(1-\rho)\E\left(\beta^{\intercal}X_{1\cdot} X_{1\cdot}^{\intercal}\beta-\beta^{\intercal}\Sigma\beta\right)^2}}\cid N(0,1)$$
By the above limiting distributions, together with the independence between the two terms on the right hand side of \eqref{eq: distribution decompostion}, we establish \eqref{eq: key distribution 2}.

\subsection{Proof of Lemma \ref{lem: key lemma random}}
\label{sec: lemma3 proof}
The proof follows from that of Lemma \ref{lem: key lemma 2}. The main change is that 
\begin{equation}
\E\left(2\left({\beta}^{\intercal}X_{i\cdot}+u_i\right)\epsilon_i+\rho_{n}\beta^{\intercal}\left(X_{i\cdot}X_{i\cdot}^{\intercal}-\Sigma\right)\beta\right)^2= 4\sigma^2\left(\beta^{\intercal}\Sigma\beta+\tau_0^2\right)+\rho_n^2\E\left(\beta^{\intercal}X_{1\cdot} X_{1\cdot}^{\intercal}\beta-\beta^{\intercal}\Sigma\beta\right)^2.
\end{equation}
In addition, we need to show that $\sqrt{n} \frac{1}{n} u^{\intercal} X^{\intercal}(\beta-\widehat{\beta})\cip 0$. Since $\frac{u^{\intercal} X^{\intercal}(\beta-\widehat{\beta})}{\|X^{\intercal}(\beta-\widehat{\beta})\|}|Z=z \sim N(0,1)$, we have 
\begin{equation}
\begin{aligned}
\PP\left(\left|\sqrt{n} \frac{1}{n} u^{\intercal} X^{\intercal}(\beta-\widehat{\beta})\right|\geq \delta_0\right)=\int \PP\left(\left|\sqrt{n} \frac{1}{n} u^{\intercal} X^{\intercal}(\beta-\widehat{\beta})\right|\geq \delta_0|z\right)f(z)dz \\=\int2\Phi^{-1}\left(\frac{\delta_0}{\sqrt{n}\|X^{\intercal}(\beta-\widehat{\beta})\|}\right) dz =\int_{z\in G_1}2\Phi^{-1}\left(\frac{\delta_0}{\sqrt{n}\|X^{\intercal}(\beta-\widehat{\beta})\|_2}\right) f(z)dz+P(G_1^c)
\end{aligned}
\end{equation}
where $\Phi^{-1}$ denotes the inverse of quantile function for the standard normal random variable.
By \eqref{eq: con 1} and $\sqrt{n}\|X^{\intercal}(\beta-\widehat{\beta})\|\lesssim k\log p/\sqrt{n} \rightarrow 0$, we show that $\PP\left(\left|\sqrt{n} \frac{1}{n} u^{\intercal} X^{\intercal}(\beta-\widehat{\beta})\right|\geq \delta_0\right)\rightarrow 0$ and hence $\sqrt{n} \frac{1}{n} u^{\intercal} X^{\intercal}(\beta-\widehat{\beta})\cip 0$.

\subsection{Proof of Lemma \ref{lem: key lemma optimal}}
\label{sec: lemma5 proof}
The proof of \eqref{eq: error bound 1 optimal} and \eqref{eq: error bound 3 optimal} in Lemma \ref{lem: key lemma optimal} follows the same arguments as those of Lemma \ref{lem: key lemma 1} and \ref{lem: key lemma 2}. 
Conditioning on $\widehat{\beta}$, we have $\{\widehat{\beta}^{\intercal}\left(\Sigma-{X}_{i\cdot}{X}_{i\cdot}^{\intercal}\right)(\widehat{\beta}-\beta)\}_{n_1+1\leq i\leq n+N}$ are i.i.d centered random variables and $\|\widehat{\beta}^{\intercal}\left(\Sigma-{X}_{i\cdot}{X}_{i\cdot}^{\intercal}\right)(\widehat{\beta}-\beta)\|_{\psi_1}\leq 2\|Z_i\|_{\psi_2}^2\|\Sigmahalf\widehat{\beta}\|_2\|\Sigmahalf(\widehat{\beta}-\beta)\|_2$ for $n_1+1\leq i\leq n$. By applying Corollary 5.17 in \citet{vershynin2010introduction}, we have 
\begin{equation}
\small 
\PP\left(\left|\widehat{\beta}^{\intercal}\left(\Sigma-\frac{1}{n_2}(\xb)^{\intercal}\xb\right)(\widehat{\beta}-\beta)\right|\geq \frac{t}{\sqrt{n_2}}\cdot2\|Z_i\|_{\psi_2}^2\|\Sigmahalf\widehat{\beta}\|_2\|\Sigmahalf(\widehat{\beta}-\beta)\|_2\right)\leq 2\exp(-ct^2).
\end{equation}
\subsection{Proof of Lemma \ref{lem: error bound unweighted}}
\label{sec: lemma unweighted}
The proof relies on the independence between $(\widehat{\Omega},\widehat{\beta})$ and $(X_i,y_i)$ for $n_1+1\leq i\leq n$.
On the event $G_5(\widehat{\Omega}\widehat{\beta},t),$ we have \eqref{eq: error 1 unweighted}. On the event $G_{6}((\widehat{\Omega}-\Omega)\widehat{\beta},\widehat{\beta}-\beta,t)$, we establish \eqref{eq: error 2 unweighted}.
On the event $G_{6}(\Omega\widehat{\beta},\widehat{\beta}-\beta,t),$ we establish \eqref{eq: error 3 unweighted}.

\subsection{Proof of Lemma \ref{lem: concentration bound}}
\label{sec: lemma4 proof}
We first establish \eqref{eq: consistent var 1} and then establish \eqref{eq: consistent var 2}. Define $\Delta_1={\widehat{\sigma}^2}/{\sigma^2}-1$ and $\Delta_2=\widehat{\beta}^{\intercal}\widehat{\Sigma}^{S}\widehat{\beta}/{\beta}^{\intercal}\Sigma{\beta}-1$ . Then we have 
\begin{equation*}
\left|\frac{\widehat{\phi}_1}{{\phi}_1}-1\right|\leq \left|\Delta_1\right|+\left|\Delta_2\right|+\left|\Delta_1\right|\cdot\left|\Delta_2\right|.
\end{equation*}
Note that 
\begin{equation}
\Delta_2=\frac{1}{{\beta}^{\intercal}\Sigma{\beta}}\left(2{\beta}^{\intercal}\widehat{\Sigma}^{S}(\widehat{\beta}-\beta)+(\widehat{\beta}-\beta)^{\intercal}\widehat{\Sigma}^{S}(\widehat{\beta}-\beta)+\beta^{\intercal}\left(\widehat{\Sigma}^{S}-\Sigma\right)\beta\right).
\label{eq: error of plugin}
\end{equation} 
The term ${\beta}^{\intercal}\widehat{\Sigma}^{S}(\widehat{\beta}-\beta)$ is decomposed as 
\begin{equation}
{\beta}^{\intercal}\widehat{\Sigma}^{S}(\widehat{\beta}-\beta)=\frac{1}{n+N}\sum_{i=1}^{n+N} X_{i\cdot}^{\intercal}(\widehat{\beta}-\beta) X_{i\cdot}^{\intercal}\beta \leq \sqrt{\frac{1}{n+N}\sum_{i=1}^{n+N} \left(X_{i\cdot}^{\intercal}(\widehat{\beta}-\beta)\right)^2}\sqrt{\frac{1}{n+N}\sum_{i=1}^{n+N}\left(X_{i\cdot}^{\intercal}\beta\right)^2},
\label{eq: CS upper}
\end{equation}
where the inequality follows from the Cauchy-Schwarz inequality.  Recall the definition of events in \eqref{eq: high prob event 1} and  \eqref{eq: high prob event 2}. On the event $G_1\cap G_4$, then $(\widehat{\beta}-\beta)^{\intercal}\widehat{\Sigma}^{S}(\widehat{\beta}-\beta)\lesssim k \log p/n$; On the event $G_6(\beta,\beta,\sqrt{\log p})$, then $\frac{\beta^{\intercal}\left(\widehat{\Sigma}^{S}-\Sigma\right)\beta}{{\beta}^{\intercal}\Sigma{\beta}}\lesssim \sqrt{\frac{\log p}{n+N}}$. Together with \eqref{eq: CS upper}, we show that on the event $G_1\cap G_4\cap G_6(\beta,\beta,\sqrt{\log p})$, $$\frac{{\beta}^{\intercal}\widehat{\Sigma}^{S}(\widehat{\beta}-\beta)}{\beta^{\intercal}\Sigma\beta}\lesssim \sqrt{\frac{k \log p/n}{\beta^{\intercal}\Sigma\beta}\cdot\left(1+\sqrt{\frac{\log p}{n+N}}\right)}.$$ Hence by the decomposition \eqref{eq: error of plugin}, we show that on the event $G_1\cap G_4\cap G_6(\beta,\beta,\sqrt{\log p})$,
$$\left|\Delta_2\right|\lesssim \frac{k \log p}{n}+\sqrt{\frac{\log p}{n}}+\sqrt{\frac{k \log p/n}{\beta^{\intercal}\Sigma\beta}\cdot\left(1+\sqrt{\frac{\log p}{n+N}}\right)}.$$
Together with the condition $\|\beta\|_2\gg {k \log p}/{\sqrt{n}}$ and Condition ${\rm (B2)}$, we establish \eqref{eq: consistent var 1}.

In the following, we present the proof of \eqref{eq: consistent var 2}. Define $\bar{\phi}_2=\frac{1}{(n+N)}\sum_{i=1}^{n+N}\left({\beta}^{\intercal} X_{i\cdot} X_{i\cdot}^{\intercal}{\beta}-{\beta}^{\intercal}\widehat{\Sigma}^{S}{\beta}\right)^2.$ Then
\begin{equation}
\frac{\widehat{\phi}_2-\phi_2}{4\phi_1+\rho \phi_2}=\frac{\widehat{\phi}_2-\bar{\phi}_2}{4\phi_1+\rho \phi_2}+\frac{\bar{\phi}_2-\phi_2}{4\phi_1+\rho \phi_2}
\end{equation}
where 
$$\widehat{\phi}_2-\bar{\phi}_2=\frac{1}{n+N}\sum_{i=1}^{n+N} \left(\left(\widehat{\beta}^{\intercal} X_{i\cdot} X_{i\cdot}^{\intercal}\widehat{\beta}-\widehat{\beta}^{\intercal}\widehat{\Sigma}^{S}\widehat{\beta}\right)^2-\left({\beta}^{\intercal} X_{i\cdot} X_{i\cdot}^{\intercal}{\beta}-{\beta}^{\intercal}\widehat{\Sigma}^{S}{\beta}\right)^2\right)$$
$$\bar{\phi}_2-\phi_2=\frac{1}{n+N}\sum_{i=1}^{n+N}\left(\left({\beta}^{\intercal} X_{i\cdot} X_{i\cdot}^{\intercal}{\beta}-{\beta}^{\intercal}\widehat{\Sigma}^{S}{\beta}\right)^2-\E\left({\beta}^{\intercal} X_{i\cdot} X_{i\cdot}^{\intercal}{\beta}-{\beta}^{\intercal}{\Sigma}{\beta}\right)^2\right)
$$
In the following, we will show 
\begin{equation}
\PP\left(\frac{1}{\phi_2}\left|\bar{\phi}_2-\phi_2\right|\geq C{\frac{\left(\log (n+N)\right)^{5/2}}{\sqrt{(n+N)}}} \frac{{\left({\beta}^{\intercal}\Sigma{\beta}\right)^2}}{\phi_2}\right)\lesssim (n+N)^{-c},
\label{eq: key equation 1}
\end{equation}
{\scriptsize
\begin{equation}
\PP\left(\frac{1}{4\phi_1+\rho \phi_2}\left|\widehat{\phi}_2-\bar{\phi}_2\right|\geq \sqrt{1+ C{\frac{\left(\log (n+N)\right)^{5/2}}{\sqrt{(n+N)}}}\frac{\left({\beta}^{\intercal}\Sigma{\beta}\right)^2}{\phi_2}} \sqrt{\frac{\Lambda(n)}{4\phi_1+\rho \phi_2}}+\frac{\Lambda(n)}{4\phi_1+\rho \phi_2}\right)\lesssim (n+N)^{-c}+p^{-c}+\gamma(n),
\label{eq: key equation 2}
\end{equation}
}
where 
\begin{equation}
\Lambda(n)=\frac{(k \log p)^2}{n+N}+{\log (n+N)}\frac{k \log p}{n}\left(\|\beta\|_2^2+\frac{k \log p}{n}\right)
\label{eq: final upper bound}
\end{equation}
Since $4\phi_1+\rho \phi_2\geq c(\|\beta\|_2^2+\rho\|\beta\|_2^4)$, under the regime $k\ll {\sqrt{n}}/{\log p}$, $\|\beta\|_2\gg k \log p/\sqrt{n}$ and $\log (N+n) k \log p\ll n$, then \eqref{eq: key equation 2} implies that $\frac{1}{4\phi_1+\rho \phi_2}\left|\widehat{\phi}_2-\bar{\phi}_2\right| \cip 0$. Together with \eqref{eq: key equation 1}, we establish  \eqref{eq: consistent var 2}. The result \eqref{eq: consistent var 3} follows from \eqref{eq: consistent var 1} and \eqref{eq: consistent var 2} and the following decomposition,
$$\left|\frac{4\widehat{\phi}_1+\widehat{\rho}\widehat{\phi}_2}{4\phi_1+\widehat{\rho}\phi_2}-1\right|=\left|\frac{4(\widehat{\phi}_1-\phi_1)}{4\phi_1+\widehat{\rho}\phi_2}+\frac{\widehat{\rho}\left(\widehat{\phi}_2-\phi_2\right)}{4\phi_1+\widehat{\rho}\phi_2}\right|\leq \left|\frac{\widehat{\phi}_1-\phi_1}{\phi_1}\right|+\frac{\left|\widehat{\phi}_2-\phi_2\right|}{\max\{\phi_1,\phi_2\}}.$$

\noindent\underline{\em Proof of Equation \eqref{eq: key equation 1}}.
 Define $A_i=X^{\intercal}_{i,\cdot}{\beta}/{\sqrt{{\beta}^{\intercal}\Sigma{\beta}}}$. Then we simply the expression of ${\phi}_2$ and $\widehat{\phi}_2$ as
$$ \frac{{\phi}_2}{\left({\beta}^{\intercal}\Sigma{\beta}\right)^2}=\E\left(A_i^2-\E A_i^2 \right)^2 \quad \text{and} \quad \frac{\bar{\phi}_2}{\left({\beta}^{\intercal}\Sigma{\beta}\right)^2}={\frac{1}{n+N}\sum_{i=1}^{n+N} \left(A_i^2-\frac{1}{n+N}\sum_{i=1}^{n+N} A_i^2\right)^2}.$$ 
Define 
$\psi_2=\frac{{\phi}_2}{\left({\beta}^{\intercal}\Sigma{\beta}\right)^2}$ and $\bar{\psi}_2=\frac{\bar{\phi}_2}{\left({\beta}^{\intercal}\Sigma{\beta}\right)^2}$ and it is sufficient to show that $\left|{\bar{\psi}_2}-{\psi_2}\right|\cip 0$, which can be proved by applying  Lemma 1 and 2 in \citet{cai2011adaptive}. To be self-contained, let's first re-state the Lemma 1 in \citet{cai2011adaptive} as Lemma \ref{lem: fund concentration lemma}. 
\begin{Lemma} 
	\label{lem: fund concentration lemma}
	Let $\xi_1,\cdots,\xi_n$ be independent random variables with mean 0. Suppose that there exists some $\eta>0$ and $M_{n}$ such that $\sum_{i=1}^{n}\E\xi_{i}^2 \exp\left(\eta|\xi_i|\right)\leq M_n^2.$ Then for $0<t\leq M_n$, 
	\begin{equation}
	\PP\left(\sum_{i=1}^{n} \xi_i\geq C_{\eta} M_n t\right)\leq \exp(-t^2),
	\end{equation}
	where $C_{\eta}=\eta+\eta^{-1}.$
\end{Lemma}
We bound $\bar{\psi}_2-\psi_2$ based on the following decomposition,
{\small
\begin{equation}
\begin{aligned}
\bar{\psi}_2-\psi_2=\frac{1}{n+N}\sum_{i=1}^{n+N} \left(A_i^{4}-\E A_i^{4}\right)+2 \E A_i^{2} \cdot \frac{1}{n+N}\sum_{i=1}^{n+N} \left(A_i^{2}-\E A_i^{2} \right)
-\left(\frac{1}{n+N}\sum_{i=1}^{n+N} \left(A_i^{2}-\E A_i^{2} \right)\right)^2
\end{aligned}
\end{equation}}
Since $\E A_i^{2}=1$, it is sufficient to establish upper bounds for  
$\frac{1}{n+N}\sum_{i=1}^{n+N} \left(A_i^{2}-\E A_i^{2} \right)$
and 
$\frac{1}{n+N}\sum_{i=1}^{n+N} \left(A_i^{4}-\E A_i^{4}\right)$. It follows from Lemma \ref{lem: product of sub-gaussian} that $A_i^2$ a sub-exponential random variable. By Remark 5.18 in \citet{vershynin2010introduction}, $A_i^{2}-\E A_i^{2} $ is a sub-exponential random variable with sub-exponential norm smaller than $2M_1\|X_{i\cdot}\|_{\psi_2}^2$. By Corollary 5.17 in \citet{vershynin2010introduction}, we have 
\begin{equation}
\PP\left(\frac{1}{n+N}\sum_{i=1}^{n+N} \left(A_i^{2}-\E A_i^{2} \right)\geq 2M_1\|X_{i\cdot}\|_{\psi_2}^2\sqrt{\frac{\log (n+N)}{n+N}}\right)\leq 2\exp\left(-c \log (n+N)\right)=2(n+N)^{-c}.
\label{eq: first concentration}
\end{equation} 
Since $A_i$ is a sub-gaussian random variable, there exists positive constants $C_1>0$ and $c>2$ such that the following concentration inequality holds,
\begin{equation}
\sum_{i=1}^{n+N}\PP\left(|A_i|\geq C_1\sqrt{\log (n+N)}\right)\leq (n+N) \max_{1\leq i\leq (n+N)}\PP\left(|A_i|\geq C_1\sqrt{\log (n+N)}\right)\lesssim (n+N)^{-c}
\label{eq: concentration of subgaussian}
\end{equation}
Define $\bar{A}_i=A_i \mathbf{1}\left(|A_i|\leq C_1\sqrt{\log (n+N)}\right)$ and $\tilde{A}_i=A_i \mathbf{1}\left(|A_i|\geq C_1\sqrt{\log (n+N)}\right)$. Then we have 
\begin{equation}
\frac{1}{n+N}\sum_{i=1}^{n+N} \left(A_i^{4}-\E A_i^{4}\right)=\frac{1}{n+N}\sum_{i=1}^{n+N} \left(\bar{A}_i^4-\E \bar{A}_i^4\right)+\frac{1}{n+N}\sum_{i=1}^{n+N} \left(\tilde{A}_i^4-\E \tilde{A}_i^4\right)
\label{eq: error decomposition}
\end{equation}
We control $\E \tilde{A}_i^4$ as follows,
\begin{equation}
\E \tilde{A}_i^4\leq  \sqrt{\E\left( A_i^8\right) \PP\left(|A_i|\geq C_1\sqrt{\log (n+N)}\right)}
\lesssim \PP\left(|A_i|\geq C_1\sqrt{\log (n+N)}\right)^{1/2}\lesssim (n+N)^{-c/2},
\label{eq: bound of expectation}
\end{equation}
where the first inequality follows from Cauchy-Schwarz inequality, the second inequality follows from the fact that $A_i$ is a sub-gaussian random variable and the last inequality follows from \eqref{eq: concentration of subgaussian}.
Now we apply Lemma \ref{lem: fund concentration lemma} to bound $\frac{1}{n+N}\sum_{i=1}^{n+N} \left(\bar{A}_i^4-\E \bar{A}_i^4\right)$. 
By taking $\eta={c_1}/{\left(C_1 \log (n+N)\right)^2}$ for some small positive constant $c_1>0$,  we have 
$$\sum_{i=1}^{n+N}\E \left(\bar{A}_i^4-\E \bar{A}_i^4\right)^2\exp\left(\eta \left|\bar{A}_i^4-\E \bar{A}_i^4\right|\right)\leq C \sum_{i=1}^{n+N}\E \left(\bar{A}_i^4-\E \bar{A}_i^4\right)^2 \leq C_2 (n+N).$$
By applying Lemma \ref{lem: fund concentration lemma} with $M_n=\sqrt{C_2 (n+N)}$, $\eta={c_1}/{\left(C_1 \log (n+N)\right)^2}$ and $t=\sqrt{\log (n+N)}$, then we have 

\begin{equation}
\PP \left(\frac{1}{n+N}\sum_{i=1}^{n+N} \left(\bar{A}_i^4-\E \bar{A}_i^4\right)\geq C {\frac{(\log (n+N))^{5/2}}{\sqrt{n+N}}}\right)\lesssim (n+N)^{-c}.
\label{eq: control of the truncated sum}
\end{equation}

By \eqref{eq: concentration of subgaussian}, \eqref{eq: error decomposition}, \eqref{eq: bound of expectation} and \eqref{eq: control of the truncated sum}, we have
{\small 
\begin{equation}
\begin{aligned}
&\PP\left(\frac{1}{n+N}\sum_{i=1}^{n+N} \left(A_i^{4}-\E A_i^{4}\right)\geq C {\frac{\left(\log (n+N)\right)^{5/2}}{\sqrt{n+N}}}\right) \leq \sum_{i=1}^{n+N}\PP\left(|A_i|\geq C\sqrt{\log (n+N)}\right)\\
&+\PP \left(\frac{1}{n+N}\sum_{i=1}^{n+N} \left(\bar{A}_i^4-\E \bar{A}_i^4\right)\geq C {\frac{\left(\log (n+N)\right)^{5/2}}{\sqrt{n+N}}}\right)+\PP\left(\E \tilde{A}_i^4 \geq C {\frac{\left(\log (n+N)\right)^{5/2}}{\sqrt{n+N}}}\right)
\lesssim(n+N)^{-c}.\\
\end{aligned}
\label{eq: second concentration}
\end{equation}
}
By \eqref{eq: first concentration} and \eqref{eq: second concentration}, then there exisits a large constant $C$ such that 
\begin{equation}
\PP\left(\left|\bar{\psi}_2-\psi_2\right|\geq C{\frac{\left(\log (n+N)\right)^{5/2}}{\sqrt{(n+N)}}}\right)\lesssim (n+N)^{-c},
\label{eq: estimation of covariance matrix variance}
\end{equation}
for some $c>0$. By the fact that $\left|\bar{\psi}_2-\psi_2\right|=\frac{1}{{\left({\beta}^{\intercal}\Sigma{\beta}\right)^2}}\left|\bar{\phi}_2-\phi_2\right|$, this implies  \eqref{eq: key equation 1}.\\

\noindent\underline{\em Proof of Equation \eqref{eq: key equation 2}}. We start with the following decomposition,
\begin{equation}
\begin{aligned}
&\frac{1}{n+N}\sum_{i=1}^{n+N} \left(\left(\widehat{\beta}^{\intercal} X_{i\cdot} X_{i\cdot}^{\intercal}\widehat{\beta}-\widehat{\beta}^{\intercal}\widehat{\Sigma}^{S}\widehat{\beta}\right)^2-\left({\beta}^{\intercal} X_{i\cdot} X_{i\cdot}^{\intercal}{\beta}-{\beta}^{\intercal}\widehat{\Sigma}^{S}{\beta}\right)^2\right)\\
&=\frac{1}{n+N}\sum_{i=1}^{n+N} \left(\widehat{\beta}^{\intercal} X_{i\cdot} X_{i\cdot}^{\intercal}\widehat{\beta}-\widehat{\beta}^{\intercal}\widehat{\Sigma}^{S}\widehat{\beta}-{\beta}^{\intercal} X_{i\cdot} X_{i\cdot}^{\intercal}{\beta}+{\beta}^{\intercal}\widehat{\Sigma}^{S}{\beta}\right)^2\\
&+\frac{1}{n+N}\sum_{i=1}^{n+N} 2\left({\beta}^{\intercal} X_{i\cdot} X_{i\cdot}^{\intercal}{\beta}-{\beta}^{\intercal}\widehat{\Sigma}^{S}{\beta}\right)\left(\widehat{\beta}^{\intercal} X_{i\cdot} X_{i\cdot}^{\intercal}\widehat{\beta}-\widehat{\beta}^{\intercal}\widehat{\Sigma}^{S}\widehat{\beta}-{\beta}^{\intercal} X_{i\cdot} X_{i\cdot}^{\intercal}{\beta}+{\beta}^{\intercal}\widehat{\Sigma}^{S}{\beta}\right)\\
\end{aligned}
\label{eq: decomposition var est}
\end{equation}
where the second term on the right hand side of \eqref{eq: decomposition var est} is further upper bounded by 
\begin{equation}
\frac{2}{n+N}\sqrt{\sum_{i=1}^{n+N} \left({\beta}^{\intercal} X_{i\cdot} X_{i\cdot}^{\intercal}{\beta}-{\beta}^{\intercal}\widehat{\Sigma}^{S}{\beta}\right)^2} \sqrt{\sum_{i=1}^{n+N}\left(\widehat{\beta}^{\intercal} X_{i\cdot} X_{i\cdot}^{\intercal}\widehat{\beta}-\widehat{\beta}^{\intercal}\widehat{\Sigma}^{S}\widehat{\beta}-{\beta}^{\intercal} X_{i\cdot} X_{i\cdot}^{\intercal}{\beta}+{\beta}^{\intercal}\widehat{\Sigma}^{S}{\beta}\right)^2}
\label{eq: cs bound}
\end{equation}
Note that \eqref{eq: estimation of covariance matrix variance} implies 
{\small
\begin{equation}
\PP\left(\sqrt{\frac{1}{n+N}\sum_{i=1}^{n+N} \left({\beta}^{\intercal} X_{i\cdot} X_{i\cdot}^{\intercal}{\beta}-{\beta}^{\intercal}\widehat{\Sigma}^{S}{\beta}\right)^2} \geq \sqrt{1+ C{\frac{\left(\log (n+N)\right)^{5/2}}{\sqrt{(n+N)}}}\frac{\left({\beta}^{\intercal}\Sigma{\beta}\right)^2}{\phi_2}} \sqrt{\phi_2}\right)\lesssim (n+N)^{-c},
\label{eq: half bound}
\end{equation}
}
Then it is sufficient to control 
$\frac{1}{n+N}\sum_{i=1}^{n+N} \left(\widehat{\beta}^{\intercal} X_{i\cdot} X_{i\cdot}^{\intercal}\widehat{\beta}-\widehat{\beta}^{\intercal}\widehat{\Sigma}^{S}\widehat{\beta}-{\beta}^{\intercal} X_{i\cdot} X_{i\cdot}^{\intercal}{\beta}+{\beta}^{\intercal}\widehat{\Sigma}^{S}{\beta}\right)^2,$ which is further decomposed as,
{\footnotesize
\begin{equation}
\begin{aligned}
&\frac{1}{n+N}\sum_{i=1}^{n+N} \left(\widehat{\beta}^{\intercal} X_{i\cdot} X_{i\cdot}^{\intercal}\widehat{\beta}-\widehat{\beta}^{\intercal}\widehat{\Sigma}^{S}\widehat{\beta}-{\beta}^{\intercal} X_{i\cdot} X_{i\cdot}^{\intercal}{\beta}+{\beta}^{\intercal}\widehat{\Sigma}^{S}{\beta}\right)^2\\
=&\frac{1}{n+N}\sum_{i=1}^{n+N} \left((\widehat{\beta}-\beta)^{\intercal} X_{i\cdot} X_{i\cdot}^{\intercal}(\widehat{\beta}-\beta)+2\beta^{\intercal} X_{i\cdot} X_{i\cdot}^{\intercal}(\widehat{\beta}-\beta)-(\widehat{\beta}-\beta)^{\intercal}\widehat{\Sigma}^{S}(\widehat{\beta}-\beta)-2\beta^{\intercal}\widehat{\Sigma}^{S}(\widehat{\beta}-\beta)\right)^2\\
\leq & \frac{4}{n+N}\sum_{i=1}^{n+N} \left((\widehat{\beta}-\beta)^{\intercal} X_{i\cdot} X_{i\cdot}^{\intercal}(\widehat{\beta}-\beta)\right)^2+4\left(\beta^{\intercal} X_{i\cdot} X_{i\cdot}^{\intercal}(\widehat{\beta}-\beta)\right)^2+2\left((\widehat{\beta}-\beta)^{\intercal}\widehat{\Sigma}^{S}(\widehat{\beta}-\beta)-2\beta^{\intercal}\widehat{\Sigma}^{S}(\widehat{\beta}-\beta)\right)^2
\end{aligned}
\label{eq: var est expression}
\end{equation}
}
Recall the definition of events in \eqref{eq: high prob event 1}. On the event $G_1\cap G_4$, $(\widehat{\beta}-\beta)^{\intercal}\widehat{\Sigma}^{S}(\widehat{\beta}-\beta)\lesssim k\log p/n$; On the event $G_1\cap G_4\cap G_{6}(\beta,\beta,\sqrt{\log p})$, $$\left|\beta^{\intercal}\widehat{\Sigma}^{S}(\widehat{\beta}-\beta)\right|\leq \sqrt{\beta^{\intercal}\widehat{\Sigma}^{S}\beta}\sqrt{(\widehat{\beta}-\beta)^{\intercal}\widehat{\Sigma}^{S}(\widehat{\beta}-\beta)}\lesssim (1+\sqrt{\frac{{\log p}}{{n+N}}}) \|\beta\|_2 \sqrt{{k \log p}/{n}}.$$ Hence,
\begin{equation}
2\left((\widehat{\beta}-\beta)^{\intercal}\widehat{\Sigma}^{S}(\widehat{\beta}-\beta)-2\beta^{\intercal}\widehat{\Sigma}^{S}(\widehat{\beta}-\beta)\right)^2\lesssim \left(\frac{k\log p}{n}\right)^2+\|\beta\|_2^2 {\frac{k\log p}{n}}.
\label{eq: var upper 1}
\end{equation}
 It remains to control $ \frac{4}{n+N}\sum_{i=1}^{n+N} \left((\widehat{\beta}-\beta)^{\intercal} X_{i\cdot} X_{i\cdot}^{\intercal}(\widehat{\beta}-\beta)\right)^2+4\left(\beta^{\intercal} X_{i\cdot} X_{i\cdot}^{\intercal}(\widehat{\beta}-\beta)\right)^2$ in the expression \eqref{eq: var est expression}, which relies on the following fact. On the event $G_1$, $\sum_{i=1}^{n}\frac{\left(X_{i\cdot}^{\intercal}(\widehat{\beta}-\beta)\right)^2}{C k \log p}\leq 1$ and hence 
\begin{equation}
\frac{1}{n+N}\sum_{i=1}^{n}{\left(X_{i\cdot}^{\intercal}(\widehat{\beta}-\beta)\right)^4}=\frac{C^2 (k\log p)^2}{n+N} \times \sum_{i=1}^{n}\frac{\left(X_{i\cdot}^{\intercal}(\widehat{\beta}-\beta)\right)^4}{C^2 (k \log p)^2}\lesssim \frac{ (k \log p)^2}{n+N}. 
\label{eq: square upper bound}
\end{equation}
Define the event $\mathcal{B}_1$ as $\mathcal{B}_1=\left\{\max_{n+1\leq i\leq n+N}\left|X_{i\cdot}^{\intercal}(\widehat{\beta}-\beta)\right|\geq C\sqrt{\log (n+N)}\|\widehat{\beta}-\beta\|_2\right\}$ and the event $\mathcal{B}_2$ as $\mathcal{B}_2=\left\{\max_{1\leq i\leq n+N}\left|X_{i\cdot}^{\intercal}\beta\right|\geq C\sqrt{\log (n+N)}\|\beta\|_2\right\}$ .
Since $X_{i\cdot}$ is sub-gaussian random variable and $\widehat{\beta}-\beta$ is independent of $X_{i\cdot}$ for $n+1\leq i\leq n+N$, then 
\begin{equation}
\max_{i=1,2}\PP\left(\mathcal{B}_i\right)\lesssim (n+N)^{-c}.
\end{equation}
On the event $\mathcal{B}_1\cap G_{6}(\widehat{\beta}-\beta,\widehat{\beta}-\beta,\sqrt{\log p})$, 
\begin{equation}
\frac{1}{N}\sum_{i=n+1}^{n+N}{\left(X_{i\cdot}^{\intercal}(\widehat{\beta}-\beta)\right)^4}\leq \frac{1}{N}\sum_{i=n+1}^{n+N}{\left(X_{i\cdot}^{\intercal}(\widehat{\beta}-\beta)\right)^2} {\log (n+N)}\|\widehat{\beta}-\beta\|_2^2 \lesssim {\log (n+N)} \left(k\log p/n\right)^2
\label{eq: var upper 2}
\end{equation}
On the event $\mathcal{B}_2\cap G_1\cap G_4$, we have
{\footnotesize
\begin{equation*}
\frac{4}{n+N}\sum_{i=1}^{n+N} 4\left(\beta^{\intercal} X_{i\cdot} X_{i\cdot}^{\intercal}(\widehat{\beta}-\beta)\right)^2\lesssim {\log (n+N)}\|\beta\|_2^2 \frac{4}{n+N}\sum_{i=1}^{n+N}\left(X_{i\cdot}^{\intercal}(\widehat{\beta}-\beta)\right)^2\lesssim {\log (n+N)}\|\beta\|_2^2\frac{k \log p}{n}
\label{eq: var upper 3}
\end{equation*}
}
Combined with \eqref{eq: var upper 1}, \eqref{eq: square upper bound} and \eqref{eq: var upper 2}, we show that on the event $\mathcal{B}_1\cap\mathcal{B}_2\cap G_{6}(\widehat{\beta}-\beta,\widehat{\beta}-\beta,\sqrt{\log p})\cap G_1\cap G_4$, 
\begin{equation}
\left|\frac{1}{n+N}\sum_{i=1}^{n+N} \left(\widehat{\beta}^{\intercal} X_{i\cdot} X_{i\cdot}^{\intercal}\widehat{\beta}-\widehat{\beta}^{\intercal}\widehat{\Sigma}^{S}\widehat{\beta}-{\beta}^{\intercal} X_{i\cdot} X_{i\cdot}^{\intercal}{\beta}+{\beta}^{\intercal}\widehat{\Sigma}^{S}{\beta}\right)^2\right|\leq \Lambda(n),
\end{equation}
where $\Lambda(n)=\frac{(k \log p)^2}{n+N}+{\log (n+N)}\frac{k \log p}{n}\left(\|\beta\|_2^2+\frac{k \log p}{n}\right).$ Together with \eqref{eq: decomposition var est}, \eqref{eq: half bound} and \eqref{eq: cs bound}, we establish \eqref{eq: key equation 2}.

\section{Additional Simulation Results}
\label{sec: additional sim}
\subsection{Inference for $\beta^{\intercal}\Sigma\beta$}
This section presents the additional inference results corresponding to Section \ref{sec: comparison}.
In addition to the significant improvement in terms of estimation, the CHIVE estimator serves as the center of confidence intervals for $\beta^{\intercal}\Sigma\beta$. The coverage and precision properties of the constructed confidence interval ${\rm CI}$ are reported in Table \ref{table: sim-table}. With a larger sample size, the empirical coverage of the proposed confidence interval achieves $95\%$ and the average lengths of the confidence intervals get shorter. The integration of the unlabelled data in the semi-supervised setting has shorten the lengths of confidence interval significantly.

\begin{table}[ht]
\centering
\begin{tabular}{|rr|rr|rr|}
  \hline  
  &&\multicolumn{2}{c}{Supervised}\vline&\multicolumn{2}{c}{Semi-Supervised}\vline\\
  \hline
&$n$ & Cov& Len & Cov & Len \\ 
  \hline
\multirow{5}{*}{Setting 1}& 200  &  0.922 & 3.750 &  0.896 & 1.796 \\ 
   & 400&  0.936 & 2.734 & 0.942 & 1.536 \\ 
   & 600&   0.946 & 2.293 &  0.950 & 1.393 \\ 
   & 800& 0.936 & 1.991 &  0.942 & 1.290 \\ 
   & 1,000& 0.966 & 1.800 &  0.960 & 1.215 \\ 
  \hline
 \multirow{5}{*}{Setting 2} &200&  0.906 & 18.218 & 0.510 & 6.217 \\ 
   &400&  0.940 & 13.444 & 0.880 & 5.930\\ 
   &600&  0.946 & 11.045 &  0.920 & 5.643 \\ 
   &800&   0.932 & 9.671 &  0.924 & 5.419 \\ 
  & 1,000&  0.966 & 8.757 &  0.956 & 5.247\\ 
 \hline
\multirow{5}{*}{Setting 3}& 200&  0.864 & 1.342 & 0.866 & 0.903  \\ 
   & 400& 0.914 & 0.982 &  0.904 & 0.721 \\ 
   & 600& 0.934 & 0.828 &  0.906 & 0.624 \\ 
   & 800& 0.944 & 0.723 &  0.924 & 0.561 \\ 
   & 1,000& 0.942 & 0.650 &  0.950 & 0.516 \\ 
   \hline
\end{tabular}
\caption{\footnotesize Coverage and precision properties of Proposed CIs. Different rows correspond to different settings (Setting 1,2,3) and different sample sizes ($n=200,400,600,800,1000$) for the given setting. Each row reports empirical coverage (indexed with ``Cov") and average lengths (indexed with ``Len") of proposed CIs. The columns indexed with ``Supervised" represent the results for the supervised setting and the columns indexed with ``Semi-Supervised" represent the results for the semi-supervised setting. 
For example, in the first row of numbers $(0.922, 3.750, 0.896,1.796)$, it corresponds to the setting 1 and sample size $n=200$, in the supervised setting, ${\rm CI}$ has empirical coverage $0.922$ and the average length is $3.750$; in the semi-supervised setting, ${\rm CI}$ has empirical coverage $0.896$ and the average length is $1.796$.}
\label{table: sim-table}
\end{table}

\subsection{Effect of Pooling over Unlabelled Data: Signal Detection}
\label{sec: detection-sim}
We generate the high-dimensional linear regression \eqref{eq: high dim linear model} with the dimension $p=400$ and the labelled data with sample size $n=100$ and unlabelled data with sample size $N=3,000$. For the linear model \eqref{eq: high dim linear model},  the covariates $\{X_{i\cdot}\}_{1\leq i\leq n}$ for the labelled data and also $\{X_{i\cdot}\}_{n+1\leq i\leq n+N}$ for the unlabelled data are generated in i.i.d. fashion to follow multivariate normal distribution with mean zero and covariance matrix $\Sigma\in \R^{p\times p}$  where $\Sigma_{ij}=0.8$ for $1\leq i\neq j \leq p$ and $\Sigma_{ii}=1$ for $1\leq i\leq p$. The errors $\{\epsilon_i\}_{1\leq i\leq n}$ are generated as i.i.d normal distribution with mean zero and standard deviation $0.2$. For the detection problem, we generate $\beta$ as $\beta_{j}=\delta$ for $1\leq j\leq 40$ and $\beta_{j}=0$ for $ j\geq 41$ and vary $\delta$ across $\frac{1}{100} \{0,1.025,1.075,1.125, 1.175, 1.225,1.275, 1.325, 1.375, 1.425 \}$. We use the randomization level $\tau_0=2$ in conducting the hypothesis testing $H_0: \beta=0$.

\begin{table}[ht]
\centering
\begin{tabular}{|r|rrr|rr|rrr|}
  \hline
  &\multicolumn{3}{c}{ERR}\vline&\multicolumn{2}{c}{Coverage}\vline&\multicolumn{3}{c}{Length}\vline\\
  \hline
 $100\delta$ & Semi-S & S &Imp& Semi-S & S  & Semi-S & S & Ratio \\ 
  \hline
 0.000 & 0.018 & 0.018 &0\%& 0.972 & 0.972 & 0.172 & 0.172 & 100\% \\ 
 1.025 & 0.662 & 0.628 &5.4\%& 0.974 & 0.974 & 0.148 & 0.153 & 97.0\% \\ 
 1.075 & 0.702 & 0.652 & 7.7\%&0.970 & 0.976 & 0.147 & 0.152 & 96.3\% \\ 
 1.125 & 0.760 & 0.734 & 3.5\%& 0.968 & 0.966 & 0.146 & 0.153 & 95.5\% \\ 
 1.175 & 0.834 & 0.804 & 3.7\%&0.948 & 0.956 & 0.145 & 0.153 & 94.7\% \\ 
 1.225 & 0.882 & 0.838 & 5.3\%&0.952 & 0.956 & 0.146 & 0.155 & 94.0\% \\ 
1.275 & 0.946 & 0.904 &  4.6\%&0.970 & 0.964 & 0.144 & 0.156 & 92.6\% \\ 
1.325 & 0.982 & 0.946 & 3.8\%& 0.972 & 0.956 & 0.143 & 0.156 & 91.1\% \\ 
1.375 & 0.982 & 0.960 & 2.3\%& 0.964 & 0.962 & 0.142 & 0.158 & 90.1\% \\ 
1.425 & 0.984 & 0.968 & 1.6\%& 0.964 & 0.966 & 0.143 & 0.161 & 88.6\% \\ 
   \hline
\end{tabular}
\caption{Signal detection $p=400$, sample size $n=100$ and $N=3000$.}
\label{tab: semi signal detection}
\end{table}

The simulations are replicated over 500 simulations and the Empirical Rejection Rate (ERR) and the coverage and length of confidence intervals are present in Table \ref{tab: semi signal detection},where the column under ``Semi-S" corresponds to the semi-supervised method and the column under ``S" corresponds to the supervised method. Regarding ERR, we observe that the incorporation of unlabelled data is of help in improving the detection rate though the improvement, reported under the column ``Imp", is only at the level of $5\%$. This small improvement for $\beta$ closed to 0 is actually predicted by the theoretical results.  Since the design matrix is generated to follow multivariate Gaussian distribution in the simulation studies, the rate of convergence related to the unlabelled data is expressed as ${\E\left(\beta^{\intercal}X_{1\cdot} X_{1\cdot}^{\intercal}\beta-\beta^{\intercal}\Sigma\beta\right)^2}/(N+n)=2 (\beta^{\intercal}\Sigma\beta)^2/(N+n)$; For the hypothesis testing problem, the interesting but challenging regime is the local alternative near $\beta=0$, that is, $\E\left(\beta^{\intercal}X_{1\cdot} X_{1\cdot}^{\intercal}\beta-\beta^{\intercal}\Sigma\beta\right)^2=2 (\beta^{\intercal}\Sigma\beta)^2$ is near zero. This explains why improvement of integrating unlabelled data in testing $H_0: \beta=0$ is not as significant as the prediction loss evaluation and also inference for the explained variance $\beta^{\intercal}\Sigma\beta$ as presented in the main paper.

The constructed confidence intervals have coverage in both the supervised and semi-supervised settings and the confidence interval constructed using the unlabelled data has a shorter length. Under the column ``Ratio", we report the ratio of the length of confidence intervals in the semi-supervised setting to that in the supervised setting and see the confidence intervals can be reduced by as much as12\% in length. With the same reasoning as the ERR, the length of the confidence interval is not significantly reduced as the simulated setting mainly focuses on the case where $\beta$ is close to zero.

\subsection{More about Signal Detection}
\label{sec: sim2}
In this section, we investigate more on the signal detection problem. We switch the focus from how to integrate the unlabelled data in the detection problem to investigating the numerical effect of the randomization levels.

\begin{figure}[ht]
\centering
\includegraphics[width=10cm]{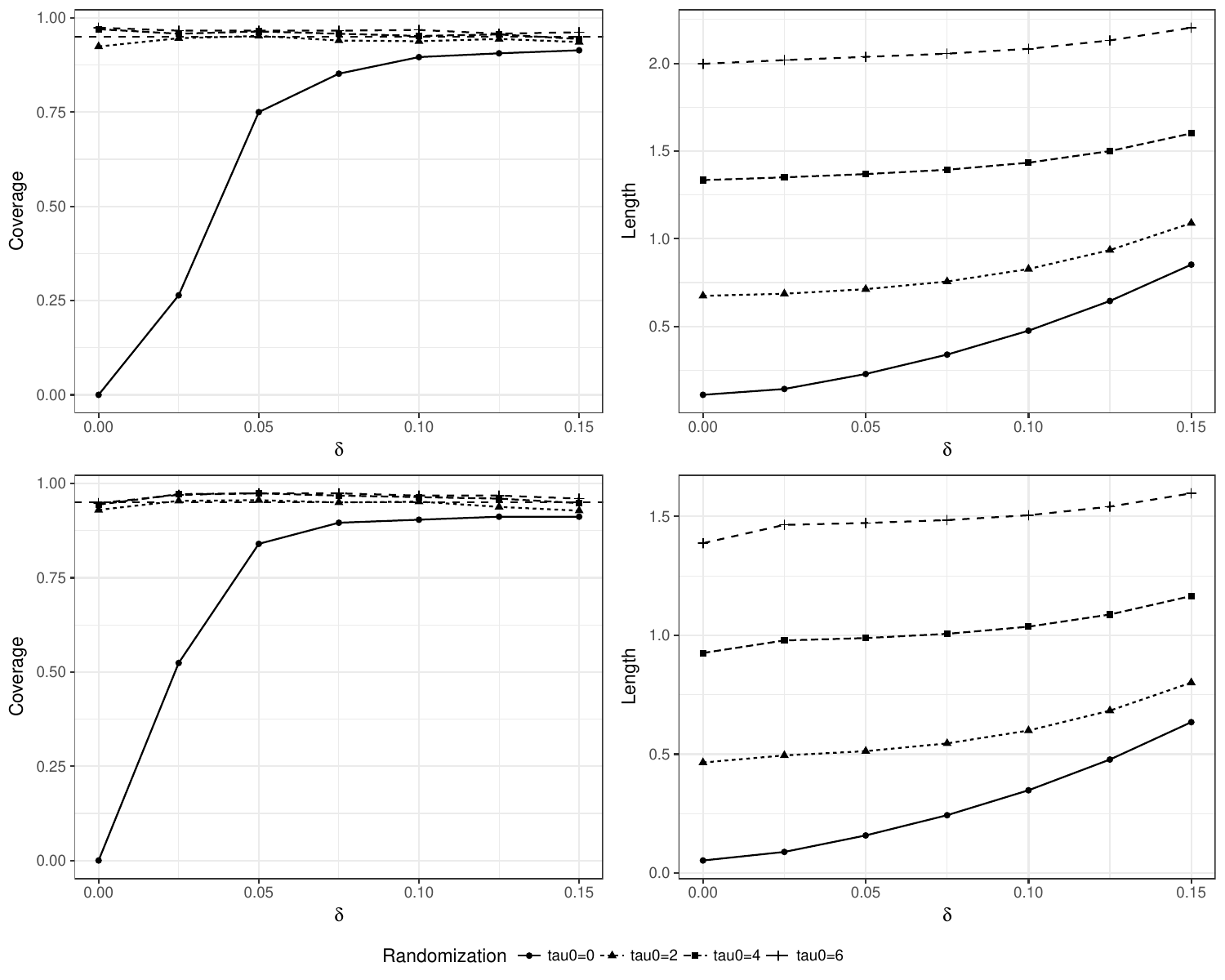}
\caption{\footnotesize Empirical coverage and average lengths of the proposed randomized confidence intervals in the supervised setting. The above two figures correspond to the sample size $n=600$ and the bottom two figures correspond to  $n=1200$ . The left hand side figures stand for the empirical coverage for different $\delta$ while the right hand side figures stand for the average lengths of CIs for different $\delta$. Different type of the curves correspond to different randomization levels $\tau_0\in \{0,2,4,6\}$. The dashed horizontal lines on the left hand figures correspond to the targeted coverage level, $0.95$.}
\label{fig: sim-figure2}
\end{figure}

For the detection problem, we generate $\beta$ as $\beta_{j}=\delta$ for $1\leq j\leq 50$ and $\beta_{j}=0$ for $51\leq j\leq 800$ and vary $\delta$ across $\{0.00,0.025,0,05,0.075,010,0.125,0.15\}$ and vary the sample size $n$ across $\{600,1200\}$. In Figure \ref{fig: sim-figure2}, we demonstrate the coverage and precision properties of the randomized confidence intervals across four methods, the non-randomized detector $D(0)$ and the three randomized detectors $D(2),D(4)$ and $D(6)$, where $D(\cdot)$ is defined in \eqref{eq: detection procedure}. 
The two plots on the top of Figure \ref{fig: sim-figure2}, corresponding to the supervised setting with $n=600$ demonstrate the effect of randomization on the empirical coverage and average lengths, where the randomization leads to a interval estimator achieving the coverage properties at the expense of wider interval estimators. With the randomization level $\tau_0$ reaching $2$, the coverage property is guaranteed while the empirical coverage for the procedure without randomization ($\tau_0=0$) is much lower than $0.95$, especially for weak signals with a small $\delta$. The bottom two plots of Figure \ref{fig: sim-figure2} corresponds to the supervised setting with $n=1,200$ and the main observation is similar to the case of $n=600$ but the  confidence intervals are much shorter than the setting with $n=600$. 

The empirical detection rate is reported in Table \ref{table: detection}, where the sample size $n$ is generated across $n=600$ and $n=1,200$ and the explained variance $\beta^{\intercal}\Sigma\beta$ is controlled via the scaler $\delta$. When $\delta=0$, it corresponds to the null case and a proper detection procedure is expected to have type I error rate $0.05$. As predicted by theory, the detection method without randomization $D(0)$ fails to give proper type I error due to presence of weak signals. With introducing the randomization procedure, the type I error rate gets closer to $0.05$. When $\delta$ moves away from zero, the detection procedure is taken as a powerful procedure as the empirical detection rate approaches $1$.   For the detection procedure with randomization level $\tau_0=2$, the setting with $\delta=0.025$ corresponds to an indistinguishable region, where it is challenging to detect the signal. However, as  $\delta$ reaches $0.05$, the detection rate reaches $0.800$ for $n=600$ and $0.944$ for $n=1200$.  As characterized by theory, a larger randomization level requires a higher value of $\delta$ such that the signal can be detected, for example, for $\tau_0=4$, until $\delta$ reaches $0.075$, the detection rate reaches $0.82$ for $n=600$ and $0.968$ for $n=1200$. The corresponding semi-supervised setting shows a similar phenomenon to the supervised setting but tends to be easier than the supervised setting due to the unlabelled data. The results are reported in the supplementary materials.  
\begin{table}[ht]
\centering
\begin{tabular}{|rr|rrrr|rrrr|}
  \hline
   && \multicolumn{4}{c}{$n=600$} \vline& \multicolumn{4}{c}{$n=1,200$}\vline\\
   \hline
  $\delta$ & $\beta^{\intercal}\Sigma\beta$ & $D(0)$ & $D(2)$ & $D(4)$ & $D(6)$ & $D(0)$ & $D(2)$ &$D(4)$&$D(6)$  \\ 
  \hline
 0.000 & 0.000 & 1.000 & 0.148 & 0.082 & 0.066 & 1.000 & 0.124 & 0.076 & 0.068 \\ 
 0.025 & 0.091 & 1.000 & 0.248 & 0.094 & 0.062 & 1.000 & 0.254 & 0.124 & 0.086 \\ 
 0.050 & 0.365 & 1.000 & 0.800 & 0.356 & 0.182 & 1.000 & 0.944 & 0.472 & 0.264 \\ 
 0.075 & 0.821 & 1.000 & 1.000 & 0.820 & 0.524 & 1.000 & 1.000 & 0.968 & 0.764 \\ 
0.100 & 1.460 & 1.000 & 1.000 & 1.000 & 0.914 & 1.000 & 1.000 & 1.000 & 0.992 \\ 
 0.125 & 2.281 & 1.000 & 1.000 & 1.000 & 0.996 & 1.000 & 1.000 & 1.000 & 1.000 \\ 
 0.150 & 3.285 & 1.000 & 1.000 & 1.000 & 1.000 & 1.000 & 1.000 & 1.000 & 1.000 \\ 
\hline
\end{tabular}
\caption{\footnotesize Empirical detection rates in the supervised setting.  The column indexed with $\delta$ represents the signal strength, where the signal is of sparsity $50$ and of the form $\delta\cdot \left(1,1,\cdots 1,0,0,\cdots,0\right)$; the column indexed with $\beta^{\intercal}\Sigma\beta$ represents the value of $\beta^{\intercal}\Sigma\beta$; the columns under ``n=600" and ``n=1,200" correspond to sample size 600 and 1,200 respectively, where the column indexed with $D(\tau_0)$ report the empirical detection rates for the detector $D(\tau_0)$. }
\label{table: detection}
\end{table}


\subsection{Prediction Loss Evaluation}
\label{sec: sim3}
In this subsection, we present additional results about prediction loss evaluation.
We generate the high-dimensional regression vector $\beta$ with sparsity $10$ where $\beta_{j}=j/5$ for $1\leq j\leq 10$ and $\beta_j=0$ for $j\geq 11$.
Let $\widehat{\beta}(\lambda)$ denote the Lasso estimator based on an independent training data $\left(X^{(0)},y^{(0)}\right)$ with sample size $n_0=300$,
$$
\widehat{\beta}\left(\lambda\right)=\arg\min_{\beta \in \R^{p}}\frac{\|y^{(0)}-X^{(0)}\beta\|_2^2}{2 n_0}+  {\lambda} \sum_{j=1}^{p} \frac{\|X^{(0)}_{\cdot j}\|_2}{\sqrt{n_0}} |\beta_j|. 
$$
 We consider the inference problem for the out-of-sample prediction accuracy $(\widehat{\beta}(\lambda)-\beta)^{\intercal}\Sigma(\widehat{\beta}(\lambda)-\beta)$. Specifically, we consider three estimators $\widehat{\beta}(\lambda_0),\widehat{\beta}(5\lambda_0)$ and $\widehat{\beta}(10\lambda_0)$ with 
$\lambda_{0}=\sqrt{\frac{Z_{(1-(0.1/p))}}{n_0}}$ and report the numerical performance of both point and interval estimators of the corresponding prediction accuracy. We consider the prediction accuracy problem across three different sample sizes, $\{600,1200,2400\}$ and introduce different randomization levels. We will use ${\rm PA}(\tau_0)$ to denote the procedure with randomization level $\tau_0$. 

Table \ref{tab: loss est} has reported the point and interval estimators of the prediction accuracy across different settings. In terms of point estimation, the sample averages get closer to the true accuracy with increasing sample sizes. Among the three estimators, the true prediction accuracy of $\widehat{\beta}(\lambda_0)$ is the smallest and also the most difficult to assess. The fundamental reason is that the small accuracy/error is hard to quantify. This phenomenon is connected to the theoretical results established in \citet{cai2016accuracy}, which showed that the estimation accuracy $\|\widehat{\beta}-\beta\|_2^2$ is hard to quantify for an accurate estimator $\widehat{\beta}$. 
The constructed confidence interval ${\rm PA(0)}$ without randomization has no coverage even for $n=2400$. 
In such a scenario with weak signals, the evaluators involved with randomized calibration, ${\rm PA}(2)$ and  ${\rm PA}(4)$  produce valid confidence intervals across different settings. 

Not only the point and interval estimators are useful, the upper limit and lower limit of the confidence intervals reported in Table \ref{tab: loss est} can also be informative in the prediction accuracy evaluation. For the estimator $\widehat{\beta}(\lambda_0)$, although the average of lower limit of confidence intervals for $(\widehat{\beta}(\lambda_0)-\beta)^{\intercal}\Sigma(\widehat{\beta}(\lambda_0)-\beta)$ is zero, the corresponding upper limits of confidence intervals are informative as they provide empirical guidance to practitioners with upper bounds for the prediction accuracy. For $\widehat{\beta}(5\lambda_0)$ and $\widehat{\beta}(10\lambda_0)$, both the upper and lower limits of the confidence intervals are informative on the size of the prediction accuracy.

\begin{table}[ht]
\centering
\resizebox{\columnwidth}{!}{
\begin{tabular}{|r|r|rrr|rrr|rrr|}
  \hline
   &&\multicolumn{3}{c}{$\widehat{\beta}(\lambda_0)$}\vline& \multicolumn{3}{c}{$\widehat{\beta}(5\lambda_0)$} \vline&\multicolumn{3}{c}{$\widehat{\beta}(10\lambda_0)$}\vline\\
&True Accuracy& \multicolumn{3}{c}{0.065 } \vline& \multicolumn{3}{c}{0.636}\vline & \multicolumn{3}{c}{2.310}\vline\\ 
  \hline
  \hline
&& PA(0) & PA(2) &PA(4)& PA(0) & PA(2) &PA(4)& PA(0) & PA(2)&PA(4) \\ 
  \hline
\multirow{3}{*}{Super, 600}   &Coverage & 0.166 & 0.938 & 0.962 & 0.778 & 0.928 & 0.956 & 0.914 & 0.934 & 0.948 \\ 
  &Est Aver&  0.157 & 0.158 & 0.160 & 0.739 & 0.743 & 0.746 & 2.417 & 2.421 & 2.424\\
  &Lower Aver & 0.090 & 0.000 & 0.000 & 0.584 & 0.373 & 0.058 & 2.062 & 1.932 & 1.665 \\ 
  &Upper Aver & 0.223 & 0.502 & 0.837 & 0.895 & 1.112 & 1.434 & 2.772 & 2.909 & 3.183 \\ 
  \hline
  \multirow{3}{*}{Semi, 600}         &Coverage & 0.180 & 0.938 & 0.962 & 0.800 & 0.936 & 0.962 & 0.930 & 0.944 & 0.958 \\ 
  &Est Aver&  0.154 & 0.155 & 0.157 & 0.726 & 0.729 & 0.732 & 2.385 & 2.388 & 2.391 \\
  &Lower Aver & 0.088 & 0.000 & 0.000 & 0.581 & 0.364 & 0.047 & 2.102 & 1.950 & 1.664 \\ 
  &Upper Aver & 0.220 & 0.499 & 0.834 & 0.871 & 1.094 & 1.418 & 2.667 & 2.827 & 3.119 \\ 
  \hline
\hline
  \multirow{3}{*}{Super, 1200}   &Coverage & 0.480 & 0.968 & 0.976 & 0.890 & 0.960 & 0.974 & 0.964 & 0.964 & 0.970 \\ 
  &Est Aver&   0.107 & 0.108 & 0.109 & 0.684 & 0.686 & 0.688 & 2.356 & 2.358 & 2.360\\
  &Lower Aver &  0.067 & 0.000 & 0.000 & 0.575 & 0.420 & 0.190 & 2.099 & 2.004 & 1.810 \\ 
  &Upper Aver & 0.146 & 0.355 & 0.599 & 0.793 & 0.952 & 1.185 & 2.613 & 2.712 & 2.909 \\ 
  \hline
    \multirow{3}{*}{Semi, 1200} &Coverage & 0.494 & 0.970 & 0.976 & 0.898 & 0.958 & 0.972 & 0.954 & 0.954 & 0.968 \\ 
    &Est Aver & 0.106 & 0.107 & 0.108 & 0.680 & 0.682 & 0.684 & 2.348 & 2.350 & 2.352 \\
    &Lower Aver &0.066 & 0.000 & 0.000 & 0.576 & 0.418 & 0.187 & 2.133 & 2.026 & 1.821 \\ 
    &Upper Aver & 0.145 & 0.354 & 0.598 & 0.783 & 0.946 & 1.180 & 2.563 & 2.675 & 2.883 \\ 
    \hline
    \hline
  \multirow{3}{*}{Super, 2400}   &Coverage & 0.738 & 0.972 & 0.978 & 0.916 & 0.954 & 0.972 & 0.948 & 0.944 & 0.958 \\ 
  &Est Aver&   0.083 & 0.084 & 0.084 & 0.663 & 0.663 & 0.662 & 2.340 & 2.340 & 2.340\\
  &Lower Aver & 0.058 & 0.000 & 0.000 & 0.585 & 0.472 & 0.306 & 2.154 & 2.085 & 1.945 \\ 
  &Upper Aver & 0.109 & 0.260 & 0.434 & 0.741 & 0.853 & 1.019 & 2.526 & 2.594 & 2.734 \\ 
  \hline
    \multirow{3}{*}{Semi, 2400} &Coverage & 0.742 & 0.972 & 0.978 & 0.912 & 0.962 & 0.976 & 0.950 & 0.950 & 0.960 \\ 
    &Est Aver & 0.083 & 0.083 & 0.083 & 0.661 & 0.661 & 0.661 & 2.337 & 2.337 & 2.336  \\
    &Lower Aver &0.058 & 0.000 & 0.000 & 0.587 & 0.472 & 0.306 & 2.173 & 2.098 & 1.952  \\ 
    &Upper Aver & 0.108 & 0.260 & 0.434 & 0.736 & 0.851 & 1.017 & 2.501 & 2.576 & 2.721\\ 
    \hline
\end{tabular}
}
\caption{\footnotesize  Inference for prediction accuracy $(\widehat{\beta}(\lambda)-\beta)^{\intercal}\Sigma(\widehat{\beta}(\lambda)-\beta)$. The table reports six settings, corresponding to three different sample sizes (600,1200, 2400) and the supervised and semi-supervised setting. For example, ``{Super, 600}" stands for the supervised setting with sample size $n=600$ and  ``{Semi, 600}" stands for the semi-supervised setting with sample size $n=600$. The true prediction accuracy of the three estimators $\widehat{\beta}(\lambda_0), \widehat{\beta}(5\lambda_0)$ and $\widehat{\beta}(\lambda_0)$ is reported as $0.065,0.636$ and $2.310$. Three prediction accuracy evaluators ${\rm PA}(0)$, ${\rm PA}(2)$ and ${\rm PA}(4)$ are reported, where ${\rm PA}(0)$ is the evaluator with no randomization, ${\rm PA}(2)$ is the evaluator with  randomization level $\tau_0=2$ and ${\rm PA}(4)$ is the evaluator with  randomization level $\tau_0=4$. For each setting, the row indexed with ``Coverage" reports the empirical coverage of the corresponding confidence intervals over 500 simulations; the row indexed with ``Est Aver" reports the sample average of the corresponding point estimators over 500 simulations; the rows indexed with ``Lower Aver" and `Upper Aver" report the sample averages of the lower and upper limits of interval estimators over 500 simulations. }
\label{tab: loss est}
\end{table}

\section{Additional Real Data Analysis}
\label{sec: add real}
\begin{figure}[ht]
 \centering
\includegraphics[scale=0.5]{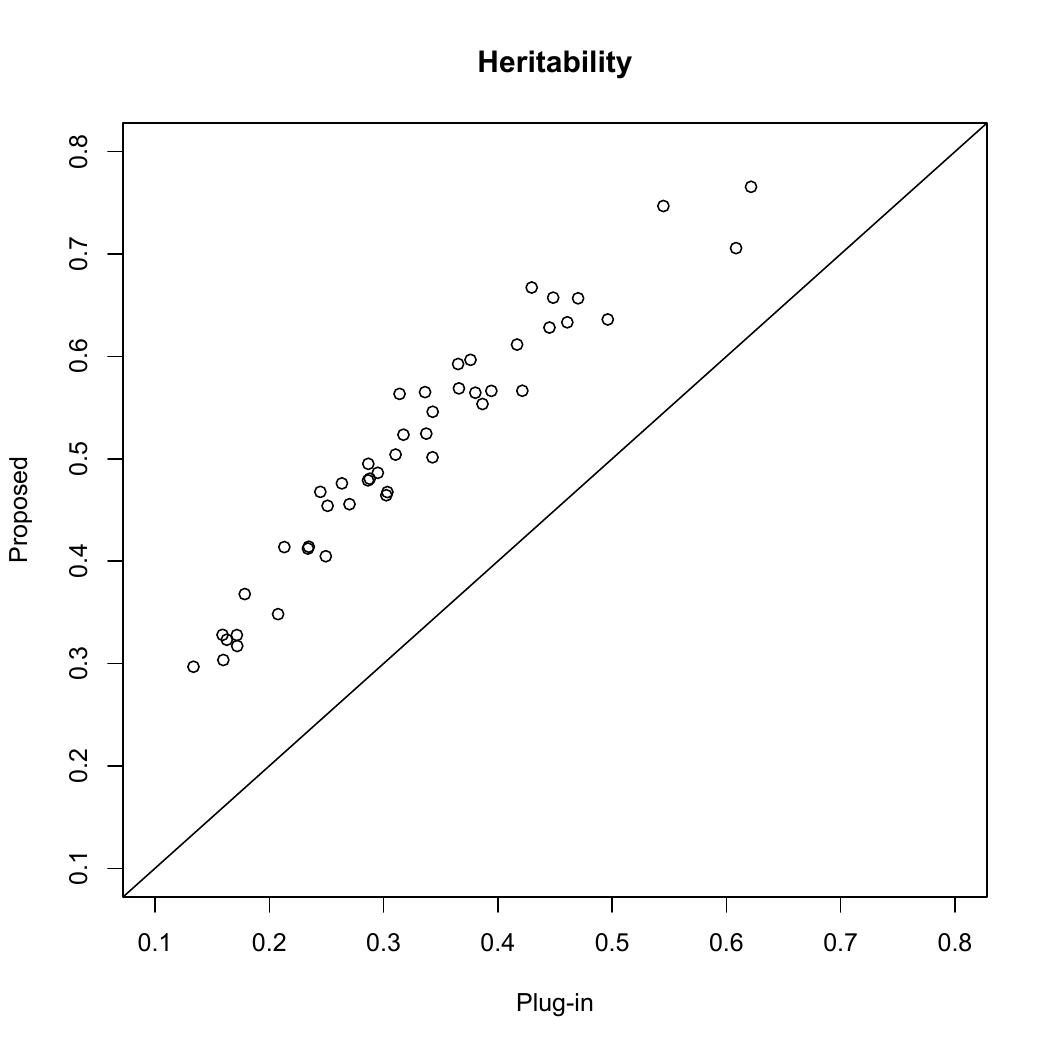}
\caption{\footnotesize  Heritability for 46 traits. The x-axis represents the heritability estimated by the plug-in estimator and the y-axis represents the heritability by the proposed CHIVE estimator; the line represents $y=x$.}
\label{fig: heritability}
\end{figure}

In the following, we compare the plug-in estimator and the CHIVE estimator to demonstrate that the CHIVE estimator adds back the missing heritability across all $46$ traits.
 We demonstrate this phenomenon in Figure \ref{fig: heritability} by comparing the CHIVE estimator and the plug-in estimators of heritability for all $46$ traits. We shall stress that all points lie above the line $y=x$ and this means that the calibration step adds back the missing heritability due to simply plugging in the Lasso estimator, where the Lasso estimator tends to ignore the genetic markers with small effects. 

